\documentclass[twoside,12pt]{report}
\usepackage{graphicx}
\usepackage{amssymb, amsmath, amsxtra}
\usepackage{epsfig,subfigure}

\usepackage{latexsym}
\usepackage{bm}

 \usepackage{ntheorem}
           \theoremstyle{plain}
                      {\theorembodyfont{\rmfamily}
                      \theoremseparator{.}
                       \newtheorem{exa}{Example}[section]
           \newtheorem{thm}{Theorem}[section]
           \theoremstyle{plain}
           
           \theoremstyle{plain} \newtheorem{lem}{Lemma}[section]
           \theoremstyle{plain}
           \newtheorem{defn}{Definition}[section]
            \newtheorem{exer}{Exercise}[section]
           \theoremstyle{plain}
           \newtheorem{rem}{Remark}[section]
            \newtheorem{prop}{Proposition}[section]}
\def\R{\mbox{I$\!$R}}
\begin{document}

\begin{center}
{\Large \textbf{NONLINEAR SELF-ADJOINTNESS IN\\[1ex]  CONSTRUCTING
  CONSERVATION LAWS}}\\[2ex]
 Nail H. Ibragimov\\
 Department of Mathematics and Science, Blekinge Institute
 of Technology,\\ 371 79 Karlskrona, Sweden
 \end{center}

 \noindent
\textbf{Abstract.} The general concept of nonlinear
 self-adjointness of differential equations is introduced.
  It includes the linear self-adjointness as a particular case. Moreover, it
 embraces the previous notions
 of self-adjoint \cite{ibr06a} and quasi self-adjoint \cite{ibr07d} nonlinear equations.
 The class of nonlinearly self-adjoint equations includes, in particular, all linear
 equations. Conservation laws associated with symmetries can be constructed for all
 nonlinearly self-adjoint differential equations and systems. The number of equations in systems can
 be different from the number of dependent variables.\\[2ex]
 \hfill

 \begin{table}[b]
  \begin{tabular}{l}\hline
 \copyright ~2010 N.H. Ibragimov.\\ First published in \textit{Archives of ALGA}, vol. 7/8, 2010-2011, pp. 1-99.\\
  \end{tabular}
 \end{table}

 \tableofcontents
  \newpage
%  \null
 % \newpage
\begin{center}
\section*{{\sc Part 1}\protect\\ Nonlinear self-adjointness}
\end{center}
\addcontentsline{toc}{chapter}{Part 1. Nonlinear self-adjointness}
 \label{concept:P1}

 \section{Preliminaries}
 %\section{Introduction}
 \label{sa:eveq.int}
  \setcounter{equation}{0}

 The concept of self-adjointness of nonlinear equations was introduced \cite{ibr06a, ibr07a}
 for constructing conservation laws associated with symmetries of differential
 equations. To extend the possibilities of the new method for constructing
 conservation laws the notion of quasi
 self-adjointness was suggested in \cite{ibr07d}.
 I introduce here the general concept of \textit{nonlinear self-adjointness}.
It embraces the previous notions of self-adjoint and quasi
self-adjoint equations and includes the linear self-adjointness as a
particular case. But the set of nonlinearly self-adjoint equations
 is essentially wider and includes, in particular, \textit{all} linear
 equations. The construction of conservation laws demonstrates a practical significance
 of the nonlinear
 self-adjointness. Namely,
 \textit{conservation laws can be associated with symmetries for all
nonlinearly self-adjoint differential equations and systems}. In
particular, this is possible for all linear equations and systems.

 \subsection{Notation}
 \label{nota}

 We will use the following notation. The independent variables are
 denoted by
 $$
 x = (x^1, \ldots, x^n).
 $$
 The dependent variables are
 $$
 u = (u^1, \ldots, u^m).
 $$
 They are used together with their first-order partial derivatives
   $u_{(1)}:$
  $$
  u_{(1)} = \{u^\alpha_i\}, \quad u^\alpha_i = D_i (u^\alpha),
  $$
 and higher-order derivatives $u_{(2)}, \ldots, u_{(s)}, \ldots,$
where
 $$
  u_{(2)} = \{u^\alpha_{ij}\}, \quad u^\alpha_{ij} = D_i D_j
  (u^\alpha),\ldots,
  $$
% up to $s$th-order  derivatives $u_{(s)}:$
  $$
  u_{(s)} = \{u^\alpha_{i_1\cdots i_s}\}, \quad u^\alpha_{i_1\cdots i_s}
  = D_{i_1} \cdots D_{i_s}(u^\alpha).
  $$
  Here $D_i$ is the total differentiation with respect to $x^i:$
 \begin{equation}
 \label{sa:cl.diff1}
  D_i = \frac{\partial}{\partial x^i} +
u^\alpha_{i}\frac{\partial}{\partial u^\alpha} +
u^\alpha_{ij}\frac{\partial}{\partial u^\alpha_j} +
 \cdots\,.
 \end{equation}

A locally analytic function $f(x, u, u_{(1)}, \ldots, u_{(k)})$ of
any finite number of the variables $x, u, u_{(1)}, u_{(2)}, \ldots$
is called a \textit{differential function}. The set of all
differential functions is denoted by ${\cal A}.$ For more details
see \cite{ibr99}, Chapter 8.

 \subsection{Linear self-adjointness}
% \subsection{Definition of the adjoint equations to nonlinear equations}
 \label{lsa}

 Recall that the adjoint operator $F^*$ to a linear operator $F$ in a Hilbert space $H$ with a scalar
 product $(u, v)$ is defined by
 \begin{equation}
 \label{ad.H}
 (Fu, v) = (u, F^*v), \quad u, v \in H.
 \end{equation}

 Let us consider, for the sake of simplicity, the case of one dependent variable $u$ and
denote by $H$ the Hilbert space of real valued functions $u(x)$ such
that
 $u^2(x)$ is integrable. The scalar product is given by
 $$
 (u, v) = \int_{\footnotesize \R^n} u(x) v(x) dx.
 $$
 Let $F$ be a linear differential operator in $H.$
 Its action on the dependent variable $u$ is denoted by $F[u].$
 The definition (\ref{ad.H}) of the adjoint
 operator $F^*$ to $F,$
 $$
 (F[u], v) = (u, F^*[v]),
 $$
can be written, using the divergence theorem, in the simple form
 \begin{equation}
 \label{clad}
 v F[u]-uF^*[v] = D_i(p^i),
  \end{equation}
where $v$ is a new dependent variable, and $p^i$ are any functions
of $x,\,u,\,v,\, u_{(1)},\,v_{(1)},\ldots .$

 It is manifest from Eq. (\ref{clad}) that the operators $F$ and $F^*$ are mutually adjoint,
 \begin{equation}
 \label{adad}
 \left(F^*\right)^* = F.
  \end{equation}
 In other words, the adjointness of linear operators is a
 \textit{symmetric relation}.

 %\begin{defn}
 %\label{sadlin.def}
 The linear operator $F$ is said to be self-adjoint if $F^* = F.$ In this case
we say that the equation $F[u] = 0$ is self-adjoint. Thus, the
self-adjointness of a linear equation $F[u] = 0$ can be expressed by
the equation
 \begin{equation}
 \label{sad-lin}
 F^*[v]\Big|_{v = u} =  F[u].
  \end{equation}
 %\end{defn}

 \subsection{Adjoint equations to nonlinear differential equations}
% \subsection{Definition of the adjoint equations to nonlinear equations}
 \label{sa:eveq.1}

 Let us consider a  system of $m$ differential equations (linear or nonlinear)
 \begin{equation}
 \label{sa:eveq.int.1}
 F_\alpha \big(x, u, u_{(1)}, \ldots, u_{(s)}\big) = 0, \quad
 \alpha = 1, \ldots, m,
 \end{equation}
 with $m$ dependent variables $u = (u^1, \ldots, u^m).$ Eqs. (\ref{sa:eveq.int.1})
 involve the partial derivatives
   $u_{(1)}, \ldots, u_{(s)}$ up to order $s.$
 \begin{defn}
 \label{sa:adj}
 The {\it adjoint equations} to Eqs. (\ref{sa:eveq.int.1}) are given by
 \begin{equation}
 \label{sa:eveq.int.2}
 F^*_\alpha \big(x, u, v, u_{(1)}, v_{(1)}, \ldots, u_{(s)}, v_{(s)}\big)  = 0, \quad
 \alpha = 1, \ldots, m,
 \end{equation}
with
 \begin{equation}
 \label{sa:eveq.int.3}
 F^*_\alpha \big(x, u, v,u_{(1)}, v_{(1)}, \ldots, u_{(s)}, v_{(s)}\big) =
  \frac{\delta {\cal L}}{\delta u^\alpha}\,,
 \end{equation}
 where ${\cal L}$ is the \textit{formal Lagrangian} for Eqs.
 (\ref{sa:eveq.int.1}) defined by \footnote{See
 \cite{ibr06a}. An approach in terms of variational principles is developed in \cite{ath-hom75}.}
 \begin{equation}
 \label{sa:Lag}
 {\cal L} = v^\beta F_\beta \equiv \sum_{\beta= 1}^m v^\beta
 F_\beta.
 \end{equation}
 Here $v = (v^1, \ldots, v^m)$ are new dependent variables,
 $v_{(1)}, \ldots, v_{(s)}$ are their derivatives, e.g.
 $v_{(1)} = \{v^\alpha_i\}, \ v^\alpha_i = D_i (v^\alpha).$
 We use $\delta/\delta u^\alpha$ for the Euler-Lagrange
operator
 $$
 \frac{\delta}{\delta u^\alpha} =
\frac{\partial}{\partial u^\alpha} + \sum_{s=1}^\infty (-1)^s
D_{i_1}\cdots D_{i_s}\,\frac{\partial}{\partial u^\alpha_{i_1\cdots
i_s}}\,,\quad \alpha=1, \ldots, m,
 $$
so that
 $$
 \frac{\delta (v^\beta F_\beta)}{\delta u^\alpha} =
 \frac{\partial (v^\beta F_\beta)}{\partial u^\alpha} -
 D_i \bigg(\frac{\partial (v^\beta F_\beta)}{\partial u^\alpha_i} \bigg)
 + D_i D_k \bigg(\frac{\partial (v^\beta F_\beta)}{\partial u^\alpha_{ik}} \bigg)
 - \cdots\,.
 $$
 The total differentiation (\ref{sa:cl.diff1}) is extended to the new dependent variables:
 \begin{equation}
 \label{sa:cl.diff2}
  D_i = \frac{\partial}{\partial x^i} +
 u^\alpha_{i}\frac{\partial}{\partial u^\alpha} +
 v^\alpha_{i}\frac{\partial}{\partial v^\alpha} +
 u^\alpha_{ij}\frac{\partial}{\partial u^\alpha_j} +
 v^\alpha_{ij}\frac{\partial}{\partial v^\alpha_j} +
 \cdots\,.
 \end{equation}
 \end{defn}

 %\begin{rem}
 %\label{nsa.rem1}
 The adjointness of \textit{nonlinear} equations is not a symmetric
 relation. In other words, nonlinear equations, unlike the linear
 ones, do not obey the condition (\ref{adad}) of mutual adjointness.
 Instead, the following equation holds:
 \begin{equation}
 \label{nsa.linearized}
 \left(F^*\right)^* = \hat F
 \end{equation}
 where $\hat F$ is the \textit{linear approximation} to $F$
 %second adjoint expression $\left(F^*\right)^*$ for $F$
 %coincides with the linearization $\hat F$ of $F$
 defined as follows.
 We use the temporary notation $F[u]$ for the left-hand side of Eq.
 (\ref{sa:eveq.int.1}) and consider $F[u + w]$ by letting $w \ll
 1.$ Then neglecting the nonlinear terms in $w$ we define $\hat F$
 by the equation
 \begin{equation}
 \label{nsa.linearize1}
 F[u + w] \approx F[u] + \hat F[w]
 \end{equation}
 For linear equations we have $\hat F = F,$ and hence Eq.
 (\ref{nsa.linearized}) is identical with Eq. (\ref{adad}).

 Let us illustrate Eq. (\ref{nsa.linearized}) by the equation
 \begin{equation}
 \label{nsa.linearize2}
 F \equiv  u_{xy} - \sin u = 0.
 \end{equation}
 Eq. (\ref{sa:eveq.int.3}) yields
 \begin{equation}
 \label{nsa.linearize3}
 F^* \equiv   \frac{\delta}{\delta u}[v(u_{xy} - \sin u)]
 = v_{xy} - v \cos u
 \end{equation}
 and
 \begin{equation}
 \label{nsa.linearize4}
 \left(F^*\right)^* \equiv  \frac{\delta}{\delta v}[w(v_{xy} - v \cos u)] = w_{xy} - w \cos u.
 \end{equation}
 Let us find $\hat F$ by using Eq. (\ref{nsa.linearize1}). Since $\sin w \approx w, \ \cos w \approx 1$
 when $w \ll 1,$ we have
\begin{align}
 F [u + w] \equiv & \ (u + w)_{xy} - \sin (u + w)\notag\\[.5ex]
 & = u_{xy} + w_{xy} - \sin u \cos w - \sin w \cos u\notag\\[.5ex]
 & \approx u_{xy} - \sin u + w_{xy} -  w \cos u,\notag\\[.5ex]
 & = F[u] + w_{xy} -  w \cos u.\notag
 \end{align}
 Hence, by (\ref{nsa.linearize1}) and
 (\ref{nsa.linearize4}), we have
 \begin{equation}
 \label{nsa.linearize5}
 \hat F[w] = w_{xy} -  w \cos u = \left(F^*\right)^*
 \end{equation}
 in accordance with Eq. (\ref{nsa.linearized}).
 %\end{rem}

 \subsection{The case of one dependent variable}
 \label{sa:eveq.2}

 Let us consider the differential equation
 \begin{equation}
 \label{saeq.1}
 F\big(x, u, u_{(1)}, \ldots, u_{(s)}\big) = 0
  \end{equation}
 with one dependent variable $u$ and any number of independent
 variables. In this case Definition \ref{sa:adj} of the adjoint equation is written
  \begin{equation}
 \label{saeq.2}
 F^*\big(x, u, v, u_{(1)}, v_{(1)}, \ldots, u_{(s)}, v_{(s)}\big)  = 0,
 \end{equation}
where
 \begin{equation}
 \label{saeq.3}
 F^*\big(x, u, v,u_{(1)}, v_{(1)},\ldots, u_{(s)}, v_{(s)}\big) =
  \frac{\delta (v F)}{\delta u}\,\cdot
 \end{equation}

 \subsection{Construction of adjoint equations to linear equations}
 %\subsection{Comparison with classical definition for linear equations}
 \label{sa:eveq.3}

 The following statement has been formulated in \cite{ibr06a, ibr07a}.
 \begin{prop}
 \label{lnad.prop1}
 In the case of linear differential equations and systems, the adjoint equations determined by
  Eq. (\ref{sa:eveq.int.3}) and by Eq. (\ref{clad})
 coincide.
 \end{prop}
 \textbf{Proof.} The proof is based on the statement
(see Proposition \ref{test:div} in Section \ref{test}) that a
function $Q(u, v)$ is a divergence, i.e.
 $Q = D_i(h^i),$ if and only if
 \begin{equation}
 \label{cladd}
 \frac{\delta Q}{\delta u^\alpha} = 0, \quad  \frac{\delta Q}{\delta v^\alpha} = 0, \quad \alpha = 1, \ldots, m.
 \end{equation}
 Let the adjoint operator $F^*$ be constructed according  to Eq.
 (\ref{clad}). Let us consider the case of many dependent variables and write Eq.
 (\ref{clad}) as follows:
 \begin{equation}
 \label{cladm}
  v^\beta F_\beta [u] = u^\beta F^*_\beta [v] + D_i(p^i).
  \end{equation}
 Applying to (\ref{cladm}) the variational differentiations and using  Eqs. (\ref{cladd})
we obtain
 $$
 \frac{\delta (v^\beta F_\beta [u])}{\delta u^\alpha} =
 \delta^\beta_\alpha F^*_\beta [v] \equiv F^*_\alpha [v].
 $$
 Hence, (\ref{sa:eveq.int.3}) coincides with $F^*_\alpha [v]$ given
 by (\ref{clad}).

 Conversely, let $F^* [v]$ be given by (\ref{sa:eveq.int.3}),
 $$
 F^*_\beta [v] = \frac{\delta (v^\gamma F_\gamma [u])}{\delta
u^\beta}\,\cdot
 $$
 Consider the expression $Q$ defined by
 $$
 Q =  v^\beta F_\beta [u] -  u^\beta F^*_\beta [v] \equiv v^\beta F_\beta [u] -
 u^\beta  \frac{\delta (v^\gamma F_\gamma [u])}{\delta u^\beta}\,\cdot
 $$
 Applying to the first expression for $Q$ the variational differentiations $\delta/\delta u^\alpha$ we obtain
 $$
 \frac{\delta Q}{\delta u^\alpha} = \frac{\delta (v^\beta F_\beta [u])}{\delta u^\alpha}
 - \delta^\beta_\alpha F^*_\beta [v] \equiv F^*_\alpha [v] - \delta^\beta_\alpha F^*_\beta [v] =
0.
 $$
 Applying $\delta/\delta v^\alpha$
 to the second expression for $Q$ we obtain
 $$
 \frac{\delta Q}{\delta v^\alpha} = \delta^\beta_\alpha F_\beta [u]
 - \frac{\delta}{\delta v^\alpha} \left[u^\beta  \frac{\delta (v^\gamma F_\gamma [u])}{\delta u^\beta}\right]
 \equiv F_\alpha [u]- \frac{\delta}{\delta v^\alpha} \left[u^\beta  \frac{\delta
(v^\gamma F_\gamma [u])}{\delta u^\beta}\right]\,\cdot
 $$
The reckoning shows that
 \begin{equation}
 \label{lnad:eq1}
 \frac{\delta}{\delta v^\alpha} \left[u^\beta  \frac{\delta (v^\gamma F_\gamma [u])}{\delta
 u^\beta}\right] = F_\alpha [u].
  \end{equation}
 Thus $Q$ solves Eq. (\ref{cladd}) and hence Eq. (\ref{cladm}) is
 satisfied. This completes the proof.
 \begin{rem}
 \label{lnad.rem1}
 Let us discuss the proof of Eq. (\ref{lnad:eq1}) in the case of a econd-order linear operator for one dependent
 variable:
 $$
 F[u] = a^{ij}(x)u_{ij} +b^i(x)u_i +c(x) u.
 $$
 Then we have:
 $$
 u \frac{\delta (v F [u])}{\delta u} = u \left[c v - v D_i(b^i) + v D_i D_j(a^{ij}) - b^i v_i
+ 2 v_i D_j(a^{ij}) + a^{ij} v_{ij}\right].
 $$
Whence, after simple calculations we obtain
 $$
 \frac{\delta}{\delta v}\left[ u \frac{\delta (v F [u])}{\delta u}\right] = \left[c u + b^i u_i + a^{ij} u_{ij}\right]
 + \left\{D_i D_j(a^{ij} u) - D_i(a^{ij} u_j) - D_i [u
D_j(a^{ij})]\right\}
 $$
 and, noting that the expression in the braces vanishes, arrive at
 Eq. (\ref{lnad:eq1}).
 \end{rem}

 Let us illustrate Proposition \ref{lnad.prop1} by the following simple example.
 \begin{exa}
 \label{sa:eveq.exa1}
 Consider the heat equation
 \begin{equation}
 \label{sa:eveq.exa1:eq1}
 F[u] \equiv u_t - u_{xx}= 0
  \end{equation}
 and construct the adjoint operator to the linear operator
 \begin{equation}
 \label{sa:eveq.exa1:eq2}
 F = D_t - D_x^2
 \end{equation}
 by using Eq. (\ref{clad}). Noting that
 \begin{align}
 & v u_t = D_t(u v) - u v_t, \notag \\[1ex]
 & v u_{xx} = D_x(v u_x) - v_x u_x
 = D_x(v u_x - u v_x) + u v_{xx}\notag
 \end{align}
 we have:
 $$
 v F[u] \equiv v(u_t - u_{xx}) = u(- v_t - v_{xx}) + D_t(u v) + D_x(u v_x - v
 u_x).
 $$
 Hence,
 $$
 v F[u] - u (- v_t - v_{xx}) = D_t(u v) + D_x(u v_x - v u_x).
 $$
 Therefore, denoting $t = x^1, \ x = x^2,$ we obtain
  Eq. (\ref{clad}) with $F^*[v] = - v_t - v_{xx}$ and
  $p^1 = u v, \ p^2 = u v_x - v u_x.$ Thus,
the adjoint operator to the linear operator (\ref{sa:eveq.exa1:eq2})
is
 \begin{equation}
 \label{sa:eveq.exa1:eq3}
 F^* = - D_t - D_x^2
 \end{equation}
 and the adjoint equation
  to the heat equation (\ref{sa:eveq.exa1:eq1}) is written $- v_t - v_{xx}=
0,$ or
 \begin{equation}
 \label{sa:eveq.exa1:eq4}
 v_t + v_{xx}= 0.
  \end{equation}

  The derivation of the adjoint equation (\ref{sa:eveq.exa1:eq4})
 and the adjoint operator (\ref{sa:eveq.exa1:eq3}) by the definition
 (\ref{saeq.3}) is much simpler.
 Indeed, we have:
  $$
   F^* = \frac{\delta (v u_t - v u_{xx})}{\delta u} =
   - D_t(v) - D_x^2(v) = - (v_t + v_{xx}).
  $$
 \end{exa}

 \subsection{Self-adjointness and quasi self-adjointness}
 %\subsection{Self-adjointness of linear and nonlinear equations}
 \label{sa:eveq.4}

 Recall that a linear differential operator $F$  is called a \textit{self-adjoint operator} if
 it is identical with its adjoint operator, $F = F^*.$
Then the equation $F[u] = 0$ is also said to be self-adjoint. Thus,
 the self-adjointness of a \textit{linear differential equation} $F[u] = 0$ means that
 the adjoint equation $F^*[v] = 0$ coincides with $F[u] = 0$ upon
 the substitution $v = u.$
 This property has been extended to nonlinear equations in \cite{ibr06a}. It will
 be called here the \textit{strict self-adjointness} and defined as follows.
  \begin{defn}
 \label{saeq:def}
 We say that the differential equation (\ref{saeq.1})
 is \textit{strictly self-adjoint} if the adjoint equation
 (\ref{saeq.2}) becomes equivalent to the original equation (\ref{saeq.1})
 upon the substitution
  \begin{equation}
  \label{saeq.4}
 v = u.
 \end{equation}
 It means that the  equation
  \begin{equation}
  \label{saeq.5}
 F^*\big(x, u, u, \ldots, u_{(s)}, u_{(s)}\big) = \lambda\, F\big(x, u,
 \ldots, u_{(s)}\big)
  \end{equation}
 holds with a certain (in general, variable) coefficient
 $\lambda.$
 \end{defn}

 \begin{exa}
 \label{sa:eveq.KdV}
 The Korteweg-de Vries (KdV) equation
 $$
 u_t = u_{xxx} + u u_x
 $$
 is strictly self-adjoint \cite{ibr07a}. Indeed, its adjoint equation (\ref{saeq.2}) has the form
 $$
 v_t = v_{xxx} + u v_x
 $$
 and coincides with the KdV equation  upon the substitution (\ref{saeq.4}).
 \end{exa}

 In the case of linear equations the strict self-adjointness is identical with the usual self-adjointness of linear
 equations.
 \begin{exa}
 \label{sa:eveq.exa2}
 Consider the linear equation
 \begin{equation}
 \label{sa:eveq.exa2:eq1}
 u_{tt} + a(x) u_{xx} + b(x) u_x + c(x) u = 0.
 \end{equation}
 According to Eqs. (\ref{saeq.2})-(\ref{saeq.3}), the adjoint
equation to Eq. (\ref{sa:eveq.exa2:eq1}) is written
   $$
 \frac{\delta}{\delta u} \{v [u_{tt} + a(x) u_{xx} + b(x) u_x + c(x)u]\}
 \equiv
   D^2_t(v) + D_x^2(a v) - D_x(b v) + c v = 0.
  $$
 %  $$
 %  \frac{\delta [v (u_{tt} + a(x) u_{xx} + b(x) u_x + c(x) u)]}{\delta u}
%\equiv
 %  D^2_t(v) + D_x^2(a v) - D_x(b v) + c v = 0.
 % $$
 Upon substituting  $v = u$ and performing the differentiations it becomes
 \begin{equation}
 \label{sa:eveq.exa2:eq2}
 u_{tt} + a u_{xx} + (2 a' - b) u_x + (a'' - b' + c) u = 0.
 \end{equation}
According to Definition \ref{saeq:def}, Eq. (\ref{sa:eveq.exa2:eq1})
is strictly self-adjoint if
 Eq. (\ref{sa:eveq.exa2:eq2}) coincides with Eq.
(\ref{sa:eveq.exa2:eq1}). This is possible if
 \begin{equation}
 \label{sa:eveq.exa2:eq3}
 b(x) = a'(x).
 \end{equation}
 \end{exa}

 Definition \ref{saeq:def} is too restrictive. Moreover, it is inconvenient in the case of systems with several
 dependent variables $u = (u^1, \ldots, u^m)$  because in this case Eq. (\ref{saeq.4})
  is not uniquely determined as it is clear from the following example.
 \begin{exa}
 \label{sa:eveq.exa3}
 Let us consider the system of two equations
 \begin{align}
 & u^1_y + u^2 u^2_x - u^2_t = 0, \notag \\[1ex]
 & u^2_y - u^1_x = 0 \label{sa:eveq.exa3.eq1}
 \end{align}
 with two  dependent  variables, $u = (u^1, u^2),$ and three
 independent variables $t, x, y.$ Using the formal Lagrangian
 (\ref{sa:Lag})
 $$
 {\cal L} = v^1 (u^1_y + u^2 u^2_x - u^2_t) + v^2 (u^2_y - u^1_x)
 $$
 and Eqs. (\ref{sa:eveq.int.3}) we write the adjoint equations
 (\ref{sa:eveq.int.2}), changing their sign, in the form
 \begin{align}
 & v^2_y + u^2 v^1_x - v^1_t = 0, \notag \\[1ex]
 & v^1_y - v^2_x = 0.\label{sa:eveq.exa3.eq2}
 \end{align}
 If we use here the substitution (\ref{saeq.4}), $v = u$ with $v = (v^1, v^2),$ i.e. let
 $$
 v^1 = u^1, \quad v^2 = u^2,
 $$
 then the adjoint system (\ref{sa:eveq.exa3.eq2}) becomes
 \begin{align}
 & u^2_y + u^2 u^1_x - u^1_t = 0, \notag \\[1ex]
 & u^1_y - u^2_x = 0,\notag
 \end{align}
 which is not connected with the system (\ref{sa:eveq.exa3.eq1}) by the equivalence relation (\ref{saeq.5}).
 But if we set
 $$
 v^1 = u^2, \quad v^2 = u^1,
 $$
 the adjoint system (\ref{sa:eveq.exa3.eq2}) coincides with the original
 system (\ref{sa:eveq.exa3.eq1}).
 \end{exa}

 The concept of
 quasi self-adjointness
   generalizes  Definition \ref{saeq:def} and  is more convenient for dealing with systems
 (\ref{sa:eveq.int.1}). This concept was formulated in \cite{ibr07d} as follows.

 The system (\ref{sa:eveq.int.1}) is \textit{quasi
 self-adjoint} if the adjoint system (\ref{sa:eveq.int.2}) becomes
equivalent to the original system (\ref{sa:eveq.int.1}) upon a
substitution
  \begin{equation}
  \label{qsa0.eq1}
 v = \varphi(u)
 \end{equation}
  such that its derivative does not vanish in a certain domain of $u,$
  \begin{equation}
  \label{qsa0.eq2}
 \varphi'(u)  \not= 0, \quad {\rm where} \quad \varphi'(u) =
 \Big|\!\Big|\frac{\partial \varphi^\alpha(u)}{\partial u^\beta}\Big|\!\Big|.
 \end{equation}
 \begin{rem}
 \label{qsa0:rem1}
 The substitution (\ref{qsa0.eq1}) defines a mapping
 $$
 v^\alpha = \varphi^\alpha(u), \quad \alpha = 1, \ldots, m,
 $$
 from the $m$-dimensional space of variables $u = (u^1, \ldots,
u^m)$
 into the $m$-dimensional space of variables $v = (v^1,\ldots, v^m).$
  It is assumed that this mapping  is continuously differentiable.
 The condition (\ref{qsa0.eq2}) guarantees that
 it is invertible, and hence Eqs. (\ref{sa:eveq.int.2}) and (\ref{sa:eveq.int.1})
 are equivalent. The equivalence means
 that the following equations hold with certain coefficients
 $\lambda^\beta_\alpha:$
 \begin{equation}
  \label{qsa.2}
 F^*_{\alpha} \big(x, u, \varphi, \ldots, u_{(s)}, \varphi_{(s)}\big) =
 \lambda^\beta_\alpha \, F_{\beta}\big(x, u,
 \ldots, u_{(s)}\big), \quad \alpha = 1, \ldots, m,
  \end{equation}
 where
  \begin{equation}
  \label{qsa.2notat}
   \varphi = \{\varphi^\alpha(u)\}, \quad \varphi_{(\sigma)}
  = \{D_{i_1} \cdots D_{i_\sigma}\big(\varphi^\alpha(u)\big)\}, \quad \sigma = 1, \ldots, s.
  \end{equation}
 It can be shown that the matrix $\|\lambda^\beta_\alpha\|$ is invertible due to the condition (\ref{qsa0.eq2}).
 \end{rem}

 \begin{exa}
 \label{qsa:exaITT}
 The quasi self-adjointness of nonlinear wave equations of the form
 $$
 u_{tt}-u_{xx} = f(t, x, u, u_t, u_x)
 $$
 is investigated in \cite{ibr-tor-tra10}. The results of
 the paper \cite{ibr-tor-tra10} show that, e.g.
 the equation
  \begin{equation}
  \label{qsa.exaITT:eq1}
 u_{tt}-u_{xx} + u_t^2-u_x^2 = 0
  \end{equation}
 is quasi self-adjoint and that in this case
 the substitution (\ref{qsa0.eq1}) has the form
  \begin{equation}
  \label{qsa.exaITT:eq2}
 v = {\rm e}^u.
 \end{equation}
 Indeed, the adjoint equation to Eq. (\ref{qsa.exaITT:eq1}) is
 written
 \begin{equation}
 \label{qsa.exaITT:eq1q}
 v_{tt}-v_{xx} - 2 v u_{tt} - 2 u_t v_t + 2 v u_{xx} + 2 u_x v_x = 0.
 \end{equation}
 After the substitution (\ref{qsa.exaITT:eq2}) the left-hand side of Eq. (\ref{qsa.exaITT:eq1q})
takes the form
 (\ref{qsa.2}):
 \begin{equation}
 \label{qsa.exaITT:eq1s}
 v_{tt}-v_{xx} - 2 v u_{tt} - 2 u_t v_t + 2 v u_{xx} + 2 u_x v_x =
 - {\rm e}^u [u_{tt}-u_{xx} + u_t^2-u_x^2].
 \end{equation}
It is manifest from Eq. (\ref{qsa.exaITT:eq1s}) that $v$ given by
(\ref{qsa.exaITT:eq2}) solves the adjoint equation
(\ref{qsa.exaITT:eq1q}) if one replaces $u$ by any solution of Eq.
(\ref{qsa.exaITT:eq1}).
 \end{exa}

 In constructing conservation laws one can relax the condition
 (\ref{qsa0.eq2}). Therefore I generalize the previous definition of quasi self-adjointness as follows.

 \begin{defn}
 \label{qsa:def}
 The system (\ref{sa:eveq.int.1}) is said to be
\textit{quasi self-adjoint} if the adjoint equations
(\ref{sa:eveq.int.2}) are satisfied for all solutions $u$ of the
original system (\ref{sa:eveq.int.1}) upon a substitution
 \begin{equation}
 \label{qsa.1}
 v^\alpha = \varphi^\alpha(u), \quad \alpha = 1, \ldots, m,
 \end{equation}
 such that
  \begin{equation}
  \label{qsa.eq3}
 \varphi(u) \not= 0.
 \end{equation}
  In other words, the equations (\ref{qsa.2}) hold after the substitution (\ref{qsa.1}),
 where not all $\varphi^\alpha(u)$ vanish
 simultaneously.
 \end{defn}
  \begin{rem}
 \label{qsa0:rem2}
  The condition (\ref{qsa.eq3}), unlike (\ref{qsa0.eq2}), does not guarantee
 the equivalence of Eqs. (\ref{sa:eveq.int.2}) and (\ref{sa:eveq.int.1}) because the matrix
 $\|\lambda^\beta_\alpha\|$ may be singular.
 \end{rem}

%%%%%%%%%%%%%%%%%%%%%%%%%%%%%%%%%%%%%%%%%%%%%%%%%%%%%%%%%
% \begin{defn}
% \label{qsa:def}
% The system (\ref{sa:eveq.int.1}) is said to be
% \textit{quasi self-adjoint} if the adjoint equations
% (\ref{sa:eveq.int.2}) are satisfied for all solutions $u$ of the
% original system (\ref{sa:eveq.int.1}) upon a substitution
% \begin{equation}
% \label{qsa.1}
% v^\alpha = \varphi^\alpha(u), \quad \alpha = 1, \ldots, m,
% \end{equation}
% with certain functions $\varphi^\alpha(u)$ such that not all $\varphi^\alpha(u)$ vanish
% simultaneously.
%  In other words, the equations (\ref{qsa.2}) hold,
% where the notation (\ref{qsa.2notat}) is used, and the following
% condition is satisfied instead of  (\ref{qsa0.eq2}):
% \begin{equation}
%  \label{qsa.eq3}
% \varphi(u) \not= 0.
% \end{equation}
% The condition (\ref{qsa.eq3}) does not guarantee that the matrix
% $\|\lambda^\beta_\alpha\|$ is invertible.
% \end{defn}

% \begin{rem}
% \label{qsa:rem1}
% It has been required in \cite{ibr07d} that the map (\ref{qsa.1})
% is convertible, i.e. its derivative does not vanish. In Definition \ref{qsa:def}
% this condition is replaced  by the relaxed
% condition that the map (\ref{qsa.1}) is not identically zero.
% \end{rem}
 \begin{exa}
 \label{qsa:exa1}
 It is well known that the linear heat equation (\ref{sa:eveq.exa1:eq1}) is not self-adjoint
(\textit{not strictly self-adjoint} in the sense of
 Definition \ref{saeq:def}). It is clear from Eqs. (\ref{sa:eveq.exa1:eq1})
 and (\ref{sa:eveq.exa1:eq4}). Let us test Eq. (\ref{sa:eveq.exa1:eq1}) for quasi
 self-adjointness. Letting $v = \varphi(u),$ we obtain
  $$
 v_t = \varphi' u_t, \ v_x = \varphi' u_x, \ v_{xx} = \varphi' u_{xx}
 + \varphi'' u_x^2,
 $$
 and the condition (\ref{qsa.2}) is written:
 $$
  \varphi'(u)[u_t + u_{xx}] + \varphi''(u) u_x^2 = \lambda [u_t - u_{xx}].
 $$
 Whence, comparing the coefficients of $u_t$ in both sides, we
obtain
 $\lambda = \varphi'(u).$ Then the above equation becomes
 $$
  \varphi'(u)[u_t + u_{xx}] + \varphi''(u) u_x^2 = \varphi'(u) [u_t - u_{xx}].
 $$
 This equation yields that $\varphi'(u) = 0.$
 Hence,
 Eq. (\ref{sa:eveq.exa1:eq1}) is quasi self-adjoint with the substitution
 $v = C,$ where $C$ is any non-vanishing constant.  This substitution
 does not satisfy the condition (\ref{qsa0.eq2}).
 \end{exa}
 \begin{exa}
 \label{qsa:exa2}
 Let us consider the Fornberg-Whitham equation \cite{for-whi}
 \begin{equation}
 \label{FW}
 u_t- u_{txx}-u u_{xxx}-3 u_x u_{xx}+u u_x+u_x=0.
 \end{equation}
 Eqs. (\ref{saeq.2})-(\ref{saeq.3}) give the following adjoint equation:
 \begin{equation}
 \label{FW:adj}
 F^* \equiv - v_t + v_{txx} + u v_{xxx} - u v_x - v_x = 0.
 \end{equation}
 It is manifest from the equations  (\ref{FW}) and (\ref{FW:adj}) that the Fornberg-Whitham equation
 is not strictly self-adjoint. Let us test it for quasi
 self-adjointness. Inserting in (\ref{FW:adj}) the substitution $v = \varphi(u)$
and its derivatives
  $$
 v_t = \varphi' u_t, \quad  v_x = \varphi' u_x, \quad  v_{xx} = \varphi' u_{xx}
 + \varphi'' u_x^2, \quad  v_{tx} = \varphi' u_{tx}
 + \varphi'' u_t u_x, \ldots\,,
 $$
 then writing the condition (\ref{qsa.2}) and comparing the coefficients for $u_t, \ u_{tx}, \ u_{xx}, \ldots$
 one can verify that $\varphi'(u) = 0.$
 Hence, Eq. (\ref{FW}) is  quasi self-adjoint but does not satisfy the condition (\ref{qsa0.eq2}).
 \end{exa}

 \section{Strict self-adjointness via multipliers}
 %\subsection{Multipliers and self-adjointness}
 \label{samul}
 \setcounter{equation}{0}

 It is commonly known that numerous linear equations used in
 practice, e.g. linear evolution equations, are not self-adjoint in the
 classical meaning of the self-adjointness. Likewise, useful
 nonlinear equations such as the nonlinear heat equation, the
 Burgers equation, etc. are not strictly self-adjoint. We will see here that
  these and  many other equations
 can be rewritten in a strictly self-adjoint equivalent form by
 using multipliers. The general discussion of this approach will be
  given in Section \ref{gsa:6}.

 \subsection{Motivating examples}
 \label{samul:1}

 \begin{exa}
 \label{Burg:mult.exa1}
 Let us consider the following second-order nonlinear equation
 \begin{equation}
 \label{Burg:mult.eq1}
 u_{xx} +  f(u) u_x - u_t = 0.
 \end{equation}
 Its adjoint equation (\ref{saeq.2}) is written
 \begin{equation}
 \label{Burg:mult.eq2}
 v_{xx} -  f(u) v_x + v_t = 0.
 \end{equation}
 It is manifest that the substitution $v = u$ does not map Eq.
 (\ref{Burg:mult.eq2})  into Eq. (\ref{Burg:mult.eq1}). Hence
 Eq. (\ref{Burg:mult.eq1}) is not strictly self-adjoint.

Let us clarify if
 Eq. (\ref{Burg:mult.eq1}) can be
written in an equivalent form
 \begin{equation}
 \label{Burg:mult.eq3}
 \mu(u) [u_{xx} +  f(u) u_x - u_t] = 0
 \end{equation}
 with a certain multiplier $ \mu(u)\not= 0$ so that
 Eq. (\ref{Burg:mult.eq3}) is strictly self-adjoint. The formal Lagrangian for Eq.
(\ref{Burg:mult.eq3}) is
 $$
 {\cal L} = v \mu(u) [u_{xx} +  f(u) u_x - u_t].
 $$
 We have:
 \begin{align}
 \frac{\delta {\cal L}}{\delta u} & = D_x^2 [\mu(u) v] - D_x [\mu(u) f(u) v]
 + D_t[\mu(u) v]\notag\\
 & + \mu'(u) v [u_{xx} +  f(u) u_x - u_t] + \mu(u) f'(u) v u_x\,,\notag
 \end{align}
 whence, upon performing the differentiations,
 $$
  \frac{\delta {\cal L}}{\delta u} = \mu v_{xx} + 2 \mu' v u_{xx}
 + 2 \mu' u_x  v_x + \mu'' v u_x^2 - \mu f v_x + \mu v_t\,.
 $$
 The strict self-adjointness requires that
 $$
 \frac{\delta {\cal L}}{\delta u}\bigg|_{v = u} = \lambda [u_{xx} +  f(u) u_x -
u_t].
 $$
 This provides the following equation for the unknown multiplier
 $\mu (u):$
 \begin{equation}
 \label{Burg:mult.eq4}
 (\mu + 2 u \mu') u_{xx}
 + (2 \mu' + u \mu'') u_x^2 - \mu f u_x + \mu u_t
 = \lambda [u_{xx} +  f(u) u_x - u_t].
 \end{equation}
 Since the right side of Eq. (\ref{Burg:mult.eq4}) does not contain $u_x^2$
 we should have $2 \mu' + u \mu'' = 0,$ whence $\mu = C_1 u^{-1} + C_2.$
Furthermore, comparing the coefficients of $u_t$ in both sides of
Eq. (\ref{Burg:mult.eq4}) we obtain $\lambda = - \mu.$ Now Eq.
(\ref{Burg:mult.eq4}) takes the form
 $$
(C_2 - C_1 u^{-1}) u_{xx} - (C_1 u^{-1} + C_2) f u_x
 = - (C_1 u^{-1} + C_2) [u_{xx} +  f(u) u_x]
 $$
 and yields $C_2 = 0.$ Thus, $\mu = C_1 u^{-1}.$ We can let
 $C_1= - 1$ and formulate the result.
 \begin{prop}
 \label{Burg:mult.prop}
  Eq. (\ref{Burg:mult.eq1}) becomes strictly self-adjoint if we rewrite it in the
 form
 \begin{equation}
 \label{Burg:mult.eq5}
 \frac{1}{u}\,[u_t - u_{xx} -  f(u) u_x] = 0.
 \end{equation}
 \end{prop}
 \end{exa}

 \begin{exa}
 \label{Burg:mult.exa4}
 One can verify that the $n$th-order nonlinear evolution equation
 \begin{equation}
 \label{Burg:mult.eq12}
 \frac{\partial u}{\partial t} - f(u) \frac{\partial^n u}{\partial x^n}
  = 0,  \quad f(u) \not= 0,
 \end{equation}
  with one spatial variable $x$ is not strictly self-adjoint.
 The following statement shows that it becomes strictly self-adjoint after using an appropriate multiplier.
 \begin{prop}
 \label{Burg:mult.prop2}
 Eq.(\ref{Burg:mult.eq12}) becomes strictly self-adjoint upon rewriting it in the following
equivalent form:
\begin{equation}
 \label{Burg:mult.eq13}
 \frac{1}{u f(u)}\,\left[ \frac{\partial u}{\partial t}
- f(u) \frac{\partial^n u}{\partial x^n}\right] = 0.
 \end{equation}
 \textbf{Proof.}
 Multiplying Eq. (\ref{Burg:mult.eq12}) by $\mu(u)$ and
 taking the formal Lagrangian
 $$
 {\cal L} = v \mu(u) [u_t - f(u) u_n],
 $$
where $u_n = D_x^n(u),$ we have:
 $$
 \frac{\delta {\cal L}}{\delta u} = - D_t[\mu(u) v] - D_x^n [\mu(u)f(u) v] +
 v \mu'(u) u_t - v [\mu(u)f(u)]' u_n.
 $$
 Noting that $- D_t[\mu(u) v] + v \mu'(u) u_t = - \mu(u) v_t$ and
letting $v = u$ we obtain
 $$
 \frac{\delta {\cal L}}{\delta u}\bigg|_{v = u} = - \mu(u) u_t
 - D_x^n [\mu(u)f(u) u] - [\mu(u)f(u)]' u u_n .
 $$
 If we take $\mu(u) = [u f(u)]^{-1},$ then $\mu(u)f(u) u = 1, \ \mu(u)f(u) = u^{-1},$ and hence
 $$
 \frac{\delta {\cal L}}{\delta u}\bigg|_{v = u}
= - \frac{1}{u f(u)}\, [u_t - f(u) u_n].
 $$
Thus, Eq. (\ref{Burg:mult.eq13}) satisfies the strict
self-adjointness condition (\ref{saeq.5})  with $\lambda = - 1.$
 \end{prop}
 \end{exa}

 \subsection{Linear heat equation}
 \label{samul:2}

 Taking in (\ref{Burg:mult.eq5}) $f (u) = 0,$ we rewrite the classical linear heat equation $u_t = u_{xx}$
in the following strictly self-adjoint form:
 \begin{equation}
 \label{Burg:mult.eq7}
 \frac{1}{u}\,[u_t - u_{xx}] = 0.
 \end{equation}
 This result can be extended to the heat equation
 \begin{equation}
 \label{Burg:mult.eq8}
  u_t - \Delta u = 0,
 \end{equation}
where $\Delta u$ is the Laplacian with $n$ variables $x = (x^1,
\ldots, x^n).$ Namely, the  strictly self-adjoint form of Eq.
(\ref{Burg:mult.eq8}) is
 \begin{equation}
 \label{Burg:mult.eq9}
 \frac{1}{u}\,[u_t - \Delta u] = 0.
 \end{equation}
 Indeed, the formal Lagrangian (\ref{sa:Lag}) for Eq.
(\ref{Burg:mult.eq9}) has the form
 $$
 {\cal L} = \frac{v}{u}\,[u_t - \Delta u].
 $$
Substituting it in (\ref{saeq.3}) we obtain
 $$
 F^* = - D_t\left(\frac{v}{u}\right) - \Delta \left(\frac{v}{u}\right)
 - \frac{v}{u^2}\,[u_t - \Delta u].
 $$
 Upon letting \ $v = u$ it becomes
 $$
 F^* = - \frac{1}{u}\,[u_t - \Delta u].
 $$
Hence, Eq. (\ref{Burg:mult.eq9}) satisfies the condition
(\ref{saeq.5})  with $\lambda = - 1.$

 \subsection{Nonlinear heat equation}
 \label{samul:3}

 Consider the nonlinear heat equation $ u_t - D_x\left(k(u) u_x\right) =
 0,$ or
 \begin{equation}
 \label{Burg:mult.eq10}
  u_t - k(u) u_{xx} - k'(u) u_x^2 = 0.
 \end{equation}
 Its adjoint equation has the form
 $$
 v_t + k(u) v_{xx} = 0.
 $$
 Therefore it is obvious that (\ref{Burg:mult.eq10}) does not satisfy Definition \ref{saeq:def}.
But it becomes strictly self-adjoint if we rewrite it in the form
 \begin{equation}
 \label{Burg:mult.eq11}
 \frac{1}{u}\,\left[u_t - k(u) u_{xx} - k'(u) u_x^2\right] = 0.
 \end{equation}
 Indeed, the formal Lagrangian (\ref{sa:Lag}) for Eq.
(\ref{Burg:mult.eq11}) is written
 $$
 {\cal L} = \frac{v}{u}\,\left[u_t - k(u) u_{xx} - k'(u) u_x^2\right].
 $$
Substituting it in (\ref{saeq.3}) we obtain
 \begin{align}
 F^* & = - D_t\left(\frac{v}{u}\right) - D_x^2 \left(\frac{v}{u}\,k(u) \right)
 + 2 D_x \left(\frac{v}{u}\,k'(u) u_x \right)\notag\\[1ex]
 & - \frac{v}{u}\,k'(u) u_{xx}
- \frac{v}{u}\,k''(u) u_x^2 - \frac{v}{u^2}\,\left[u_t - k(u) u_{xx}
- k'(u) u_x^2\right].\notag
 \end{align}
Letting here \ $v = u$ we have:
 $$
 F^* = - \frac{1}{u}\,\left[u_t - k(u) u_{xx} - k'(u) u_x^2\right].
 $$
Hence, Eq. (\ref{Burg:mult.eq9}) satisfies the strict
self-adjointness condition (\ref{saeq.5})  with $\lambda = - 1.$

 \subsection{The Burgers equation}
 \label{samul:4}

Taking in (\ref{Burg:mult.eq5}) $f (u) = u$ we obtain the
 strictly self-adjoint form
 \begin{equation}
 \label{Burg:mult.eq6}
 \frac{1}{u}\,[u_t - u_{xx}] -  u_x = 0
 \end{equation}
 of the Burgers equation $u_t = u_{xx} +  u u_x.$

 \subsection{Heat conduction in solid hydrogen}
 \label{samul:5}

 According to \cite{ros79}, the heat conduction in solid crystalline
 molecular hydrogen at low pressures is governed by the nonlinear equation (up-to positive constant coefficient)
 \begin{equation}
 \label{hydrogen.eq1}
 u_t =  u^2 \Delta u.
 \end{equation}
It is derived from the Fourier equation
 $$
 \rho\,c_* \frac{\partial T}{\partial t}
 = \nabla\cdot (k\,\nabla T)
 $$
 using the empirical information that the density $\rho$ at low
 pressures has a constant value, whereas the specific heat $c_*$ and
 the thermal conductivity $k$ have the  estimations
 $$
 c_* \cong T^3, \quad k \cong T^3 \left(1 + T^4\right)^{-2}.
 $$

 It is also shown in \cite{ros79} that the one-dimensional equation
 (\ref{hydrogen.eq1}),
 \begin{equation}
 \label{Burg:mult.eq14}
 u_t = u^2 u_{xx}\,,
 \end{equation}
 is related to the linear heat equation by a non-point transformation
 (Eq. (5) in \cite{ros79}). A similar relation was found in
\cite{blu-kum80}
 for another representation of Eq. (\ref{Burg:mult.eq14}). The non-point
 transformation of Eq. (\ref{Burg:mult.eq14}) to the linear heat equation
 \begin{equation}
 \label{hydrogen.eq2}
 w_s =  w_{\xi \xi}
 \end{equation}
  is written in \cite{ibr83} as the
 differential substitution
 \begin{equation}
 \label{hydrogen.eq3}
 t = s, \quad x = w, \quad  u =  w_\xi.
 \end{equation}
 It is also demonstrated in \cite{ibr83}, Section  20,
 that Eq. (\ref{Burg:mult.eq14}) is the unique equation
 with nontrivial Lie-B\"{a}cklund symmetries among the equations of
 the form
 $$
 u_t = f(u) + h (u, u_x), \quad f'(u) \not= 0.
 $$
The connection between Eq. (\ref{Burg:mult.eq14}) and the
 heat equation is treated in \cite{rog85} as a reciprocal transformation \cite{rog85}.
It is shown in \cite{rog86} that this connection, together with its
extensions, allows the analytic solution of certain moving boundary
problems in nonlinear heat conduction.

 Our Example \ref{Burg:mult.exa4} from Section \ref{samul:1}
 reveals one more remarkable
 property of Eq. (\ref{Burg:mult.eq14}). Namely,
taking $n = 2$ and $f(u) = u^2$ in Eq. (\ref{Burg:mult.eq13}) we
 see  that \textit{Eq.} (\ref{Burg:mult.eq14})  \textit{becomes strictly self-adjoint if we rewrite it in the form}
 \begin{equation}
 \label{Burg:mult.eq15}
 \frac{u_t}{u^3} = \frac{u_{xx}}{u}\,\cdot
 \end{equation}

 \subsection{Harry Dym equation}
 \label{samul:6}

 Taking in  Example \ref{Burg:mult.exa4} from Section \ref{samul:1} $n = 3$ and $f(u) = u^3$
we see that the Harry Dym equation
 \begin{equation}
 \label{Burg:mult.eq16}
 u_t - u^3 u_{xxx} = 0
 \end{equation}
 becomes strictly elf-adjoint upon rewriting it in the form
 $$
 \frac{u_t}{u^4} - \frac{u_{xxx}}{u} = 0.
 $$

\subsection{Kompaneets equation}
 \label{samul:7}

 The equations considered in Sections
 \ref{samul:1} - \ref{samul:6} are quasi self-adjoint. For example,
 for Eq. (\ref{Burg:mult.eq12}) we have
 $$
 F^* = - v_t - D_x^n(f(u) v) -  v f'(u)  u_n,
 $$
 whence making the substitution
 $$
 v = \frac{1}{f(u)}
 $$
 we obtain
  $$
 F^* = \frac{f'}{f^2}\, u_t  -  \frac{ f'}{f}\,  u_n
= \frac{f'}{f^2} [u_t  - f(u)\,  u_n].
 $$
Hence, Eq. (\ref{Burg:mult.eq12}) is quasi self-adjoint.

 \begin{exa}
 \label{Burg:mult.exa7}
 The Kompaneets equation
 \begin{equation}
 \label{Burg:mult.Komp1}
 u_t = \frac{1}{x^2}\, D_x\left[x^4 (u_x + u + u^2)\right]
 \end{equation}
 provides an example of an equation that is not quasi
self-adjoint. Indeed, Eq. (\ref{Burg:mult.Komp1})
 has the formal Lagrangian
 $$
 {\cal L} = v [- u_t + x^2 u_{xx} + (x^2 + 4 x + 2 x^2 u) u_x
 + 4 x (u + u^2)].
 $$
 The calculation yields the following adjoint equation to (\ref{Burg:mult.Komp1}):
  \begin{equation}
 \label{Burg:mult.Komp2}
 \frac{\delta {\cal L}}{\delta u} \equiv v_t + x^2 v_{xx} -x^2(1 + 2 u)v_x + 2 (x + 2 x u - 1)v = 0.
 \end{equation}
 Letting $v = \varphi(u)$ one obtains:
 \begin{align}
 \frac{\delta {\cal L}}{\delta u}\bigg|_{v = \varphi(u)}
 & = \varphi'(u) [u_t + x^2 u_{xx} - x^2 (1 + 2 u) u_x]\notag\\[1.5ex]
 & +
\varphi''(u)  x^2 u_x^2 + 2 (x + 2 x u - 1) \varphi(u).\notag
 \end{align}
Writing the quasi self-adjointness condition (\ref{qsa.2}) in the
form
 $$
 \frac{\delta {\cal L}}{\delta u}\bigg|_{v = \varphi(u)}
 = \lambda [- u_t + x^2 u_{xx} + (x^2 + 4 x + 2 x^2 u) u_x
 + 4 x (u + u^2)]
 $$
and comparing the coefficients for $u_t$ in both sides one obtains
$\lambda = - \varphi'(u),$ so that the quasi self-adjointness
condition takes the form
 \begin{align}
 & \varphi'(u) [u_t + x^2 u_{xx} - x^2 (1 + 2 u) u_x] +
\varphi''(u)  x^2 u_x^2 + 2 (x + 2 x u - 1) \varphi(u)\notag\\[1.5ex]
 & = \varphi'(u) [u_t - x^2 u_{xx} - (x^2 + 4 x + 2 x^2 u) u_x
 - 4 x (u + u^2)].\notag
 \end{align}
 Comparing the coefficients for $u_{xx}$ in both sides we obtain
 $\varphi'(u) = 0.$ Then the above equation becomes
 $(x + 2 x u - 1) \varphi(u) = 0$ and yields
 $\varphi(u) = 0.$
 Hence the Kompaneets equation
 is not quasi self-adjoint because the condition (\ref{qsa.eq3}) is not satisfied.

 But
 we can rewrite  Eq. (\ref{Burg:mult.Komp1}) in the strictly self-adjoint form  by using
 a more general
multiplier than above, namely, the multiplier
 \begin{equation}
 \label{Burg:mult.Komp3}
 \mu = \frac{x^2}{u}\,\cdot
 \end{equation}
 Indeed, upon multiplying by this $\mu$ \ Eq. (\ref{Burg:mult.Komp1}) is written
 $$
 \frac{x^2}{u}\,u_t = \frac{1}{u}\,D_x\left[x^4 (u_x + u +
 u^2)\right].
 $$
 Its  formal Lagrangian
 $$
 {\cal L} = \frac{v}{u}\,\left\{- x^2 u_t + D_x\left[x^4 (u_x + u +
u^2)\right]\right\}
 $$
  satisfies the strict self-adjointness condition (\ref{saeq.5}) with $\lambda = - 1:$
$$
  \frac{\delta {\cal L}}{\delta u}\bigg|_{v = u} = - \frac{1}{u}\,\left\{- x^2 u_t + D_x\left[x^4 (u_x + u +
u^2)\right]\right\}.
 $$
 \end{exa}
 \begin{rem}
 \label{Burg:mult.Komp2.rem}
 Note that $v = x^2$ solves Eq. (\ref{Burg:mult.Komp2}) for
 any $u.$ The connection of this solution with the multiplier
(\ref{Burg:mult.Komp3}) is discussed in Section \ref{gsa:6}. See
 also Section \ref{gsa:7}.
 \end{rem}

 \section{General concept of nonlinear self-adjointness}
%\section{Introduction of generalized self-adjointness}
 %\section{Definition and main properties}
 %\section{Main statements on generalized self-adjointness}
 \setcounter{equation}{0}
 \label{gsa}
  Motivated by the examples discussed in Sections \ref{sa:eveq.int} and  \ref{samul}
 as well as other similar  examples, I suggest here the general concept
 of \textit{nonlinear self-adjointness} of systems consisting of any number of equations
 with $m$ dependent variables. This concept encapsulates Definition \ref{saeq:def}
 of strict self-adjointness  and
 Definition \ref{qsa:def} of  quasi self-adjointness.
 The new concept has two different features. They are expressed below
by two different but equivalent definitions.
% These two definitions are equivalent but .

 %\subsection{Definition}
 \subsection{Two definitions and their equivalence}
 \label{gsa:1}
 \begin{defn}
 \label{gsa:def1}
 The system of $\overline m$ differential equations (compare with Eqs. (\ref{sa:eveq.int.1}))
 \begin{equation}
 \label{gsa:eqs}
 F_{\bar \alpha} \big(x, u, u_{(1)}, \ldots, u_{(s)}\big) = 0, \quad
 \bar \alpha = 1, \ldots, \overline m,
 \end{equation}
with $m$ dependent variables $u= (u^1, \ldots, u^m)$ is said to be
\textit{nonlinarly self-adjoint} if the \textit{adjoint
 equations}
 \begin{equation}
 \label{gsa:adeqs}
 F^*_\alpha \big(x, u, v, u_{(1)}, v_{(1)}, \ldots, u_{(s)}, v_{(s)}\big)\equiv
 \frac{\delta (v^{\bar \beta} F_{\bar \beta})}{\delta u^\alpha}  = 0, \quad
 \alpha = 1, \ldots, m,
 \end{equation}
 are satisfied for all solutions $u$
of the original system (\ref{gsa:eqs}) upon a substitution
  \begin{equation}
  \label{gsa:eq1}
 v^{\bar \alpha} = \varphi^{\bar \alpha}(x, u), \quad \bar \alpha = 1, \ldots, \overline m,
 \end{equation}
such that
  \begin{equation}
  \label{gsa:eq2}
 \varphi(x, u) \not= 0.
 \end{equation}
 In other words,  the following equations hold:
  \begin{equation}
  \label{gsa:eq3}
 F^*_{\alpha} \big(x, u, \varphi(x, u), \ldots, u_{(s)}, \varphi_{(s)}\big) =
 \lambda^{\bar \beta}_\alpha \, F_{\bar \beta}\big(x, u,
 \ldots, u_{(s)}\big), \quad \alpha = 1, \ldots, m,
  \end{equation}
  where $\lambda^{\bar \beta}_\alpha$ are undetermined
 coefficients, and $ \varphi_{(\sigma)}$ are derivatives of (\ref{gsa:eq1}),
 $$
  \varphi_{(\sigma)}
  = \{D_{i_1} \cdots D_{i_\sigma}\big(\varphi^{\bar \alpha}(x, u)\big)\}, \quad \sigma = 1, \ldots, s.
  $$
 Here $v$ and $\varphi$ are the $\overline m$-dimensional vectors
  $$
  v = (v^1, \ldots, v^{\overline m}), \quad \varphi = (\varphi^1, \ldots, \varphi^{\overline m}),
  $$
  and
  Eq. (\ref{gsa:eq2}) means that not all components $\varphi^{\bar \alpha}(x, u)$ of  $\varphi$ vanish
 simultaneously.
 \end{defn}

 %\subsection{Main properties}
% \subsection{Second definition}
 %\subsection{Generalized self-adjointness and solution of adjoint equation}
 %\label{sa:eveq.gsa:2}
 \begin{rem}
 \label{gsa:rem}
 If the system (\ref{gsa:eqs}) is \textit{over-determined}, i.e. $\overline m
 > m,$ then the adjoint system (\ref{gsa:adeqs}) is \textit{sub-definite} since it contains
 $m < \overline m$ equations for
 $\overline m$ new dependent variables $v.$  Vise versa, if $\overline m <
 m,$ then the system (\ref{gsa:eqs}) is sub-definite and the adjoint system (\ref{gsa:adeqs}) is
 over-determined.
  \end{rem}
 \begin{rem}
 \label{gsa:rem1}
 The adjoint system (\ref{gsa:adeqs}), upon substituting there any solution $u(x)$ of Eqs.
 (\ref{gsa:eqs}),
 becomes a linear homogeneous system for the new dependent variables $v^{\bar \alpha}.$ The essence of Eqs.
 (\ref{gsa:eq3}) is that for the self-adjoint system (\ref{gsa:eqs})
  there  exist functions (\ref{gsa:eq1})
  that
 provide a non-trivial (not identically zero) solution to the adjoint system (\ref{gsa:adeqs})
 \textit{for all solutions of the original system}
 (\ref{gsa:eqs}).
 This property can be taken as the following alternative definition
 of the nonlinear self-adjointness.
  \end{rem}
 \begin{defn}
 \label{gsa:def2}
 The system (\ref{gsa:eqs})  is nonlinearly self-adjoint if
 there exist functions $v^{\bar \alpha}$ given by  (\ref{gsa:eq1})
 that solve the adjoint system (\ref{gsa:adeqs})
 for all solutions $u(x)$ of Eqs.
(\ref{gsa:eqs}) and satisfy the condition (\ref{gsa:eq2}).
 %%%%%%%
 % The system (\ref{sa:eveq.int.1})  is said to be self-adjoint if
 % there exist functions $v^\alpha$ given by  (\ref{sa:eveq.int.4})
 % and satisfying the condition (\ref{gsa:4a})
 % that provide a  solution to the adjoint system (\ref{sa:eveq.int.2})
 % for all solutions $u(x)$ of the original system
 % (\ref{sa:eveq.int.1}).
 \end{defn}
 \begin{prop}
 \label{gsa:prop1}
 The above two definitions are equivalent.
 \end{prop}
 \textbf{Proof.} Let the system (\ref{gsa:eqs})
 be nonlinearly self-adjoint by Definition \ref{gsa:def1}. Then,
 according to Remark \ref{gsa:rem1},  the system
(\ref{gsa:eqs}) satisfies the condition of Definition
\ref{gsa:def2}.

 Conversely, let the system (\ref{gsa:eqs}) be nonlinearly self-adjoint
 by Definition \ref{gsa:def2}. Namely, let  the functions $v^{\bar \alpha}$ given by  (\ref{gsa:eq1})
 and satisfying the condition (\ref{gsa:eq2}) solve the adjoint system (\ref{gsa:adeqs})
 for \textit{all} solutions $u(x)$ of Eqs. (\ref{gsa:eqs}).
 This is possible if and only if
 Eqs.  (\ref{gsa:eq3}) hold. Then the system (\ref{gsa:eqs}) is
 nonlinearly  self-adjoint by Definition \ref{gsa:def1}.
 \begin{exa}
 \label{gsa:KdV}
 It has been mentioned in Example \ref{sa:eveq.KdV} that the KdV  equation
 \begin{equation}
 \label{gsa:KdV.eq1}
 u_t = u_{xxx} + u u_x
 \end{equation}
 is strictly self-adjoint. In terms of Definition \ref{gsa:def2} it means
 that  $v = u$ solves the adjoint equation
 \begin{equation}
 \label{gsa:KdV.eq2}
 v_t = v_{xxx} + u v_x
 \end{equation}
 for all solutions of the KdV equation (\ref{gsa:KdV.eq1}). One can verify  that
 the general substitution of the form (\ref{gsa:eq1}), $v = \varphi (t, x, u),$
 satisfying Eq. (\ref{gsa:eq3}) is given by
 \begin{equation}
 \label{gsa:KdV.eq3}
 v = A_1 + A_2 u + A_3 (x + t u),
 \end{equation}
 where $A_1, A_2, A_3$ are arbitrary constants. One can also check that $v$ given by Eq. (\ref{gsa:KdV.eq3})
 solves the adjoint equation (\ref{gsa:KdV.eq2}) for all solutions $u$ of the KdV equation. The
 solution $v = x + t u$ is an invariant of the Galilean transformation
 of the KdV equation and appears in different
 approaches (see \cite{ibr83}, Section 22.5, and \cite{blu-che-anc10}).
 Thus, the KdV equation is nonlinearly self-adjoint with the
 substitution (\ref{gsa:KdV.eq3}).
 \end{exa}

 \begin{prop}
 \label{gsa:prop2}
 Any linear equation  is nonlinearly self-adjoint.
 \end{prop}
 \textbf{Proof.} This property is the direct consequence of Definition
 \ref{gsa:def2} because the adjoint equation $F^*[v] = 0$ to a
 linear equation $F[u] = 0$ does not involve the variable $u.$

 \subsection{Remark on differential substitutions}
 \label{gsa:rem2}
 One can further extend the concept of self-adjointness by replacing the \textit{point-wise} substitution (\ref{gsa:eq1})
 with \textit{differential} substitutions of the form
  \begin{equation}
  \label{gsa:eq4}
 v^{\bar \alpha} = \varphi^{\bar \alpha}(x, u, u_{(1)}, \ldots, u_{(r)}),
 \quad {\bar \alpha} = 1, \ldots, \overline m.
 \end{equation}
 Then Eqs. (\ref{gsa:eq3}) will be written, e.g. in the case
 $r = 1,$ as follows:
  \begin{equation}
  \label{gsa:eq5}
 F^*_{\alpha} \big(x, u, \varphi, \ldots, u_{(s)}, \varphi_{(s)}\big) =
 \lambda^{\bar \beta}_\alpha \, F_{\bar \beta} + \lambda^{j \bar \beta}_\alpha \,D_j(F_{\bar \beta}).
  \end{equation}

 \begin{exa}
 \label{gsa:rem2.exa}
  The reckoning shows that the equation
  \begin{equation}
  \label{gsa:eq5.r1}
  u_{xy} = \sin u
  \end{equation}
  is not self-adjoint via a point-wise substitution $v = \varphi (x, y, u),$ but it is
  self-adjoint in the sense of Definition \ref{gsa:def1} with the following differential substitution:
  \begin{equation}
  \label{gsa:eq5.r2}
 v = \varphi (x, y, u_x, u_y) \equiv A_1 [x u_x - y u_y] + A_2 u_x + A_3 u_y,
  \end{equation}
 where $A_1,\ A_2,\ A_3$ are arbitrary constants.  The adjoint equation to Eq. (\ref{gsa:eq5.r1}) is
  $$
  v_{xy} - v \cos u = 0,
  $$
  and  the self-adjointness condition (\ref{gsa:eq5}) with the function $\varphi$ given by (\ref{gsa:eq5.r2}) is satisfied in the form
  \begin{equation}
  \label{gsa:eq5.r3}
 \varphi_{xy} - \varphi \cos u = (A_1x + A_2) D_x (u_{xy} - \sin u) +
 (A_3 - A_1y) D_y (u_{xy} - \sin u).
  \end{equation}
 \end{exa}

 %However, in what follows (excepting Section \ref{difsub}) we identify the term \textit{nonlinear self-adjointness}
 %either with Definition \ref{gsa:def1} or with the  equivalent Definition \ref{gsa:def2}.
 %These definitions are illustrated in Sections
 %\ref{gsa:2}-\ref{gsa:5}.

 \subsection{Nonlinear heat equation}
 %\subsection{Example 1: Nonlinear heat equation}
 \label{gsa:2}

 \subsubsection{One-dimensional case}
 \label{gsa:2.1}

 Let us apply the new viewpoint to the nonlinear heat equation
(\ref{Burg:mult.eq10}), $u_t = (k(u) u_x)_x,$  discussed in
 Section \ref{samul:3}. We will take it in the
 expanded form
 \begin{equation}
 \label{gsa:eq6}
  u_t - k(u) u_{xx} - k'(u) u_x^2 = 0, \quad k(u) \not= 0.
 \end{equation}
 The adjoint equation
 (\ref{saeq.2}) to Eq. (\ref{gsa:eq6}) is
 \begin{equation}
 \label{gsa:eq7}
 v_t + k(u) v_{xx} = 0.
 \end{equation}
 We take the substitution (\ref{gsa:eq1}) written together with the necessary
 derivatives:
 \begin{equation}
 \label{gsa:eq8}
 \begin{split}
 & v = \varphi (t, x, u),\\[1ex]
 & v_t = \varphi_u u_t + \varphi_t,
 \quad v_x = \varphi_u u_x + \varphi_x,\\[1ex]
 & v_{xx} = \varphi_u u_{xx} + \varphi_{uu} u_x^2 + 2 \varphi_{xu} u_x + \varphi_{xx},
 \end{split}
 \end{equation}
 and arrive at the following self-adjointness condition (\ref{gsa:eq3}):
 \begin{equation}
 \label{gsa:eq9}
 \begin{split}
 & \varphi_u u_t + \varphi_t + k(u)[\varphi_u u_{xx} + \varphi_{uu} u_x^2 + 2 \varphi_{xu} u_x + \varphi_{xx}]
 \\[1ex]
 &  = \lambda [u_t - k(u) u_{xx} - k'(u) u_x^2].
 \end{split}
 \end{equation}
 The comparison of the coefficients of $u_t$ in both sides of Eq.
 (\ref{gsa:eq9}) yields $\lambda = \varphi_u.$ Then, comparing the
 terms with $u_{xx}$  we see that $\varphi_u = 0.$ Hence Eq.
 (\ref{gsa:eq9}) reduces to
 \begin{equation}
 \label{gsa:eq10}
 \varphi_t + k(u) \varphi_{xx} = 0
 \end{equation}
 and yields $\varphi_t = 0, \ \varphi_{xx} = 0,$ whence $\varphi = C_1 x +
 C_2,$ where $C_1, C_2 = {\rm const.}$
 We have demonstrated that Eq. (\ref{gsa:eq6}) is nonlinearly self-adjoint
 by Definition \ref{gsa:def1}  and that the substitution (\ref{gsa:eq1}) has the form
 \begin{equation}
 \label{gsa:eq11}
 v = C_1\, x + C_2.
 \end{equation}

 The same result can be easily obtained by using Definition
 \ref{gsa:def2}. We look for the solution of the adjoint equation (\ref{gsa:eq7}) in
the form $v = \varphi (t, x).$ Then  Eq. (\ref{gsa:eq7}) has  the
form (\ref{gsa:eq10}). Since it should be satisfied
 for all solutions $u$ of Eq. (\ref{gsa:eq6}), we obtain $\varphi_t = 0, \ \varphi_{xx} = 0,$
 and hence Eq. (\ref{gsa:eq11}).

 \subsubsection{Multi-dimensional case}

 The similar analysis can be applied to the nonlinear heat equation
 with several variables $x = (x^1, \ldots, x^n):$
 \begin{equation}
 \label{gsa:eq12}
  u_t = \nabla \cdot \left(k(u)\nabla u\right),
 \end{equation}
 or
 \begin{equation}
 \label{gsa:eq13}
  u_t - k(u)\, \Delta u - k'(u) |\nabla u|^2 = 0.
 \end{equation}
 The reckoning shows that the adjoint equation  (\ref{saeq.2}) to Eq. (\ref{gsa:eq13}) is
 written
 \begin{equation}
 \label{gsa:eq14}
  v_t + k(u)\, \Delta v = 0.
 \end{equation}
 It is easy to verify the nonlinear elf-adjointness by Definition
 \ref{gsa:def2}. Namely, searching the solution of the adjoint equation (\ref{gsa:eq14}) in
 the form $v = \varphi (t, x^1, \ldots, x^n),$ one obtains
 $$
  \varphi_t + k(u)\, \Delta \varphi = 0,
 $$
 whence
 $$
  \varphi_t  = 0, \quad \Delta \varphi = 0.
 $$
 We conclude that Eq. (\ref{gsa:eq13}) is self-adjoint and that the substitution (\ref{gsa:eq1})
is given by
 \begin{equation}
 \label{gsa:eq15}
 v = \varphi (x^1, \ldots, x^n),
 \end{equation}
where $\varphi (x^1, \ldots, x^n)$ is any solution of the
$n$-dimensional Laplace equation $\Delta \varphi = 0.$

 %\subsection{Korteweg-de Vries equation}
 %\label{gsa:3a}
 %For the Korteweg-de Vries equation
 % $$
 % u_t - uu_x - u_{xxx} = 0
 % $$
 %the adjoint equation (\ref{saeq.2}) is written
 % $$
 % v_t - uv_x - v_{xxx} = 0.
 % $$
 %Proceeding as in Section \ref{gsa:2.1} one can verify that
 %the Korteweg-de Vries equation  is self-adjoint in the
 %sense of Definition \ref{gsa:def1} with the substitution (\ref{gsa:eq1})
 % $$
 % v = C_1 + C_2\,u  + C_3 (x + tu),
 % $$
 %where $C_1, C_2, C_3$ are arbitrary constants. Taking $C_1 = C_3 = 0, C_2 =
 %1$ we obtain the self-adjointness in the restricted sense of
 %Definition \ref{saeq:def} as discussed in Example \ref{sa:eveq.KdV}.

\subsection{Anisotropic nonlinear heat equation}
 %\subsection{Example 2: Anisotropic nonlinear heat equation}
 \label{gsa:3}

 \subsubsection{Two-dimensional case}

 Consider the heat diffusion equation
 \begin{equation}
 \label{gsa:eq16}
  u_t = (f(u) u_x)_x + (g(u) u_y)_y
 \end{equation}
 in an anisotropic two-dimensional  medium (see \cite{ibr94-96}, vol. 1, Section 10.8)
  with arbitrary functions $f(u)$ and $g(u).$  The adjoint equation is
 \begin{equation}
 \label{gsa:eq17}
 v_t + f(u) v_{xx} + g(u) v_{yy} = 0.
 \end{equation}
 Using Definition \ref{gsa:def2} we obtain the following equations for
 nonlinear self-adjointness of Eq. (\ref{gsa:eq16}):
 \begin{equation}
 \label{gsa:eq18}
 \varphi_t = 0, \quad \varphi_{xx} = 0, \quad \varphi_{yy} = 0.
 \end{equation}
 Integrating Eqs. (\ref{gsa:eq18}) we obtain the
following substitution (\ref{gsa:eq1}):
 \begin{equation}
 \label{gsa:eq19}
 v = C_1\,xy + C_2\,x + C_3\,y + C_4.
 \end{equation}

 \subsubsection{Three-dimensional case}

 The three-dimensional anisotropic nonlinear heat diffusion equation
 has the following form (see \cite{ibr94-96}, vol. 1, Section 10.9):
 \begin{equation}
 \label{gsa:eq20}
  u_t = (f(u) u_x)_x + (g(u) u_y)_y + (h(u) u_z)_z\,.
 \end{equation}
 Its adjoint equation is
 \begin{equation}
 \label{gsa:eq21}
 v_t + f(u) v_{xx} + g(u) v_{yy} + h(u) v_{zz} = 0.
 \end{equation}
 Eq. (\ref{gsa:eq20}) is nonlinearly self-adjoint. In this case the substitution (\ref{gsa:eq19}) is replaced by
 \begin{equation}
 \label{gsa:eq22}
 v = C_1\,xyz + C_2\,xy + C_3\,xz + C_4\,y z + C_5\,x + C_6\,y + C_7\,z + C_8.
 \end{equation}

 \subsection{Nonlinear wave equations}
 %\subsection{Example 3: Nonlinear wave equation}
 \label{gsa:4}

 \subsubsection{One-dimensional case}

 Consider the following one-dimensional nonlinear wave equation:
 \begin{equation}
 \label{gsa:eq23}
 u_{tt} = (k(u) u_x)_x, \quad k(u) \not= 0,
 \end{equation}
 or in the expanded form
 \begin{equation}
 \label{gsa:eq24}
  u_{tt} - k(u) u_{xx} - k'(u) u_x^2 = 0.
 \end{equation}
 The adjoint equation (\ref{saeq.2}) to Eq. (\ref{gsa:eq23}) is
written
 \begin{equation}
 \label{gsa:eq25}
 v_{tt} - k(u) v_{xx} = 0.
 \end{equation}

 Proceeding as in Section \ref{gsa:2.1}
 or applying Definition \ref{gsa:def2} to Eqs. (\ref{gsa:eq24}), (\ref{gsa:eq25})
 by letting  $v = \varphi (t, x),$ we obtain the following
equations that guarantee the nonlinear
 self-adjointness of Eq. (\ref{gsa:eq23}):
 \begin{equation}
 \label{gsa:eq26}
 \varphi_{tt} = 0, \quad \varphi_{xx} = 0.
 \end{equation}
 Integrating Eqs. (\ref{gsa:eq26}) we obtain the
following substitution:
 \begin{equation}
 \label{gsa:eq27}
 v = C_1\,tx + C_2\,t + C_3\,x + C_4.
 \end{equation}

 \subsubsection{Multi-dimensional case}

 The multi-dimensional version of Eq. (\ref{gsa:eq23}) with $x = (x^1, \ldots, x^\nu)$ is written
 \begin{equation}
 \label{gsa:eq28}
  u_{tt} = \nabla \cdot \left(k(u)\nabla u\right),
 \end{equation}
 or
 \begin{equation}
 \label{gsa:eq29}
  u_{tt} - k(u)\, \Delta u - k'(u) |\nabla u|^2 = 0.
 \end{equation}
 The adjoint equation is
 \begin{equation}
 \label{gsa:eq30}
  v_{tt} - k(u)\, \Delta v = 0.
 \end{equation}
 Using Definition \ref{gsa:def2} and searching the solution of the adjoint equation (\ref{gsa:eq30}) in
 the form $v = \varphi (t, x^1, \ldots, x^\nu),$ we obtain the
 equations
 $$
  \varphi_{tt}  = 0, \quad \Delta \varphi = 0.
 $$
 Solving them we arrive at the following substitution (\ref{gsa:eq1}):
 \begin{equation}
 \label{gsa:eq31}
 v = a (x) t + b(x),
 \end{equation}
 where $a (x)$ and $b(x)$ solve the
 $\nu$-dimensional Laplace equation,
 $$
 \Delta a (x^1, \ldots, x^\nu) = 0, \quad  \Delta b (x^1, \ldots, x^\nu) = 0.
 $$
 Hence Eq. (\ref{gsa:eq28}) is nonlinearly self-adjoint.

 \subsubsection{Nonlinear vibration of membranes}

 Vibrations of a uniform membrane whose tension varies during
deformations are described by the following Lagrangian:
 \begin{equation}
 \label{gsa:eq32}
 L = \frac{1}{2}\, \left[u_t^2 - k(u)
 \Big(u_x^2 + u_y^2\Big)\right], \quad k'(u) \not= 0.
 \end{equation}
The corresponding Euler -Lagrange equation
 $$
 \frac{\partial L}{\partial u} -
 D_t \left(\frac{\partial L}{\partial u_t}\right) -
 D_x \left(\frac{\partial L}{\partial u_x}\right) - D_y
 \left(\frac{\partial L}{\partial u_y}\right)= 0
 $$
 provides the  nonlinear wave equation
 \begin{equation}
 \label{gsa:eq33}
 u_{tt} =  k(u)\,(u_{xx} + u_{yy}) +
  \frac{1}{2}\,k'(u) (u_x^2 + u_y^2).
 \end{equation}
 Note that Eq. (\ref{gsa:eq33}) differs from the two-dimensional nonlinear wave
 equation (\ref{gsa:eq29}) by the coefficient $1/2.$ Let us
 find out if this difference affects  self-adjointness.

 By applying (\ref{gsa:adeqs}) to the formal Lagrangian of Eq. (\ref{gsa:eq33}) we
 obtain:
 $$
 F^* = v_{tt} -  k(u)\,(v_{xx} + v_{yy}) - k'(u)(u_x v_x + u_y v_y + v u_{xx} + v u_{yy})
 - \frac{v}{2}\,k''(u) (u_x^2 + u_y^2).
 $$
 We take the substitution (\ref{gsa:eq1}) together with the necessary
 derivatives (see Eqs. (\ref{gsa:eq8})):
 \begin{equation}
 \label{gsa:eq34}
 \begin{split}
 & v = \varphi (t, x, y, u), \quad  v_t = \varphi_u u_t + \varphi_t,\\[1ex]
 & v_x = \varphi_u u_x + \varphi_x,
 \quad v_y = \varphi_u u_y + \varphi_y,\\[1ex]
 & v_{xx} = \varphi_u u_{xx} + \varphi_{uu} u_x^2 + 2 \varphi_{xu} u_x
 + \varphi_{xx},\\[1ex]
 & v_{yy} = \varphi_u u_{yy} + \varphi_{uu} u_y^2 + 2 \varphi_{yu} u_y + \varphi_{yy},\\[1ex]
 & v_{tt} = \varphi_u u_{tt} + \varphi_{uu} u_t^2 + 2 \varphi_{tu} u_t +
 \varphi_{tt},
 \end{split}
 \end{equation}
 and substitute the expressions (\ref{gsa:eq34}) in the self-adjointness condition (\ref{gsa:eq3}):
 $$
 F^*\big|_{v = \varphi} = \lambda [ u_{tt} -  k(u)\,(u_{xx} + u_{yy})
 - \frac{1}{2}\,k'(u) (u_x^2 + u_y^2)].
 $$
 Comparing the coefficients of $u_{tt}$ we obtain $\lambda =
 \varphi_u.$ Then we compare the coefficients of $u_{xx}$ and obtain
 $\varphi \, k'(u) = 0.$ This equation yields $\varphi  = 0$ because $k'(u) \not= 0.$
 Thus, the condition (\ref{gsa:eq2}) is not satisfied for the point-wise substitution
 (\ref{gsa:eq1}). Further investigation of
  Eq. (\ref{gsa:eq33})  for the nonlinear self-adjointness requires differential substitutions.

 \subsection{Anisotropic nonlinear wave equation}
 %\subsection{Example 4: Anisotropic nonlinear wave equation}
 \label{gsa:5}

 \subsubsection{Two-dimensional case}

 The two-dimensional anisotropic nonlinear wave  equation is (see \cite{ibr94-96}, vol. 1,
 Section 12.6)
 \begin{equation}
 \label{gsa:eq35}
  u_{tt} = (f(u) u_x)_x + (g(u) u_y)_y\,.
 \end{equation}
Its adjoint equation has the form
 \begin{equation}
 \label{gsa:eq36}
  v_{tt} - f(u) v_{xx} - g(u) v_{yy} = 0.
 \end{equation}
 Proceeding as in Section \ref{gsa:3} we obtain the following
 equations that guarantee the self-adjointness of Eq. (\ref{gsa:eq35}):
 \begin{equation}
 \label{gsa:eq37}
 \varphi_{tt} = 0, \quad \varphi_{xx} = 0, \quad \varphi_{yy} = 0.
 \end{equation}
 Integrating Eqs. (\ref{gsa:eq37}) we obtain the
following substitution (\ref{gsa:eq1}):
 \begin{equation}
 \label{gsa:eq38}
 v = C_1\,txy + C_2\,t x + C_3\,ty + C_4\,xy + C_5\,t + C_6\,x + C_7\,y + C_8.
 \end{equation}
 \begin{rem}
 \label{gsa:rem3}
 I provide here detailed calculations in integrating Eqs.
 (\ref{gsa:eq37}). The general solution to the linear second-order equation $\varphi_{tt} = 0$
 is given by
 \begin{equation}
 \label{gsa:eq37S}
 \varphi = A(x, y) t + B(x, y)
 \end{equation}
 with arbitrary functions $A(x, y)$ and $B(x, y).$  Substituting
 this expression for $\varphi$ in the second and third equations
 (\ref{gsa:eq37}) and splitting with respect to $t$ we obtain the
 following equations for $A(x, y)$ and $B(x, y):$
 $$
 A_{xx} = 0, \quad A_{yy} = 0,
 $$
 $$
  B_{xx} = 0, \quad B_{yy} = 0.
 $$
 Substituting the general solution $$A = a_1(y) x + a_2(y)$$ of the
 equation $A_{xx} = 0$ in the equation $ A_{yy} = 0$ and splitting
 with respect to $x,$ we obtain $a_1'' = 0, \ a_2'' = 0,$ whence
 $$
 a_1 = c_{11} y  + c_{12}, \quad a_2 = c_{21} y  + c_{22},
 $$
where $c_{11}, \ldots, c_{22}$ are arbitrary constants. Substituting
these in the above expression for $A$ we obtain
 $$
 A = c_{11} xy  + c_{12} x + c_{21} y  + c_{22}.
 $$
 Proceeding likewise with the equations for $B(x, y),$ we have
 $$
 B = d_{11} xy  + d_{12} x + d_{21} y  + d_{22}
 $$
with arbitrary constant coefficients $d_{11}, \ldots, d_{22}.$
 Finally, we substitute the resulting $A$ and $B$ in the expression
 (\ref{gsa:eq37S}) for $\varphi$ and, changing the notation, arrive at (\ref{gsa:eq38}).
 \end{rem}

 \subsubsection{Three-dimensional case}

 The three-dimensional anisotropic nonlinear wave  equation
 \begin{equation}
 \label{gsa:eq39}
  u_{tt} = (f(u) u_x)_x + (g(u) u_y)_y + (h(u) u_z)_z
 \end{equation}
 has the following adjoint equation
 \begin{equation}
 \label{gsa:eq40}
  v_{tt} - f(u) v_{xx} - g(u) v_{yy} - h(u) v_{zz} = 0.
 \end{equation}
 In this case Eqs. (\ref{gsa:eq37}) are replaced by
 $$
 \varphi_{tt} = 0, \quad \varphi_{xx} = 0, \quad \varphi_{yy}, \quad \varphi_{zz} = 0
 $$
 and yield the
 following substitution (\ref{gsa:eq1}):
 \begin{equation}
 \label{gsa:eq41}
 \begin{split}
 v & = C_1\,txyz + C_2\,txy + C_3\,txz + C_4\,tyz + C_5\,tx + C_6\,ty +
  C_7\,tz\\[1ex]
 & + C_8\,xy + C_9\,xz
 + C_{10}\,yz + C_{11}\,t + C_{12}\,x + C_{13}\,y + C_{14}\,z + C_{15}.
 \end{split}
 \end{equation}

 \subsection{Nonlinear self-adjointness and multipliers}
 %\subsection{Generalized self-adjointness and multipliers}
 \label{gsa:6}

 The approach of this section is not used for constructing
conservation laws. But it may be useful for other applications of
 the nonlinear self-adjointness.

 \begin{thm}
 \label{gsa:6.thm}
 The differential equation (\ref{saeq.1}),
 \begin{equation}
 \label{gsa:6.eq1}
 F\big(x, u, u_{(1)}, \ldots, u_{(s)}\big) = 0,
  \end{equation}
  is nonlinearly self-adjoint (Definition \ref{gsa:def1})
 if and only if it  becomes strictly self-adjoint (Definition \ref{saeq:def}) upon rewriting in the
 equivalent form
 \begin{equation}
 \label{gsa:6.eq2}
 \mu (x, u) F\big(x, u, u_{(1)}, \ldots, u_{(s)}\big) = 0,
\quad  \mu (x, u) \not=  0,
 \end{equation}
with an appropriate multiplier $\mu (x, u).$
 \end{thm}
 \textbf{Proof.}
%Let us denote the left side of Eq. (\ref{gsa:6.eq2}) by $\widetilde F:$
% \begin{equation}
% \label{gsa:6.eq3}
% \widetilde F =  \mu (x, u) F\big(x, u, u_{(1)}, \ldots, u_{(s)}\big).
% \end{equation}
 We will write the condition (\ref{gsa:eq3}) for nonlinear self-adjointness of Eq. (\ref{gsa:6.eq1})  in the form
 \begin{equation}
 \label{gsa:6.eq3}
 \frac{\delta (v F)}{\delta u}\bigg|_{v = \varphi(x, u)} =  \lambda (x, u) F\big(x, u, u_{(1)}, \ldots, u_{(s)}\big).
 \end{equation}
 Furthermore, invoking that the equations (\ref{gsa:6.eq2}) and (\ref{gsa:6.eq1})
 are equivalent, we will write the condition (\ref{saeq.5}) for strict self-adjointness of  Eq.
(\ref{gsa:6.eq2}) in the form
 \begin{equation}
 \label{gsa:6.eq4}
 \frac{\delta (w \mu F)}{\delta u}\bigg|_{w = u} =  \tilde \lambda (x, u) F\big(x, u, u_{(1)}, \ldots, u_{(s)}\big).
 \end{equation}
  Since $w$ is a dependent variable and $\mu = \mu (x, u)$ is a certain function of $x, u,$
  the variational derivative in the left-hand side of (\ref{gsa:6.eq4}) can be written as follows:
 \begin{align}
% \label{gsa:6.eq4}
 \frac{\delta (w \mu F)}{\delta u} & = w  \frac{\partial \mu}{\partial u}\, F
  + \mu w \frac{\partial F}{\partial u} - D_i \left(\mu w \frac{\partial F}{\partial u_i}\right)
  + D_i D_j\left(\mu w \frac{\partial F}{\partial u_{ij}}\right) -
 \cdots \notag\\[1ex]
  & = w  \frac{\partial \mu}{\partial u}\, F + \frac{\delta (v F)}{\delta u}\,,\notag
 \end{align}
 where $v$ is the new dependent variable defined by
 \begin{equation}
 \label{gsa:6.eq5}
 v =  \mu (x, u) w.
 \end{equation}
 is the new dependent variable instead of $w.$ Now the left side of Eq. (\ref{gsa:6.eq4}) is
written
 \begin{equation}
 \label{gsa:6.eq6}
 \frac{\delta (w \mu F)}{\delta u}\bigg|_{w = u} =  u  \frac{\partial \mu}{\partial u}\, F
 + \frac{\delta (v F)}{\delta u}\bigg|_{v = u \mu (x, u)}\,\cdot
 \end{equation}

%%%%%%%%%%%%%%%%%%%%%%%%%%
 %Let us denote the left side of Eq. (\ref{gsa:6.eq2}) by $\widetilde F:$
 %\begin{equation}
 %\label{gsa:6.eq3}
 %\widetilde F =  \mu (x, u) F\big(x, u, u_{(1)}, \ldots, u_{(s)}\big).
 %\end{equation}
 %We will write the condition (\ref{gsa:eq3}) of Definition
 %\ref{gsa:def1} for Eq. (\ref{gsa:6.eq1})  in the form
 %\begin{equation}
 %\label{gsa:6.eq4}
 %\frac{\delta (v F)}{\delta u}\bigg|_{v = \varphi(x, u)} =  \lambda (x, u) F\big(x, u, u_{(1)}, \ldots, u_{(s)}\big).
 %\end{equation}
 %Using the notation (\ref{gsa:6.eq3}) and invoking that the equations (\ref{gsa:6.eq2}) and (\ref{gsa:6.eq1})
 %are equivalent, we can write the condition (\ref{saeq.5}) of Definition \ref{saeq:def} for Eq.
 %(\ref{gsa:6.eq2}) in the form
 %\begin{equation}
 %\label{gsa:6.eq5}
 %\frac{\delta (w \widetilde F)}{\delta u}\bigg|_{w = u} =  \tilde \lambda (x, u) F\big(x, u, u_{(1)}, \ldots, u_{(s)}\big).
 %\end{equation}
%%%%%%%%%%%%%%%%%%%%%%%%%%%%

 Let us assume that Eq. (\ref{gsa:6.eq1}) is nonlinearly self-adjoint.
Then Eq. (\ref{gsa:6.eq3}) holds with a certain given function
$\varphi (x, u).$ Therefore, we
 take the multiplier
 \begin{equation}
 \label{gsa:6.eq7}
 \mu (x, u) = \frac{\varphi (x, u)}{u}
 \end{equation}
 and reduce Eq. (\ref{gsa:6.eq6}) to the following form:
 $$
 \frac{\delta (w \mu F)}{\delta u}\bigg|_{w = u} =
 \left(\lambda + \frac{\partial \varphi}{\partial u} - \frac{\varphi}{u} \right) F.
 $$
 This proves that Eq. (\ref{gsa:6.eq4}) holds with
 $$
 \tilde \lambda = \frac{\partial \varphi}{\partial u} - \frac{\varphi}{u} +
 \lambda.
 $$
 Hence, Eq. (\ref{gsa:6.eq2}) with the multiplier $\mu$ given by
(\ref{gsa:6.eq7}) is strictly self-adjoint.

 Let us assume  now that Eq. (\ref{gsa:6.eq2}) with a certain multiplier $\mu (x, u)$ is strictly self-adjoint.
Then Eq. (\ref{gsa:6.eq4}) holds. Therefore, if we
 take the function $\varphi$ defined  by (see (\ref{gsa:6.eq7}))
 \begin{equation}
 \label{gsa:6.eq8}
  \varphi (x, u) = u \mu (x, u),
 \end{equation}
 Eq. (\ref{gsa:6.eq6}) yields:
 $$
 \frac{\delta (v F)}{\delta u}\bigg|_{v = \varphi (x, u)}=
 \left(\tilde \lambda - u  \frac{\partial \mu}{\partial u}\right) F.
 $$
 It follows that Eq. (\ref{gsa:6.eq3}) holds with
 $$
 \lambda = \tilde \lambda - u  \frac{\partial \mu}{\partial u}.
 $$
 We conclude that Eq. (\ref{gsa:6.eq1}) is nonlinearly self-adjoint, thus completing the
 proof.
 \begin{exa}
 \label{gsa.exa1}
 The multiplier (\ref{Burg:mult.Komp3}) used in Example \ref{Burg:mult.exa7}
 and the function $\varphi = x^2$ that provides a solution of the adjoint equation
 (\ref{Burg:mult.Komp2}) to the Kompaneets equation are related by Eq.
 (\ref{gsa:6.eq8}).
 \end{exa}
 \begin{exa}
 \label{gsa.exa2}
 Let us consider the one-dimensional nonlinear wave equation
 (\ref{gsa:eq24}),
 $$
  u_{tt} - k(u) u_{xx} - k'(u) u_x^2 = 0.
 $$
 If we substitute in (\ref{gsa:6.eq7}) the function $\varphi$ given by the right-hand side of (\ref{gsa:eq27})
 we will obtain the multiplier that maps Eq. (\ref{gsa:eq24}) into
 the strictly self-adjoint equivalent form. For example, taking
 (\ref{gsa:eq27})  with$C_1 = C_3 = C_4 = 0, \ C_2 = 1$ we obtain the multiplier
 $$
 \mu = \frac{t}{u}\,\cdot
 $$
 The corresponding equivalent equation to Eq. (\ref{gsa:eq24}) has the formal Lagrangian
 $$
 {\cal L} =  \frac{t v}{u}[u_{tt} - k(u) u_{xx} - k'(u) u_x^2].
 $$
 We have
 \begin{align}
 \frac{\delta {\cal L}}{\delta u} & = D_t^2\left(\frac{t v}{u}\right) - \frac{tv}{u^2}\,u_{tt}
 - D_x^2 \left(\frac{tv}{u}\,k(u)\right) - \frac{tv}{u}\,k'(u) u_{xx}
 + \frac{tv}{u^2}\,k(u) u_{xx}\notag\\[1ex]
 & + 2 D_x \left(\frac{tv}{u}\,k'(u) u_x \right)
- \frac{tv}{u}\,k''(u) u_x^2 + \frac{tv}{u^2}\,k'(u) u_x^2.\notag
 \end{align}
 Letting here $v = u$ we see that the strict self-adjointness condition is satisfied in the following form:
 $$
 \frac{\delta {\cal L}}{\delta u}\bigg|_{v = u} = - \frac{t}{u}[u_{tt} - k(u) u_{xx} - k'(u) u_x^2].
 $$
 \end{exa}

 \section{Generalized Kompaneets equation}
 \label{gsa:7}
 \setcounter{equation}{0}

 \subsection{Introduction}
  \label{gsa:7.1}

 The equation
 \begin{equation}
 \label{Komp.eq1}
 \frac{\partial n}{\partial t} = \frac{1}{x^2}\,
 \frac{\partial}{\partial x}\bigg[x^4 \bigg(\frac{\partial
 n}{\partial x} + n + n^2\bigg)\bigg],
 \end{equation}
 known as the Kompaneets equation or the \textit{photon
diffusion equation}, was derived independently
 by A.S. Kompaneets\footnote{He mentions in his paper that the work has been done in
 1950 and published in \textit{Report N. 336} of the Institute
 of Chemical Physics of the USSR Acad. Sci.} \cite{kom56}
 and R. Weymann \cite{wey65}. They take as a starting point
 the kinetic equations for the distribution function of a photon
 gas\footnote{Weymann uses Dreicer's kinetic equation \cite{dre64} for a photon gas
 interacting with a plasma which is slightly different from the equation used by Kompaneets.}
  and arrive, at certain idealized conditions,
 at Equation (\ref{Komp.eq1}). This equation
provides a mathematical model for describing the time development of
the energy spectrum
 of a low energy homogeneous photon gas interacting with a rarefied electron gas
 via the Compton scattering.
  Here $n$ is the density of the photon gas (photon number density), $t$ is time and $x$
  is connected with the photon frequency $\nu$ by the formula
  \begin{equation}
 \label{Komp.notation}
 x = \frac{h \nu}{kT_e}\,,
 \end{equation}
    where $h$ is Planck's constant and $kT_e$ is the
    \textit{electron temperature} with the standard notation $k$ for Boltzmann's
    constant. According to this notation, $h \nu$
    has the meaning of the \textit{photon energy}.
    The nonrelativistic approximation is used,
    i.e. it is assumed that the electron temperatures satisfy the condition $kT_e\ll m c^2,$ where $m$ is
    the electron mass and $c$ is the light velocity.
    The term \textit{low energy photon gas} means that $h \nu \ll mc^2.$

 The question arises if the idealized conditions assumed in deriving Eq. (\ref{Komp.eq1})
 may be satisfied in the real world. For discussions of theoretical and observational evidences for
 such possibility in astrophysical environments, for example in intergalactic gas,
 see e.g. \cite{wey66}, \cite{syu71} and the references
 therein. See also the recent publication \cite{chl-sun08}.

 \subsection{Discussion of self-adjointness of the Kompaneets equation}
 \label{gsa:7.2}

 For unifying the notation, the dependent variable $n$ in Eq. (\ref{Komp.eq1}) will be denoted
 by $u$ and Eq. (\ref{Komp.eq1}) will be written  further in the
 form
 \begin{equation}
 \label{PP.eq1}
 u_t = \frac{1}{x^2}\, D_x\big[x^4 (u_x + u + u^2)\big].
 \end{equation}
 Writing it in the expanded form
 \begin{equation}
 \label{PP.eq2}
 u_t = x^2 u_{xx} + (x^2 + 4 x + 2 x^2 u) u_x + 4 x (u + u^2),
 \end{equation}
 we have the following formal Lagrangian for Eq. (\ref{PP.eq1}):
 $$
 {\cal L} = v [- u_t + x^2 u_{xx} + (x^2 + 4 x + 2 x^2 u) u_x
 + 4 x (u + u^2)].
 $$
 Working out the variational derivative of this formal Lagrangian,
 $$
 \frac{\delta {\cal L}}{\delta u} = D_t(v) + D_x^2(x^2 v) - D_x[(x^2 + 4 x + 2 x^2 u) v]
 + 2 x^2v u_x + 4 x (1 + 2 u)v,
 $$
  we obtain the adjoint equation to Eq. (\ref{PP.eq1}):
 \begin{equation}
 \label{PP.eq3}
 \frac{\delta {\cal L}}{\delta u} \equiv v_t + x^2 v_{xx} -x^2(1 + 2 u)v_x + 2 (x + 2 x u - 1)v = 0.
 \end{equation}

 If $v = \varphi(u),$ then
 $$
 v_t = \varphi'(u) u_t, \quad  v_x = \varphi'(u) u_x, \quad v_{xx} = \varphi'(u)
 u_{xx} + \varphi''(u) u_x^2.
 $$
It follows that the quasi self-adjointness condition (\ref{qsa.2}),
 $$
 \frac{\delta {\cal L}}{\delta u}\bigg|_{v = \varphi(u)}
 = \lambda [- u_t + x^2 u_{xx} + (x^2 + 4 x + 2 x^2 u) u_x
 + 4 x (u + u^2)],
 $$
 is not satisfied.

 Let us check if this condition is satisfied in
 the more general form (\ref{gsa:eq3}):
 \begin{equation}
 \label{PP.eq4}
 \frac{\delta {\cal L}}{\delta u}\bigg|_{v = \varphi(t, x, u)}
 = \lambda [- u_t + x^2 u_{xx} + (x^2 + 4 x + 2 x^2 u) u_x
 + 4 x (u + u^2)].
 \end{equation}
 In this case
 \begin{equation}
 \label{PP.eq5}
 \begin{split}
  v_t & = D_t[\varphi(t, x, u)] = \varphi_u u_t +  \varphi_t,\\
  v_x & = D_x[\varphi(t, x, u)] = \varphi_u u_x +  \varphi_x,\\
  v_{xx} & =D_x(v_x) = \varphi_u u_{xx} + \varphi_{uu} u_x^2 +  2 \varphi_{xu} u_x
  + \varphi_{xx}.
 \end{split}
 \end{equation}
 Inserting (\ref{PP.eq5}) in the expression for the variational
 derivative given by (\ref{PP.eq3})  and singling out in Eq.
 (\ref{PP.eq4}) the terms containing $u_t$ and $u_{xx},$ we obtain
 the following equation:
 $$
 \varphi_u [u_t + x^2 u_{xx}]
 = \lambda [- u_t + x^2 u_{xx}].
 $$
 Since this equation should be satisfied identically in $u_t$ and $u_{xx},$ it yields $\lambda = \varphi_u = 0.$
 Hence $\varphi = \varphi(t, x)$ and Eq. (\ref{PP.eq4}) becomes:
 \begin{equation}
 \label{PP.eq6}
 \varphi_t + x^2 \varphi_{xx} - x^2 (1 + 2 u) \varphi_x + 2(x + 2
 x u - 1) \varphi = 0.
 \end{equation}
 This equation should be satisfied identically in $t, x$ and $u.$
 Therefore we nullify the coefficient for $u$ and obtain
 $$
 x \varphi_x - 2 \varphi = 0,
 $$
 whence
 $$
 \varphi (t, x)= c(t) x^2.
 $$
Substitution in Eq. (\ref{PP.eq6}) yields $c'(t) = 0.$ Hence,
 $v = \varphi (t, x)= C x^2$ with arbitrary constant $C.$ Since $\lambda = 0$ in (\ref{PP.eq4})
 and the adjoint equation (\ref{PP.eq3}) is linear and homogeneous in $v,$ one can let $C = 1.$
 Thus, we have demonstrated the following statement.
 \begin{prop}
 \label{Kom:sa1}
 The adjoint equation (\ref{PP.eq3}) has the solution
 \begin{equation}
 \label{PP.eq7}
 v = x^2
 \end{equation}
 for any solution $u$ of Equation (\ref{PP.eq1}). In another
words, \textit{the Kompaneets equation} (\ref{PP.eq1}) \textit{is
 nonlinearly self-adjoint} with the substitution (\ref{gsa:eq1}) given by
(\ref{PP.eq7}).
 \end{prop}
 \begin{rem}
The substitution (\ref{PP.eq7}) does not depend on $u.$ The question
arises on existence of a substitution $v = \varphi (t, x, u )$
involving $u$ if we rewrite Eq. (\ref{PP.eq1}) in an equivalent form
 \begin{equation}
 \label{PP.eq2'}
 \tag{\ref{PP.eq2}$'$}
 \alpha(t, x, u)[- u_t + x^2 u_{xx} + (x^2 + 4 x + 2 x^2 u) u_x
 + 4 x (u + u^2)] = 0
 \end{equation}
with an appropriate multiplier $\alpha \not= 0.$ This question is
investigated
 in next section  for a more general model.
 \end{rem}

 \subsection{The generalized model}
 \label{gsa:7.3}

In the original derivation of Eq. (\ref{Komp.eq1}) the following
more general equation appears accidentally (see \cite{kom56}, Eqs.
(9), (10) and their discussion):
 \begin{equation}
 \label{Komp.eq1:gen}
 \frac{\partial n}{\partial t} = \frac{1}{g(x)}\,
 \frac{\partial}{\partial x}\bigg[g^2(x) \bigg(\frac{\partial
 n}{\partial x} + f(n)\bigg)\bigg]
 \end{equation}
with undetermined functions $f(u)$ and $g(x).$ Then, using a
physical reasoning, Kompaneets takes $f(u)= n(1 + n)$ and $g(x) =
x^2.$ This choice restricts the symmetry properties of the model
significantly. Namely,
 Equation (\ref{Komp.eq1}) has only the time-translational symmetry with
 the generator
 \begin{equation}
 \label{PP.eq1:sym}
 X = \frac{\partial}{\partial t}\,\cdot
 \end{equation}
 The symmetry (\ref{PP.eq1:sym}) provides only one invariant
 solution, namely the stationary solution $n = n(x)$ defined by
 the Riccati equation
 $$
 \frac{d n}{d x} + n^2 + n = \frac{C}{x^4}\,\cdot
 $$
 The generalized model (\ref{Komp.eq1:gen}) can be used for extensions of symmetry properties
 via the methods of \textit{preliminary group classification} \cite{akh-gaz89, ibr-tor91}.
 In this way, exact solutions
  known for particular approximations to the Kompaneets equation can be obtained. This may
 also lead to new approximations of the solutions by taking into
 account various  inevitable perturbations of the idealized
 situation assumed in the Kompaneets model (\ref{Komp.eq1}).

 So, we will take with minor changes in notation the generalized model (\ref{Komp.eq1:gen}):
  \begin{equation}
 \label{PP.eq8}
 u_t = \frac{1}{h(x)}\, D_x\big\{h^2(x) [u_x + f(u)]\big\}, \quad h'(x) \not= 0.
 \end{equation}
 It is written in the expanded form as follows:
 \begin{equation}
 \label{PP.eq9}
 u_t = h(x)\big(u_{xx} + f'(u)u_x\big) + 2 h'(x)\big(u_x + f(u)\big).
 \end{equation}
 We will write  Eq. (\ref{PP.eq9}) in the equivalent form similar to (\ref{PP.eq2'}):
 \begin{equation}
 \label{PP.eq9'}
 %\tag{\ref{PP.eq9}$'$}
  \alpha(t, x, u)\big[- u_t + h(x)\big(u_{xx} + f'(u)u_x\big) + 2 h'(x)\big(u_x +
  f(u)\big)\big] = 0,
 \end{equation}
 where $\alpha \not= 0.$ This provides the following formal Lagrangian:
 \begin{equation}
 \label{PP.eq10}
 {\cal L} = v\,\alpha(t, x, u) \big[- u_t + h(x)\big(u_{xx} + f'(u)u_x\big) + 2 h'(x)\big(u_x +
  f(u)\big)\big],
 \end{equation}
 where $v$ is a new dependent variable. For this Lagrangian, we
 have
 \begin{align}
 \frac{\delta {\cal L}}{\delta u} & = D_t(v \alpha) + D_x^2[h(x) v
 \alpha] - D_x[h(x)f'(u) v \alpha + 2 h'(x) v \alpha]\notag\\[1ex]
 & + h(x) f''(u)\, v\, \alpha\, u_x + 2 h'(x) f'(u) v \alpha  \notag\\[1ex]
 & + v \alpha_u\big[- u_t + h(x)\big(u_{xx} + f'(u)u_x\big) + 2 h'(x)\big(u_x +
  f(u)\big)\big].\notag
 \end{align}
The reckoning shows that
 \begin{align}
  \label{PP.eq11}
 \frac{\delta {\cal L}}{\delta u} & = D_t(v \alpha) + h D_x^2(v \alpha)
  - hf' D_x(v \alpha) + (h'f' - h'') v \alpha\notag\\[1ex]
 & + v \alpha_u\big[- u_t + (u_{xx} + f'u_x)h  + 2 (u_x + f) h'].
 \end{align}

 Now we write the condition for the
self-adjointness of Eq. (\ref{PP.eq9}) in the form
 \begin{equation}
 \label{PP.eq12}
 \frac{\delta {\cal L}}{\delta u}\Big|_{v = \varphi (t, x, u)}
 = \lambda \big[- u_t + (u_{xx} + f'u_x) h + 2 (u_x + f) h']
 \end{equation}
 with an undetermined coefficient $\lambda.$ Substituting (\ref{PP.eq11}) in (\ref{PP.eq12}) we have:
 \begin{equation}
 \label{PP.eq13}
 \begin{split}
 & D_t(\varphi \alpha) + h D_x^2(\varphi \alpha)
  - hf' D_x(\varphi \alpha) + (h'f' - h'') \varphi \alpha\\[1ex]
 & + \varphi \alpha_u\big[- u_t + (u_{xx} + f'u_x)h  + 2 (u_x + f) h']\\[1ex]
 &= \lambda \big[- u_t + (u_{xx} + f'u_x) h + 2 (u_x + f) h'].
 \end{split}
 \end{equation}
Here $\varphi  = \varphi (t, x, u), \ \alpha  = \alpha (t, x, u)$
and consequently (see (\ref{PP.eq5}))
 \begin{equation}
 \label{PP.eq14}
 \begin{split}
   & D_t(\varphi \alpha) = (\varphi \alpha)_u\, u_t + (\varphi \alpha)_t,\\
   & D_x(\varphi \alpha) = (\varphi \alpha)_u\, u_x + (\varphi \alpha)_x,\\
   & D_x^2(\varphi \alpha) = (\varphi \alpha)_u\, u_{xx} + (\varphi \alpha)_{uu}\, u_x^2
   +  2 (\varphi \alpha)_{xu}\, u_x + (\varphi \alpha)_{xx}.
 \end{split}
 \end{equation}
  We substitute (\ref{PP.eq11}) in Eq. (\ref{PP.eq13}), equate the coefficients for $u_t$
 in both sides of the resulting equation and obtain
 $
 (\varphi \alpha)_u - \varphi \alpha_u = - \lambda.
 $
 Hence,
 $$
 \lambda = - \alpha \varphi_u.
 $$
 Using this expression
 for $\lambda$  and equating the coefficients for $h u_{xx}$
 in in both sides of Eq. (\ref{PP.eq13}) we get $(\varphi \alpha)_u + \varphi \alpha_u = -
 \alpha \varphi_u.$ It follows that
 $
 (\varphi \alpha)_u = 0
 $
 and hence
 $$
 \alpha \varphi = k(t, x).
 $$
 Now Eq. (\ref{PP.eq13}) becomes:
 $$
 k_t + h(x) k_{xx} - h''(x) k + f'(u) [h'(x) k - h(x) k_x] = 0.
 $$
 If $f''(u) \not= 0,$ the above equation splits into two
 equations:
 $$
 h'(x) k - h(x) k_x = 0, \quad k_t + h(x) k_{xx} - h''(x) k.
 $$
 The first of these equations yields $k (t, x)= c(t)h(x),$ and
 then the second equation shows that $c'(t) = 0.$ Hence,
 $k = C\, h(x)$ with $C = {\rm const.}$ Letting $C= 1,$ we have:
 \begin{equation}
 \label{PP.eq15}
 \alpha \varphi = h(x).
 \end{equation}
Eq. (\ref{PP.eq15}) can be satisfied by taking, e.g.
 \begin{equation}
 \label{PP.eq16}
  \alpha = \frac{h(x)}{u}\,, \quad \varphi = u.
 \end{equation}
 Thus, we have proved the following statement.
  \begin{prop}
  \label{PP.prop}
 Eq. (\ref{PP.eq8}) written in the equivalent form
 \begin{equation}
 \label{PP.eq17}
 \frac{h(x)}{u}\,u_t = \frac{1}{u}\, D_x\big\{h^2(x) [u_x + f(u)]\big\}
 \end{equation}
 is strictly self-adjoint. In another words, the adjoint equation to Eq. (\ref{PP.eq17})
 coincides with (\ref{PP.eq17}) upon the substitution
 \begin{equation}
 \label{PP.eq18}
 v = u.
 \end{equation}
  \end{prop}

In particular, let us verify by direct calculations that the
original equation (\ref{PP.eq1}) becomes strictly self-adjoint if we
rewrite it in the equivalent form
 \begin{equation}
 \label{PP.eq19}
 \frac{x^2}{u}\, u_t = \frac{1}{u}\, D_x\big[x^4 (u_x + u + u^2)\big].
 \end{equation}
 Eq. (\ref{PP.eq19}) reads
  \begin{equation}
 \label{PP.eq20}
 - \frac{x^2}{u}\, u_t + \frac{x^4}{u}\, u_{xx} + \Big[(x^4 + 4 x^3)\frac{1}{u} + 2 x^4\Big] u_x
 + 4 x^3 (1 + u)= 0
 \end{equation}
 and has the formal Lagrangian
 $$
 {\cal L} = - x^2 \frac{v}{u}\, u_t + x^4 \frac{v}{u}\, u_{xx}
  + \Big[(x^4 + 4 x^3)\frac{v}{u} + 2 x^4 v\Big] u_x
 + 4 x^3 (v + uv).
 $$
Accordingly, the adjoint equation to Eq. (\ref{PP.eq20}) is written
 \begin{align}
 & D_t\Big(x^2 \frac{v}{u}\Big) + D_x^2 \Big(x^4 \frac{v}{u}\Big)
 - D_x \Big[(x^4 + 4 x^3) \frac{v}{u}
  + 2 x^4 v\Big]\notag\\[1.5ex]
 & + x^2 \frac{v}{u^2}\,u_t - x^4 \frac{v}{u^2}\,u_{xx} - (x^4 + 4 x^3) \frac{v}{u^2}\,u_x
 + 4 x^3 v = 0.\notag
 \end{align}
 Letting here $v = u$ one has $v/u = 1$ and after simple calculations arrives at Eq.
 (\ref{PP.eq20}).

%%%%%%%%%%%%%%%%%%%%%%%%%%%%%%%%%%%%%%%%%%%%%%%%%%%%%%%%%%%%%%%%%%%%%%%%%%%%%%%%%%%%%%%%%%%%%%
 %  \subsection{Third definition}
 % \subsection{Generalized self-adjointness and multipliers}
 %\label{sa:eveq.gsa:3}

 %\begin{defn}
 %\label{gsa:def3}
 %The system (\ref{sa:eveq.int.1}) is said to be self-adjoint if
 %there exist multipliers
 %\begin{equation}
 %\label{sa:eveq.int.8}
 %\mu^\alpha_\beta(x, u), \quad \alpha, \beta = 1, \ldots, m,
 %\end{equation}
 %satisfying the condition
 %\begin{equation}
 %\label{sa:eveq.int.9}
 %\|\mu^\alpha_\beta(x, u)\| \not= 0
 %\end{equation}
 %of invertibility of the matrix $\|\mu^\alpha_\beta(x, u)\|,$
 %such the system
 %\begin{equation}
 %\label{sa:eveq.int.10}
 %\widetilde F_\beta \equiv \mu^\alpha_\beta(x, u) F_\alpha \big(x, u, u_{(1)}, \ldots, u_{(s)}\big) = 0, \quad
 %\beta = 1, \ldots, m,
 %\end{equation}
 %is quasi self-adjoint in the sense of Definition \ref{qsa0:def}.
 %Note that the system (\ref{sa:eveq.int.8}) is equivalent to the
 %original system (\ref{sa:eveq.int.1}).
 %\end{defn}

 % \subsection{Equivalence of two definitions}
 %\label{sa:eveq.gsa:4}

 \section{Quasi self-adjoint reaction-diffusion models}
 \label{gqsa:cross}
 \setcounter{equation}{0}

 Let us consider the one-dimensional reaction-diffusion model
  described by the following system (see, e.g. \cite{tsyg07}):

 \begin{equation}
 \label{gqsa:cross.eq1}
 \begin{split}
 & \frac{\partial u}{\partial t} = f(u, v)+A\frac{\partial^2 u}{\partial x^2}
 + \frac{\partial}{\partial x}\left(\phi(u, v) \frac{\partial v}{\partial x}\right), \\[1ex]
 & \frac{\partial v}{\partial t} = g(u, v)+B\frac{\partial^2 v}{\partial x^2}
 +  \frac{\partial}{\partial x}\left(\psi(u, v) \frac{\partial u}{\partial
 x}\right).
 \end{split}
 \end{equation}
It is convenient to write Eqs. (\ref{gqsa:cross.eq1}) in the form
 \begin{equation}
 \label{gqsa:cross.eq2}
 \begin{split}
 & D_t(u) = A D_x^2(u) + D_x\left[\phi(u, v) D_x(v)\right] + f(u, v), \\[1.5ex]
 & D_t(v) = B D_x^2(v) + D_x\left[\psi(u, u) D_x(u)\right] + g(u,
 v).
 \end{split}
 \end{equation}
 The total differentiations have the form
 \begin{equation}
 \label{gqsa:cross.eq3}
 \begin{split}
 & D_t = \frac{\partial}{\partial t} + u_t \frac{\partial}{\partial u}
  + v_t \frac{\partial}{\partial v}
 + u_{tt} \frac{\partial}{\partial u_t} + u_{tx} \frac{\partial}{\partial u_x}
 + v_{tt} \frac{\partial}{\partial v_t} + v_{tx} \frac{\partial}{\partial v_x}
 + \cdots,\\[1.5ex]
 & D_x = \frac{\partial}{\partial x} + u_x \frac{\partial}{\partial u}
  + v_x \frac{\partial}{\partial v}
 + u_{tx} \frac{\partial}{\partial u_t} + u_{xx} \frac{\partial}{\partial u_x}
 + v_{tx} \frac{\partial}{\partial v_t} + v_{xx} \frac{\partial}{\partial v_x}
 + \cdots
 \end{split}
 \end{equation}
 and Eqs. (\ref{gqsa:cross.eq2}) are written
 \begin{equation}
 \label{gqsa:cross.eq4}
 \begin{split}
 & u_t = A u_{xx} + \phi v_{xx} + \left[\phi_u u_x + \phi_v v_x\right]v_x + f, \\[1.5ex]
 & v_t = B v_{xx} + \psi u_{xx} + \left[\psi_u u_x + \psi_v v_x\right]u_x + g.
 \end{split}
 \end{equation}

 The formal Lagrangian for the system (\ref{gqsa:cross.eq4}) is
 \begin{equation}
 \label{gqsa:cross.eq5}
 \begin{split}
 & {\cal L} = z (A u_{xx} - u_t + \phi v_{xx}
 + \phi_u u_x v_x + \phi_v v_x^2 + f) \\[0.5ex]
 & + w (B v_{xx} - v_t + \psi u_{xx} + \psi_u u_x^2
 + \psi_v u_x v_x + g),
 \end{split}
 \end{equation}
 where $z$ and $w$ are new dependent variables. Eqs.
(\ref{sa:eveq.int.3}) are written:
 $$
 F_1^* = \frac{\delta {\cal L}}{\delta u} = D^2_x\left(\frac{\partial {\cal L}}{\partial u_{xx}}\right)
  - D_t\left(\frac{\partial {\cal L}}{\partial u_t}\right)
 - D_x\left(\frac{\partial {\cal L}}{\partial u_x}\right)
 + \frac{\partial {\cal L}}{\partial u}\,,
 $$
 $$
 F_2^* = \frac{\delta {\cal L}}{\delta v} = D^2_x\left(\frac{\partial {\cal L}}{\partial v_{xx}}\right)
  - D_t\left(\frac{\partial {\cal L}}{\partial v_t}\right)
 - D_x\left(\frac{\partial {\cal L}}{\partial v_x}\right)
 + \frac{\partial {\cal L}}{\partial v}\,\cdot
 $$
 Substituting here the expression (\ref{gqsa:cross.eq5}) for ${\cal L}$ we obtain
 after simple calculations the following adjoint
 equations (\ref{gsa:adeqs}) to the system (\ref{gqsa:cross.eq4}):
 \begin{equation}
 \label{gqsa:cross.eq6}
 A z_{xx} + z_t + \psi_v v_x w_x - \phi_u v_x z_x + \psi w_{xx}
 + z f_u + w g_u = 0,
 \end{equation}
 \begin{equation}
 \label{gqsa:cross.eq7}
 B w_{xx} + w_t + \phi_u u_x z_x - \psi_v u_x w_x  + \phi z_{xx}
 + z f_v + w g_v = 0.
 \end{equation}

Let us investigate the system (\ref{gqsa:cross.eq4}) for quasi
self-adjointness (Definition \ref{qsa:def}). We
 write the left-hand sides of Eqs. (\ref{gqsa:cross.eq6}) and  (\ref{gqsa:cross.eq7})
 as linear combinations of the left-hand sides of Eqs.
 (\ref{gqsa:cross.eq4}):
 \begin{align}
 \label{gqsa:cross.eq8}
  &  A z_{xx} + z_t + \psi_v v_x w_x - \phi_u v_x z_x + \psi w_{xx}
 + z f_u + w g_u\\[1ex]
 & = (A u_{xx} - u_t + \phi v_{xx} + \phi_u u_x v_x + \phi_v v_x^2 + f) P \notag \\[0.5ex]
 & + ( B v_{xx} - v_t + \psi u_{xx} + \psi_u u_x^2 + \psi_v u_x v_x + g) Q,\notag
 \end{align}
 \begin{align}
 \label{gqsa:cross.eq9}
  & B w_{xx} + w_t + \phi_u u_x z_x - \psi_v u_x w_x  + \phi z_{xx}
 + z f_v + w g_v\\[1ex]
 & = (A u_{xx} - u_t + \phi v_{xx} + \phi_u u_x v_x + \phi_v v_x^2 + f) M \notag \\[0.5ex]
 & + ( B v_{xx} - v_t + \psi u_{xx} + \psi_u u_x^2 + \psi_v u_x v_x + g) N,\notag
 \end{align}
 where $P, Q, M$ and $N$ are undetermined coefficients.
 We write the substitution (\ref{qsa.1}) in the form
 \begin{equation}
 \label{gqsa:cross.eq10}
 z = Z(u, v), \quad w = W(u, v)
 \end{equation}
 and insert in the left-hand sides of Eqs.
 (\ref{gqsa:cross.eq8})-(\ref{gqsa:cross.eq9}) these expressions for $z, w$ together with their
 derivatives
 \begin{align}
 %\label{gqsa:cross.eq12}
 %\begin{split}
 & z_t = Z_u u_t + Z_v v_t, \quad z_x = Z_u u_x + Z_v v_x,\notag\\[1ex]
 & z_{xx} = Z_u u_{xx} + Z_v v_{xx} + Z_{uu} u^2_x + 2 Z_{uv} u_x v_x
 + Z_{vv} v^2_x,\notag\\[1ex]
 & w_t = W_u u_t + W_v v_t, \quad w_x = W_u u_x + W_v v_x,\notag\\[1ex]
 & w_{xx} = W_u u_{xx} + W_v v_{xx} + W_{uu} u^2_x + 2 W_{uv} u_x v_x
 + W_{vv} v^2_x.\notag
 %\end{split}
 \end{align}
 Equating the coefficients for $u_t$ and $v_t$
 in both sides of Eqs.
(\ref{gqsa:cross.eq8})-(\ref{gqsa:cross.eq9}) we obtain
 \begin{equation}
 \label{gqsa:cross.eq11}
 \begin{split}
 & P = - Z_u, \quad Q = - Z_v,\\
 & N = - W_v, \quad M = - W_u.
 \end{split}
 \end{equation}
 Now we calculate the coefficients for $u_{xx}$ and $v_{xx},$ take
 into account Eqs. (\ref{gqsa:cross.eq11}) and arrive at the
 following equations:
 \begin{equation}
 \label{gqsa:cross.eq12}
 \begin{split}
 & 2 A Z_u + \psi Z_v + \psi W_u = 0,\\[1ex]
 & (A + B) Z_v + \phi Z_u + \psi W_v = 0,\\[1ex]
 & 2 B W_v + \phi Z_v + \psi W_u = 0,\\[1ex]
 & (A + B) W_u + \phi Z_u + \psi W_v = 0.
 \end{split}
 \end{equation}

 Eqs. (\ref{gqsa:cross.eq12}) provide a linear homogeneous
 algebraic equations for the quantities
 $$
 Z_u, \quad Z_v, \quad W_u, \quad W_u
 $$
 with the matrix
 $$
  \begin{pmatrix}
 2 A & \psi & \psi & 0\\
 \phi & A + B & 0 & \psi\\
 0 & \phi & \phi & 2 B\\
 \phi & 0 & A + B & \psi
 \end{pmatrix}.
 $$
 This matrix has the inverse because its determinant is equal to
 $$
 4 (A + B)^2 (\phi \psi - AB)
 $$
 and does not vanish in the case of arbitrary $A, B, \phi$
 and $\psi.$  Hence, Eqs. (\ref{gqsa:cross.eq12}) yield:
 \begin{equation}
 \label{gqsa:cross.eq13}
  Z_u = Z_v = W_u = W_u= 0.
 \end{equation}
 It follows that $Z(u, v) = C_1, \ W(u, v) = C_2.$
 Thus, the substitution (\ref{qsa.1}) has the form
 \begin{equation}
 \label{gqsa:cross.eq14}
 z = C_1, \quad w = C_2
 \end{equation}
 with arbitrary constants \ $C_1, \ C_2.$ Then Eqs.
 (\ref{gqsa:cross.eq8})-(\ref{gqsa:cross.eq9}) become
 $$
 (C_1 f + C_2 g)_u = 0, \quad (C_1 f + C_2 g)_v =0
 $$
 and yield $$\tilde f + \tilde g = C,$$  where $\tilde f = C_1 f, \ \tilde g = C_2 g,$ and $C =$ const.
 Since $\tilde f$ and $\tilde g,$ along with $f$ and $g,$ are
arbitrary functions, we can omit the ``tilde" and write
 \begin{equation}
 \label{gqsa:cross.eq15}
  f + g = C.
 \end{equation}
 Eq. (\ref{gqsa:cross.eq15}) provides the necessary and sufficient
 condition for the quasi self-adjointness of the system (\ref{gqsa:cross.eq1}).
  Thus, we have proved the following statement.

 \begin{thm}
 \label{gqsa:cross.thm}
 The system (\ref{gqsa:cross.eq1}) is quasi self-adjoint if and only
 if it has the form
  \begin{equation}
 \label{gqsa:cross.eq16}
 \begin{split}
 & \frac{\partial u}{\partial t} = f(u, v)+A\frac{\partial^2 u}{\partial x^2}
 + \frac{\partial}{\partial x}\left(\phi(u, v) \frac{\partial v}{\partial x}\right), \\[1.5ex]
 & \frac{\partial v}{\partial t} = C - f(u, v)+B\frac{\partial^2 v}{\partial x^2}
 +  \frac{\partial}{\partial x}\left(\psi(u, v) \frac{\partial u}{\partial
 x}\right),
 \end{split}
 \end{equation}
 where $\phi(u, v), \ \psi(u, v), \ f(u, v)$ are arbitrary functions and $A, B,
 C$ are arbitrary constants. The substitution (\ref{qsa.1}) is given
by (\ref{gqsa:cross.eq14}).
 \end{thm}

 \begin{rem}
 If we replace (\ref{gqsa:cross.eq10}) by the general substitution
(\ref{gsa:eq1}), i.e. take
 \begin{equation}
 \label{gqsa:cross.eq17}
 z = Z(t, x, u, v), \quad w = W(t, x, u, v),
 \end{equation}
 then Eqs. (\ref{gqsa:cross.eq14}) will be replaced by
 \begin{equation}
 \label{gqsa:cross.eq18}
 z = Z(t, x), \quad w = W(t, x),
 \end{equation}
 with functions $Z(t, x), \ W(t, x)$ satisfying the following
 equations:
 \begin{equation}
 \label{gqsa:cross.eq19}
  \left(\psi_v W - \phi_u Z\right)_x  =0,
 \end{equation}
  \begin{equation}
 \label{gqsa:cross.eq20}
 \begin{split}
 & A Z_{xx} + Z_t + \psi W_{xx} + (f Z + g W)_u = 0, \\[1.5ex]
 & B W_{xx} + W_t + \phi Z_{xx} + (f Z + g W)_v = 0.
 \end{split}
 \end{equation}
 \end{rem}

 \section{A model of an irrigation system}
 \label{irrig}
 \setcounter{equation}{0}

 Let us consider the second-order nonlinear partial differential
 equation
 \begin{equation}
 \label{irrig.eq1}
 C(\psi) \psi_t= \left[ K(\psi)\psi_{x}\right]_x
  + \left[K(\psi) \left( \psi_z - 1 \right)\right]_z - S(\psi).
\end{equation}
It serves as a mathematical model for investigating certain
irrigation systems (see \cite{ibr94-96}, vol. 2, Section 9.8 and the
references therein). The dependent variable $\psi $ denotes the soil
moisture pressure head, $C(\psi)$ is the specific water capacity,
$K(\psi)$ is the unsaturated hydraulic conductivity, $S(\psi)$ is a
source term. The independent variables are the time  $t,$ the
horizontal axis $x$ and the vertical axis $z$ which is taken to be
positive downward.

The adjoint equation (\ref{gsa:adeqs}) to Eq. (\ref{irrig.eq1}) has
the form
 \begin{equation}
 \label{irrig.eq2}
 C(\psi) v_t + K(\psi)\left[v_{xx} + v_{zz}\right] + K'(\psi)v_z - S\,'(\psi) v = 0.
\end{equation}
It follows from (\ref{irrig.eq2}) that Eq. (\ref{irrig.eq1}) is not
nonlinearly self-adjoint if $C(\psi), K(\psi)$ and $S(\psi)$ are
arbitrary functions. Indeed, using Definition \ref{gsa:def2} of the
nonlinear self-adjointness and nullifying in (\ref{irrig.eq2}) the
term with
 $S\,'(\psi)$ we obtain $v = 0.$ Hence, the condition
(\ref{gsa:eq2}) of the nonlinear self-adjointness is not satisfied.

 However, Eq. (\ref{irrig.eq1}) can be nonlinearly self-adjoint
 if there are certain relations between the functions
$C(\psi), K(\psi)$ and $S(\psi).$ For example, let us suppose that
the specific water capacity $C(\psi)$ and the hydraulic conductivity
$K(\psi)$ are arbitrary, but the source term $S(\psi)$ is related
with $C(\psi)$ by the equation
 \begin{equation}
 \label{irrig.eq3}
 S\,'(\psi) = a C(\psi), \quad a = {\rm const.}
\end{equation}
Then Eq. (\ref{irrig.eq2}) becomes
 $$
  C(\psi) [v_t - a v] + K(\psi)\left[v_{xx} + v_{zz}\right] + K'(\psi)v_z = 0
 $$
 and yields:
 \begin{equation}
 \label{irrig.eq4}
 v_z = 0, \quad  v_{xx} = 0, \quad v_t - a v = 0.
 \end{equation}
 We solve the first two equations (\ref{irrig.eq4}) and obtain
 $$
 v = p(t) x + q(t).
 $$
 We substitute this in the third equation (\ref{irrig.eq4}),
 $$
 [p'(t) - a p(t)] x + q'(t) - a q(t) = 0,
 $$
 split it with respect to $x$ and obtain:
 $$
 p'(t) - a p(t) = 0, \quad q'(t) - a q(t) = 0,
 $$
whence
 $$
 p(t) = b {\rm e}^{at},  \quad q(t) = l {\rm e}^{at}, \quad b, l = {\rm const.}
 $$
 Thus, Eq. (\ref{irrig.eq1}) satisfying the condition Eq.
 (\ref{irrig.eq3}) is nonlinearly self-adjoint, and the substitution
 (\ref{gsa:eq1}) has the form
 \begin{equation}
 \label{irrig.eq5}
 v = (b x + l) {\rm e}^{at}.
 \end{equation}

 One can obtain various nonlinearly self-adjoint Equations (\ref{irrig.eq1}) by
 considering other relations between $C(\psi), K(\psi)$ and
$S(\psi)$ different from (\ref{irrig.eq3}).

% \section{Krichever-Novikov (KN) equation}
% \label{KN}
% \setcounter{equation}{0}

% \begin{equation}
% \label{KN.eq1}
% u_t - u_{xxx} + \frac{3}{2}\,\frac{u^2_{xx}}{u_x} - \frac{P(u)}{u_x} =
% 0,
% \end{equation}
%where $P(u)$ a polynomial  of degree four with distinct roots.

 \newpage
 \begin{center}
 \section*{{\sc Part 2}\protect\\ Construction of conservation laws\\ using symmetries}
 \end{center}
 \addcontentsline{toc}{chapter}{Part 2. Construction of
 conservation laws using symmetries}
 \label{concept:P2}

 %\section{Preliminaries}
 \section{Discussion of the operator identity}
 %\section{Utilization of an operator identity}
 \label{opid}
 \setcounter{equation}{0}

 \subsection{Operator identity and alternative proof of Noether's theorem}

Let us discuss some consequences of the operator
identity\footnote{Recently I learned that the identity
(\ref{opid.eq1}) was proved in \cite{ros72}. Namely, Eq.
(\ref{opid.eq1}) is the same (except for notation) as Eq. (19) from
\cite{ros72}. The operator identity (\ref{opid.eq1}) was
rediscovered in \cite{ibr79a} and used for simplifying
 the proof of Noether's theorem. Accordingly, Eq. (\ref{opid.eq1})
 was called in \cite{ibr79a}
 the Noether identity. See also \cite{ibr99}, Section 8.4.}
 \begin{equation}
 \label{opid.eq1}
X + D_i(\xi^i) = W^\alpha \frac{\delta}{\delta u^\alpha} + D_i {\sf
N}^i\;.
\end{equation}
 Here
 \begin{equation}
 \label{opid.eq2}
 X=\xi^i \frac{\partial}{\partial x^i} + \eta^\alpha
 \frac{\partial}{\partial u^\alpha} + \zeta_i^\alpha
 \frac{\partial}{\partial u_i^\alpha} + \zeta_{i_1i_2}^\alpha
 \frac{\partial}{\partial u_{i_1i_2}^\alpha} + \cdots\,,
 \end{equation}
 \begin{equation}
 \label{opid.eq3}
 W^\alpha = \eta^\alpha - \xi^j u_j^\alpha, \quad \alpha = 1,
 \ldots, m,
 \end{equation}
 \begin{equation}
 \label{opid.eq4}
 \frac{\delta}{\delta u^\alpha} = \frac{\partial}{\partial u^\alpha}
+ \sum_{s=1}^\infty (-1)^s D_{i_1}\cdots
D_{i_s}\,\frac{\partial}{\partial u^\alpha_{i_1\cdots i_s}}\,,\quad
\alpha=1, \ldots, m,
 \end{equation}
and
 \begin{equation}
 \label{opid.eq5}
 {\sf N}^i = \xi^i + W^\alpha\;\frac{\delta}{\delta u^\alpha_i} +
 \sum_{s=1}^{\infty} D_{i_1}\cdots D_{i_s}
 (W^\alpha)\;\frac{\delta}{\delta u^\alpha_{ii_1\cdots i_s}}\,, \quad
 i = 1, \ldots, n,
\end{equation}
where the Euler-Lagrange operators with respect to derivatives of
$u^\alpha$ are obtained from (\ref{opid.eq4}) by replacing
$u^\alpha$ by the corresponding derivatives, e.g.
 \begin{equation}
 \label{opid.eq6}
 \frac{\delta}{\delta u^\alpha_i} =
 \frac{\partial}{\partial u^\alpha_i} + \sum_{s=1}^\infty (-1)^s
 D_{j_1}\cdots D_{j_s}\;\frac{\partial}{\partial u^\alpha_{ij_1\cdots
 j_s}}\,\cdot
 \end{equation}
The coefficients $\xi^i, \ \eta^\alpha$ in (\ref{opid.eq2}) are
arbitrary \textit{differential functions} (see Section \ref{nota})
and the other coefficients are determined by the prolongation
formulae
 \begin{equation}
 \label{opid.eq7}
\zeta_i^{\alpha} =  D_i(W^{\alpha}) + \xi ^ju_{ij}^{\alpha},\quad
\zeta _{i_1i_2}^{\alpha} =  D_{i_1}D_{i_2}(W^{\alpha}) + \xi^j
u_{ji_1i_2}^{\alpha}, \ldots \,.
 \end{equation}
 The derivation of Eq. (\ref{opid.eq1}) is essentially based on
 Eqs. (\ref{opid.eq7}).

 Recall that Noether's theorem, associating conservation laws with
symmetries of differential equations obtained from variational
principles, was originally proved by calculus of variations. The
alternative proof of this theorem given in \cite{ibr79a}
 (see also \cite{ibr83, ibr99}) is
 based on the identity (\ref{opid.eq1}) and is
 simple. Namely, let us consider the
Euler-Lagrange equations
 \begin{equation}
 \label{opid.eq8}
 \frac{\delta {\cal L}}{\delta u^\alpha} = 0, \quad \alpha = 1, \ldots,
 m.
 \end{equation}
 If we assume that the operator (\ref{opid.eq2}) is admitted by
 Eqs. (\ref{opid.eq8}) and that the variational integral $$\int
{\cal L}(x, u, u_{(1)}, \ldots) dx$$ is invariant under the
transformations of the group with the generator $X$ then the
following equation holds:
 \begin{equation}
 \label{opid.eq9}
 X({\cal L}) + D_i(\xi^i){\cal L}= 0.
 \end{equation}
Therefore, if we act on ${\cal L}$ by both sides of the identity
(\ref{opid.eq1}),
 $$
 X({\cal L}) + D_i(\xi^i){\cal L} = W^\alpha
 \frac{\delta {\cal L}}{\delta u^\alpha} + D_i [{\sf N}^i({\cal L})]\;,
 $$
 and take into account  Eqs. (\ref{opid.eq8}), (\ref{opid.eq9}), we see
 that the vector with the components
 \begin{equation}
 \label{opid.eq10}
 C^i = {\sf N}^i({\cal L}), \quad i = 1, \ldots, n,
 \end{equation}
 satisfies the conservation equation
 \begin{equation}
 \label{opid.eq11}
 \left.D_i(C^i)\right|_{(\ref{opid.eq8})} = 0.
 \end{equation}
 For practical applications, when we deal with law order Lagrangians ${\cal L},$ it is convenient to
 restrict the operator (\ref{opid.eq5}) on the derivatives involved
 in ${\cal L}$  and write the expressions (\ref{opid.eq10}) in the
expanded form
 \begin{align}
 \label{opid.eq12}
 & C^i  = \xi^i {\cal L}+W^\alpha\,
 \left[\frac{\partial {\cal L}}{\partial u_i^\alpha} -
 D_j \left(\frac{\partial {\cal L}}{\partial u_{ij}^\alpha}\right)
 + D_j D_k\left(\frac{\partial {\cal L}}{\partial u_{ijk}^\alpha}\right) -
 \ldots\right]\\[1.5ex]
 &+D_j\left(W^\alpha\right)\,
 \left[\frac{\partial {\cal L}}{\partial u_{ij}^\alpha} -
 D_k \left(\frac{\partial {\cal L}}{\partial
 u_{ijk}^\alpha}\right) + \ldots\right]
 + D_j D_k\left(W^\alpha\right)\left[\frac{\partial {\cal L}}{\partial u_{ijk}^\alpha} -
 \ldots\right].\notag
  \end{align}
Thus, Noether's theorem can be formulated as follows.
 \begin{thm}
 \label{Noether}
 If the operator (\ref{opid.eq2}) is admitted by
 Eqs. (\ref{opid.eq8}) and satisfies the condition (\ref{opid.eq9})
of the invariance of the variational integral, then the vector
(\ref{opid.eq12}) constructed by Eqs. (\ref{opid.eq12}) satisfies
the conservation law (\ref{opid.eq11}).
 \end{thm}

 \begin{rem}
 \label{opid.rem1}
 The identity (\ref{opid.eq1}) is valid also in the case when the coefficients
$\xi^i, \ \eta^\alpha$ of the operator $X$
 involve not only the \textit{local variables}  $x, u, u_{(1)},
u_{(2)}, \ldots $ but also \textit{nonlocal
 variables} (see  Section \ref{gas:5}). Accordingly, the formula (\ref{opid.eq12})
associates conserved vectors with \textit{nonlocal symmetries} as
well.
 \end{rem}
 \begin{rem}
 \label{opid.rem2}
 If the invariance condition (\ref{opid.eq9}) is replaced by the divergence condition
 $$
 X({\cal L}) + D_i(\xi^i){\cal L}= D_i(B^i),
 $$
then the identity (\ref{opid.eq1}) leads to the conservation law
(\ref{opid.eq11}) where the conserved vector (\ref{opid.eq10}) is
replaced with
 \begin{equation}
 \label{opid.eq13}
 C^i = {\sf N}^i({\cal L}) - B^i, \quad i = 1, \ldots, n.
 \end{equation}
 \end{rem}
 \begin{rem}
 \label{opid.rem3}
 If we write the operator (\ref{opid.eq2}) in the equivalent form
 \begin{equation}
 \label{opid.eq14}
 X=W^\alpha
 \frac{\partial}{\partial u^\alpha} + \zeta_i^\alpha
 \frac{\partial}{\partial u_i^\alpha} + \zeta_{i_1i_2}^\alpha
 \frac{\partial}{\partial u_{i_1i_2}^\alpha} + \cdots\,,
 \end{equation}
 then the prolongation formulae (\ref{opid.eq7}) become simpler:
 \begin{equation}
 \label{opid.eq15}
 \zeta_i^{\alpha} =  D_i(W^{\alpha}),\quad
\zeta _{i_1i_2}^{\alpha} =  D_{i_1}D_{i_2}(W^{\alpha}), \ldots \,.
 \end{equation}
 \end{rem}

 \subsection{Test for total derivative and for for divergence}
 \label{test}
 %\subsection{Divergence test. Integrating factor for linear ODEs}
 %\subsection{Integrating factors and adjoint equations to linear ODEs}

I recall here the well-known necessary and sufficient condition for
a differential function to be divergence, or total derivative in the
case of one independent variable.

 One can easily derive from the definition (\ref{sa:cl.diff1}) of the
 total differentiation $D_i$ the following lemmas (see also \cite {ibr99},
 Section 8.4.1).
 \begin{lem}
 \label{opid:lem1}
 The following infinite series of equations hold:
 \begin{align}
  \frac{\partial}{\partial u^\alpha}\,D_i & =
 D_i\,\frac{\partial}{\partial u^\alpha}\,,\notag\\[1ex]
  D_j \frac{\partial}{\partial u^\alpha_j}\,D_i & =
 D_i\,\frac{\partial}{\partial u^\alpha} + D_i D_j \frac{\partial}{\partial u^\alpha_j}\,,\notag\\[1ex]
  D_j D_k \frac{\partial}{\partial u^\alpha_{jk}}\,D_i & =
 D_i D_k\,\frac{\partial}{\partial u^\alpha_k} + D_i D_j D_k\frac{\partial}{\partial u^\alpha_{jk}}\,,\notag\\[1ex]
 \cdots \cdots \cdots \cdots \cdots & \cdots \cdots \cdots \cdots \cdots \cdots \cdots \cdots \cdots \cdots \notag
 \end{align}
 \end{lem}
 \begin{lem}
 \label{opid:lem2}
 The following operator identity holds for every $i$ and $\alpha:$
 $$
 \frac{\delta}{\delta u^\alpha}\, D_i = 0.
 $$
\end{lem}
{\it Proof.} Using Lemma \ref{opid:lem1} and manipulatinng with
summation indices  we obtain:
 \begin{align}
 \frac{\delta}{\delta u^\alpha} D_i & =
 \left(\frac{\partial}{\partial u^\alpha} - D_j
 \frac{\partial}{\partial u^\alpha_j} + D_j D_k
 \frac{\partial}{\partial u^\alpha_{jk}} - D_j D_k D_l
 \frac{\partial}{\partial u^\alpha_{jkl}} + \cdots \right)
 D_i\notag\\[1ex]
& =
  \frac{\partial}{\partial u^\alpha} D_i  - D_i\,\frac{\partial}{\partial u^\alpha}
 - D_i D_j \frac{\partial}{\partial u^\alpha_j}
 +  D_i D_k\,\frac{\partial}{\partial u^\alpha_k}
 + D_i D_j D_k\frac{\partial}{\partial u^\alpha_{jk}}\notag\\[1ex]
&
 - D_i D_k D_l \frac{\partial}{\partial u^\alpha_{kl}} - \cdots = 0. \notag
 \end{align}

 \begin{prop}
 \label{test:div}
A differential function ~$f(x, u, u_{(1)}, \ldots, u_{(s)}) \in
{\cal A}$  is divergence,
\begin{equation}
 \label{opid.eq16}
 f = D_i (h^i), \quad h^i(x, u, \ldots, u_{(s - 1)})
\in {\cal A},
\end{equation}
if and only if the following equations hold identically in $x, u,
u_{(1)}, \ldots\,:$
\begin{equation}
 \label{opid.eq17a}
 \frac{\delta f}{\delta u^\alpha} = 0, \quad \alpha = 1,
\ldots, m.
\end{equation}
 \end{prop}

 The statement that (\ref{opid.eq16}) implies (\ref{opid.eq17a}) follows immediately from
Lemma \ref{opid:lem2}.
 For the proof of the inverse statement that (\ref{opid.eq17a}) implies (\ref{opid.eq16}),
 see  \cite{cou-hil89}, Chapter 4, \S~3.5, and \cite{ros72}. See also  \cite {ibr99}, Section 8.4.1.

We will use Proposition \ref{test:div} also in the particular case
of one independent variable $x$ and one dependent variable $u = y.$
Then it is formulated as follows.

 \begin{prop}
 \label{total:der}
 A differential function ~$f(x, y, y', \ldots, y^{(s)}) \in {\cal A}$
 is the total derivative,
\begin{equation}
 \label{opid.eq18}
 f = D_x (g), \quad g(x, y, y', \ldots, y^{(s - 1)})
\in {\cal A},
\end{equation}
if and only if the following equation holds identically in $x, y,
y', \ldots\,:$
\begin{equation}
 \label{opid.eq17}
 \frac{\delta f}{\delta y} = 0.
\end{equation}
 \end{prop}
 Here ${\delta f}/{\delta y}$ is the Euler-Lagrange operator
 (\ref{opid.eq6}):
\begin{equation}
 \label{opid.eq19}
 \frac{\delta}{\delta y} =
\frac{\partial}{\partial y} - D_x \frac{\partial}{\partial y'} +
D_x^2 \frac{\partial}{\partial y''} - D_x^3 \frac{\partial}{\partial
y'''} + \cdots \,.
\end{equation}

 \subsection{Adjoint equation to linear ODE}
 \label{ttd}

Let us consider an arbitrary $s$th-order linear ordinary
differential operator
 \begin{equation}
 \label{opid.eq20}
 L[y] = a_0 y^{(s)} + a_1 y^{(s-1)} + \cdots + a_{s-2} y''
 + a_{s-1} y' + a_s y,
 \end{equation}
 where $a_i= a_i(x).$ We know from Section \ref{sa:eveq.3} that the adjoint operator to
(\ref{opid.eq20}) can be calculated by using Eq.
(\ref{sa:eveq.int.3}). I give here the independent proof  based on
the operator identity (\ref{opid.eq1}).
 \begin{prop}
 \label{total:prop2}
The adjoint operator to (\ref{opid.eq20}) can be calculated by the
formula
 \begin{equation}
 \label{opid.eq21}
 L^*[z] = \frac{\delta (z L[y])}{\delta y}\,\cdot
 \end{equation}
 \end{prop}
 \textbf{Proof.} Let
 \begin{equation}
 \label{opid.eq22}
 X=w \frac{\partial}{\partial y} + w'
 \frac{\partial}{\partial y'} + w''
 \frac{\partial}{\partial y''} + \cdots
 \end{equation}
 % We will write the operator (\ref{opid.eq14}) with one independent
 % variable $x$ and one dependent variable $u = y$  in the form
be the operator (\ref{opid.eq14}) with one independent variable $x$
and one dependent variable $u = y,$ where the prolongation formulae
(\ref{opid.eq15}) are written using the notation
 \begin{equation}
 \label{opid.eq23}
 w' = D_x(w), \quad  w''= D^2_x(w), \ldots\,.
 \end{equation}
In this notation the operator (\ref{opid.eq5}) is written
 $$
 {\sf N} = w\;\frac{\delta}{\delta y'} +
 w'\;\frac{\delta}{\delta y''} + w''\;\frac{\delta}{\delta y'''} +
\cdots\,.
 $$
 Having in mind its application to the differential function $L[y]$ given by (\ref{opid.eq20})
we consider the following restricted form of ${\sf N}:$
 \begin{equation}
 \label{opid.eq24}
 {\sf N} = w\;\frac{\delta}{\delta y'} +
 w'\;\frac{\delta}{\delta y''}  +
\cdots + w^{(s-1)}\;\frac{\delta}{\delta y^{(s)}}\,.
\end{equation}
The identity (\ref{opid.eq1}) has the form
 \begin{equation}
 \label{opid.eq25}
 X = w \frac{\delta}{\delta y} + D_x {\sf N}\;.
 \end{equation}
We act  by both sides of this identity on $z L[y],$ where $z$ is a
new dependent variable:
 \begin{equation}
 \label{opid.eq26}
 X (z L[y]) = w \frac{\delta (z L[y])}{\delta y}
 + D_x {\sf N}(z L[y])\;.
 \end{equation}
Since the operator (\ref{opid.eq22}) does not act on the variables
$x$ and $z,$ we have
 \begin{equation}
 \label{opid.eq27}
 X (z L[y]) = z X (L[y]).
 \end{equation}
Furthermore we note
 that
 \begin{equation}
 \label{opid.eq28}
 X (L[y]) = L[w].
 \end{equation}
 Inserting (\ref{opid.eq27}) and (\ref{opid.eq28}) in Eq.
(\ref{opid.eq26}) we obtain
 \begin{equation}
 \label{opid.eq29}
 z L[w] - w \frac{\delta (z L[y])}{\delta y} = D_x (\Psi)\;,
 \end{equation}
 where $\Psi$ is a quadratic form $\Psi = \Psi [w, z]$ defined by
 \begin{equation}
 \label{opid.eq30}
\Psi = {\sf N}(z L[y]).
 \end{equation}
 After replacing $w$ with $y$  Eq. (\ref{opid.eq29}) coincides
 with Eq. (\ref{clad}) for the adjoint operator,
 \begin{equation}
 \label{opid.eq31}
 z L[y] - y L^*[z] = D_x (\psi),
 \end{equation}
 where $L^*[z]$ is given by the formula (\ref{opid.eq21}) and $\psi = \psi[y, z]$
 is defined by
 \begin{equation}
 \label{opid.eq32}
  \psi[y, z] = \Psi[w, z]\big|_{w = y} \equiv {\sf N}(z L[y])\big|_{w = y}.
\end{equation}
 \begin{rem}
 \label{opid.rem4}
 Let us find the explicit formula for $\psi$ in Eq. (\ref{opid.eq31})
  We write the
operator ${\sf N}$ given by Eq. (\ref{opid.eq24}) in the expanded
form
 \begin{align}
 {\sf N} & = w\left[\frac{\partial}{\partial y'} - D_x \frac{\partial}{\partial y''} + \cdots
 + (- D_x)^{s-1} \frac{\partial}{\partial y^{(s)}}\right]\notag\\
 & + w'\left[\frac{\partial}{\partial y''} - D_x \frac{\partial}{\partial y'''} + \cdots
 + (- D_x)^{s-2} \frac{\partial}{\partial y^{(s)}}\right] + \cdots \notag\\
 & + w^{(s-2)}\left[\frac{\partial}{\partial y^{(s-1)}} - D_x \frac{\partial}{\partial y^{(s)}}\right]
 + w^{(s-1)}\;\frac{\delta}{\delta y^{(s)}}\,,\notag
\end{align}
 act on $z L[y]$ written in the form
 $$
 z L[y] = a_s y z + a_{s-1} y' z + a_{s-2} y'' z + \cdots + a_1 y^{(s-1)} z + a_0 y^{(s)} z,
 $$
 and obtain $\Psi.$  We replace $w$ with $y$  in $\Psi = \Psi [w, z]$ and  $\psi = \psi[y, z]:$
 \begin{equation}
 \label{opid.eq33}
 \begin{split}
 \psi[y, z] & = y\left[a_{s-1}\, z - (a_{s-2}\, z)' + \cdots + (- 1)^{s-1} (a_0\, z)^{(s-1)} \right]\\
 & + y'\left[a_{s-2}\, z - (a_{s-3}\, z)' + \cdots + (- 1)^{s-2} (a_0\, z)^{(s-2)}\right] + \cdots \\
 & + y^{(s-2)}\left[a_1\, z - (a_0\, z)'\right]
 + y^{(s-1)}\;a_0\, z.
\end{split}
\end{equation}
%This is precisely the classical expression given in \cite{step58},
%Chapter 5, \S 4, Eq. (31$'$).

The expression (\ref{opid.eq33})  is obtained in the classical
literature using integration by parts (see, e.g. \cite{step58},
Chapter 5, \S 4, Eq. (31$'$)).
 \end{rem}

 \subsection{Conservation laws and integrating factors for linear ODEs}
 \label{int.f}

 Consider an $s$th-order homogeneous linear ordinary differential equation
 \begin{equation}
 \label{opid.eq34}
 L[y] = 0,
 \end{equation}
 where $L[y]$ is the operator defined by Eq. (\ref{opid.eq20}). If $L[y]$ is a total derivative,
 \begin{equation}
 \label{opid.eq35}
 L[y]  = D_x \left(\psi (x, y, y', \ldots, y^{(s-1)})\right),
 \end{equation}
 Eq. (\ref{opid.eq34}) is written as a conservation law
 $$
 D_x \left(\psi (x, y, y', \ldots, y^{(s-1)})\right) = 0,
 $$
 whence upon integration one obtains a linear equation of order
$s-1:$
 \begin{equation}
 \label{opid.eq36}
 \psi (x, y, y', \ldots, y^{(s-1)}) = C_1.
 \end{equation}
 We can also reduce the order of the non-homogeneous equation
 \begin{equation}
 \label{opid.eq37}
 L[y] = f(x)
 \end{equation}
 by rewriting it in the the conservation form
 \begin{equation}
 \label{opid.eq38}
  D_x\left[\psi (x, y, y', \ldots, y^{(s-1)}) - \int f(x) dx\right] = 0.
 \end{equation}
Integrating it once we obtain the non-homogeneous linear equation of
order $s-1:$
 $$
 \psi (x, y, y', \ldots, y^{(s-1)}) = C_1 + \int f(x) dx.
 $$
 \begin{exa}
 \label{opid.exa1}
 Consider the second-order equation
$$
 y'' + y' \sin x  + y \cos x = 0.
 $$
We have
$$
y'' + y' \sin x  + y \cos x = D_x(y' + y \sin x).
$$
Therefore the second-order equation in question reduces to the
first-order equation
 $$
 y' + y \sin x = C_1.
 $$
Integrating the latter equation we obtain the  general solution
$$
 y = \left[C_2 +  C_1 \int {\rm e}^{- \cos x}\,dx\right] {\rm e}^{\cos x}
$$
to our second-order equation. Dealing likewise with the
non-homogeneous equation
$$
y'' + y' \sin x  + y \cos x = 2 x
$$
we obtain its general solution
$$
 y = \left[C_2 +  \int \left(C_1 + x^2\right){\rm e}^{- \cos x}\,dx\right] {\rm e}^{\cos  x}.
$$
 \end{exa}

If $L[y]$ in Eq. (\ref{opid.eq34}) is not a total derivative, one
can find an appropriate factor $\phi (x) \not= 0,$ called an
\textit{integrating factor}, such that $\phi (x) L[y]$ becomes a
total derivative:
 \begin{equation}
 \label{opid.eq39}
 \phi (x) L[y]  = D_x \left(\psi (x, y, y', \ldots,
y^{(s-1)})\right).
 \end{equation}
 A connection between integrating factors
and the adjoint equations for linear equations is well known in the
classical literature (see, e.g. \cite{step58}, Chapter 5, \S 4).
Proposition \ref{total:der} gives a simple way to establish this
connection and prove the following statement.
 \begin{prop}
 \label{total:prop3}
A function ~$\phi (x)$ is an integrating factor for Eq.
(\ref{opid.eq34}) if and only if
 \begin{equation}
 \label{opid.eq40}
 z = \phi (x), \quad \phi (x) \not= 0,
 \end{equation}
 is a solution of
the adjoint equation \footnote{This statement is applicable to
nonlinear ODEs as well, see \cite{ibr08f}.} to Eq.
(\ref{opid.eq34}):
 \begin{equation}
 \label{opid.eq41}
 L^*[z] = 0.
\end{equation}
 Knowledge of a solution (\ref{opid.eq40}) to the adjoint
 equation (\ref{opid.eq41}) allows to reduce the order of Eq. (\ref{opid.eq34})
 by integrating Eq. (\ref{opid.eq39}):
 \begin{equation}
 \label{opid.eq42}
 \psi (x, y, y', \ldots, y^{(s-1)}) = C_1.
 \end{equation}
 Here $C_1$ is an arbitrary constants and $\psi$ defined
 according to Eqs. (\ref{opid.eq30})-(\ref{opid.eq31}), i.e.
 \begin{equation}
 \label{opid.eq43}
 \psi = {\sf N}(z L[y])\big|_{w = y}.
 \end{equation}
 \end{prop}
 \textbf{Proof.} If (\ref{opid.eq40}) is a solution of
 the adjoint equation (\ref{opid.eq41}), we substitute it in Eq.
(\ref{opid.eq31}) and arrive at Eq. (\ref{opid.eq39}). Hence $\phi
(x)$ is an integrating factor for Eq. (\ref{opid.eq34}). Conversely,
if $\phi (x)$ is an integrating factor for Eq. (\ref{opid.eq34}),
then Eq. (\ref{opid.eq39}) is satisfied. Now Proposition
\ref{total:der} yields
 $$
 \frac{\delta (\phi (x) L[y])}{\delta y} = 0.
 $$
Hence (\ref{opid.eq40}) is a solution of
 the adjoint equation (\ref{opid.eq41}).
Finally, Eq. (\ref{opid.eq43}) follows from (\ref{opid.eq31}).
 \begin{exa}
 \label{opid.exa2}
 Let us apply the above approach to the  first-order equation
 \begin{equation}
 \label{opid.eq44}
 y' + P(x) y = Q(x).
 \end{equation}
 Here $L[y] = y' + P(x) y.$ The adjoint equation (\ref{opid.eq41}) is written
 $$
 z' - P(x) z = 0.
 $$
Solving it we obtain the integrating factor
  $$
  z = {\rm e}^{\int P(x) d x}.
 $$
 % Multiplying by $z$ both sides of Eq.
 %(\ref{opid.eq44}) we obtain the equivalent equation
 Therefore we rewrite Eq. (\ref{opid.eq44}) in the equivalent form
 \begin{equation}
 \label{opid.eq45}
 \left[y' + P(x) y\right]{\rm e}^{\int P(x) d x} = Q(x) {\rm e}^{\int P(x) d x},
 \end{equation}
% where the
 and compute the function $\Psi$ given by Eq. (\ref{opid.eq30}):
 $$
 \Psi = {\sf N}(z L[y]) = w \frac{\partial}{\partial y'} [z (y' + P(x) y)] = w z = w {\rm e}^{\int P(x) d x}.
 $$
 Eq.  (\ref{opid.eq43}) yields
 \begin{equation}
 \label{opid.eq46}
 \psi = y {\rm e}^{\int P(x) d x}.
 \end{equation}
 Now we can take (\ref{opid.eq45}) instead of Eq. (\ref{opid.eq37})
 and write it in the form (\ref{opid.eq38}) with $\psi$ given by
(\ref{opid.eq46}). Then we obtain
 $$
 D_x\left[y {\rm e}^{\int P(x) d x} - \int Q(x) {\rm e}^{\int P(x) d x} dx \right] = 0,
 $$
whence
 $$
 y {\rm e}^{\int P(x) d x} = C_1 + \int Q(x) {\rm e}^{\int P(x) d x} dx.
 $$
 Solving the latter equation for $y$ we obtain the general solution  of Eq.
(\ref{opid.eq44}):
 \begin{equation}
 \label{opid.eq47}
 y = \left[C_1 + \int Q(x) {\rm e}^{\int P(x) d x} dx \right] {\rm e}^{- \int P(x) d x}.
 \end{equation}
 \end{exa}
 \begin{exa}
 \label{opid.exa3}
 Let us consider the second-order homogeneous equation
 \begin{equation}
 \label{opid.eq48}
 y'' + \frac{\sin x}{x^2}\,y' + \left(\frac{\cos x}{x^2} - \frac{\sin x}{x^3}\right) y = 0.
 \end{equation}
 Its left-hand side does not satisfy the total derivative condition
(\ref{opid.eq17}) because
 $$
 \frac{\delta}{\delta y} \left[y'' + \frac{\sin x}{x^2}\,y' + \left(\frac{\cos x}{x^2} - \frac{\sin x}{x^3}\right) y\right] = \frac{\sin x}{x^2}\,\cdot
 $$
Therefore we will apply Proposition \ref{total:prop3}. The adjoint
equation to Eq. (\ref{opid.eq48}) is written
 $$
 z'' - \frac{\sin x}{x^2}\,z' + \frac{\sin x}{x^3}\, z = 0.
 $$
 We take its obvious solution $z = x,$ substitute it
 in Eq. (\ref{opid.eq30}) and using (\ref{opid.eq32}) find
 $$
 \Psi = {\sf N}\left[x y'' + \frac{\sin x}{x}\,y' + \left(\frac{\cos x}{x} - \frac{\sin x}{x^2}\right) y\right]
 = \frac{\sin x}{x}\, w - w + x w'.
 $$
 Therefore Eq. (\ref{opid.eq42}) is written
 $$
 x y' + \left(\frac{\sin x}{x}\, - 1 \right) y = C_1.
 $$
 Integrating this first-order linear equation we obtain the  general  solution
of Eq. (\ref{opid.eq48}):
\begin{equation}
 \label{opid.eq49}
 y = \left(C_2 + C_1 \int \frac{1}{x^2}\,{\rm e}^{\int \frac{\sin x}{x^2}\,dx} dx\right) x {\rm e}^{-\int \frac{\sin x}{x^2}\,dx}.
 \end{equation}
 \end{exa}

 \subsection{Application of the operator identity to linear PDEs}
% \subsection{Divergence test. Integrating factor for linear ODEs}
  \label{opi.lpde}

 Using the operator identity (\ref{opid.eq1}) one can easily extend
 the equations (\ref{opid.eq31})-(\ref{opid.eq32}) for linear ODEs
 to linear partial differential equations and systems. Let us
 consider the second-order linear operator
 \begin{equation}
 \label{opid.eq50}
 L[u] = a^{ij}(x)u_{ij} +b^i(x)u_i +c(x) u
 \end{equation}
 considered in Section \ref{sa:eveq.3}, Remark \ref{lnad.rem1}. The
 adjoint operator is
 \begin{equation}
 \label{opid.eq51}
  L^*[v] \equiv \frac{\delta (v F [u])}{\delta u} = D_i D_j (a^{ij} v)
 - D_i (b^i v) +c v.
 \end{equation}
 Let us take the operator identity (\ref{opid.eq1}),
 \begin{equation}
 \label{opid.eq52}
 X = W \frac{\delta}{\delta u} + D_i {\sf N}^i,
 \end{equation}
 where $X$ is the operator (\ref{opid.eq14}) with one dependent variable $u,$
 $$
  X = W \frac{\partial}{\partial u} + W_i \frac{\partial}{\partial u_i}
 + W_{ij} \frac{\partial}{\partial u_{ij}}\,,
 $$
 and ${\sf N}^i$ are the operators (\ref{opid.eq5}),
 $$
 {\sf N}^i = W\;\frac{\delta}{\delta u_ i} +
 W_j\;\frac{\delta}{\delta u_{ij}} =
 W\left[\frac{\partial}{\partial u_i} -
 D_j \frac{\partial}{\partial u_{ij}}\right] +
  W_j\;\frac{\partial}{\partial u_{ij}}\,\cdot
 $$
 We use above the notation $ W_i = D_i (W), \ W_{ij} = D_i D_j (W).$
 Now we proceed as in Section \ref{ttd}. Namely, we act on $v L[u]$ by both sides of the
 identity (\ref{opid.eq52}),
 $$
 X (v L[u])= W \frac{\delta (v L[u])}{\delta u} + D_i {\sf N}^i (v
 L[u]),
 $$
 take into account that $X$ does not act on the variables $x^i, \ v,$ and that
 $X(L[u]) = L[W],$ use Eq. (\ref{opid.eq51}) and obtain:
 $$
 v L[W]) - W L^*[v] = D_i {\sf N}^i (v L[u]).
 $$
 Letting here $W = u$ we arrive at the following  generalization of
 the equation (\ref{opid.eq31}):
\begin{equation}
 \label{opid.eq53}
 v L[u] - u L^*[v] = D_i(\psi^i),
 \end{equation}
 where $\psi^i$ are defined as in (\ref{opid.eq32})-(\ref{opid.eq33}):
 \begin{equation}
 \label{opid.eq51a}
  \psi^i = {\sf N}^i(v L[u])\big|_{W = u} \equiv a^{ij}(x)[v u_i - u v_ i] + [b^i(x) - D_i\big(a^{ij}(x)\big)] uv.
\end{equation}

  \subsection{Application of the operator identity to nonlinear equations}
  \label{opi.nlpde}
 %\subsection{The operator identity and integrating factors to ordinary differential equations}

 Let us apply the constructions of Section \ref{opi.lpde} to
 nonlinear equations (\ref{sa:eveq.int.1}),
 \begin{equation}
 \label{opid.eq56}
 F_\alpha \big(x, u, u_{(1)}, \ldots, u_{(s)}\big) = 0, \quad
 \alpha = 1, \ldots, m.
 \end{equation}
 We write the operator (\ref{opid.eq14}) in the form
 $$
  X = W^\alpha \frac{\partial}{\partial u^\alpha} + W^\alpha_i \frac{\partial}{\partial u^\alpha_i}
 + W^\alpha_{ij} \frac{\partial}{\partial u^\alpha_{ij}} + \cdots\,,
 $$
 where $ W^\alpha_i = D_i (W^\alpha), \ W^\alpha_{ij} = D_i D_j (W^\alpha), \ldots\,.$
 Then the operator (\ref{opid.eq5}) is written
 $$
 {\sf N}^i = W^\alpha_j\;\frac{\delta}{\delta u^\alpha_i} +
 W^\alpha\;\frac{\delta}{\delta u^\alpha_{ij}} + \cdots\,.
 $$
 We act on $v^\beta F_\beta$ by both sides of the operator identity (\ref{opid.eq1})
 $$
 X  = W^\alpha \frac{\delta}{\delta u^\alpha} + D_i {\sf N}^i\,,
 $$
 denote by $F^*_\alpha[v]$  the adjoint operator
 defined by Eq. (\ref{sa:eveq.int.3}) and obtain
 \begin{equation}
 \label{opid.eq57}
 v^\beta \hat F_\beta [W] - W^\alpha F^*_\alpha [v] = D_i (\Psi^i),
\end{equation}
 where
 $$
 \Psi^i = {\sf N}^i (v^\beta F_\beta)
 $$
 and
 $\hat F_\beta [W]$ is the \textit{linear approximation} to
 $F_\beta$ defined by (see also Section \ref{sa:eveq.1})
 $$
 \hat F_\beta [W] = X (F_\beta) \equiv
W^\alpha \frac{\partial F_\beta}{\partial u^\alpha} + W^\alpha_i
\frac{\partial F_\beta}{\partial u^\alpha_i}
 + W^\alpha_{ij} \frac{\partial F_\beta}{\partial u^\alpha_{ij}} +
\cdots\,.
 $$
 \begin{rem}
 \label{lin:nonlin}
 Eq. (\ref{opid.eq57}) shows that $F^*_\alpha [v] = \hat F^*_\beta [W],$ i.e. the adjoint
 operator $F^*_\alpha$ to \textit{nonlinear}
 Eqs. (\ref{opid.eq56}) is the usual adjoint operator $\hat F^*_\beta$
 to the \textit{linear} operator $\hat F_\beta [W]$  (see also \cite{anc-blu97}).
 But the \textit{linear self-adjointness} of $\hat F_\beta [W]$ is not
 identical with the \textit{nonlinear self-adjointness} of Eqs.
(\ref{opid.eq56}). For example, the KdV equation
 $
 F \equiv u_t - u_{xxx} - u u_x = 0
 $
 is nonlinearly self-adjoint (see Example \ref{sa:eveq.KdV} in Section \ref{sa:eveq.4}).
 But its linear approximation
 $
 \hat F [W] = W_t - W_{xxx} - u W_x - W u_x
 $
is not a self-adjoint linear operator. Moreover, \textit{all linear
equations are nonlinearly self-adjoint}.
 \end{rem}

 \section{Conservation laws: Generalities and explicit formula}
 %\section{Description of the method}
 %\section{Main statements}
 %\section{Main constructions}
 \label{mainth}
 \setcounter{equation}{0}

 \subsection{Preliminaries}
 %\subsection{Generalities}
 \label{cons.gen}

 Let us consider a system of $\overline m$ differential equations
 \begin{equation}
 \label{main:eq.1}
 F_{\bar \alpha} \left(x, u, u_{(1)}, \ldots, u_{(s)}\right) = 0, \quad
 \bar \alpha = 1, \ldots, \overline m,
 \end{equation}
 with $m$ dependent variables $u^1, \ldots, u^m$ and $n$
 independent variables $x^1, \ldots, x^n.$

  A conservation law
 for Eqs. (\ref{main:eq.1}) is written
 \begin{equation}
 \label{main:eq.4}
 \left[D_i (C^i)\right]_{(\ref{main:eq.1})} = 0.
 \end{equation}
 The subscript $|_{(\ref{main:eq.1})}$ means that the left-hand side
 of (\ref{main:eq.4}) is restricted on the solutions of Eqs. (\ref{main:eq.1}). In practical calculations
 this restriction can be achieved by solving Eqs. (\ref{main:eq.1}) with respect to certain derivatives of $u$
 and eliminating these derivatives from the left-hand side
 of (\ref{main:eq.4}). For example, if (\ref{main:eq.1}) is an evolution
 equation
 $$
 u_t = \Phi (t, x, u, u_x, u_{xx}),
 $$
  the restriction $|_{(\ref{main:eq.1})}$  can be understood as
 the elimination of $u_t.$
 The $n$-dimensional vector
 \begin{equation}
 \label{cons:vec.1}
  C= (C^1, \ldots, C^n)
 \end{equation}
 satisfying Eq. (\ref{main:eq.4}) is
 called  a \textit{conserved vector} for the system
 (\ref{main:eq.1}).
 If its components are functions $C^i = C^i(x, u, u_{(1)}, \ldots)$
 of $x, u$ and derivatives $u_{(1)}, \ldots$ of a finite order, the conserved
 vector (\ref{cons:vec.1}) is called a \textit{local conserved vector.}

 %The subscript $|_{(\ref{main:eq.1})}$ in Eq. (\ref{main:eq.4})
 % conservation equation (\ref{main:eq.4})
 %means that the divergence of a conserved vector
 %vanishes on the solutions of the system (\ref{main:eq.1}).
 %In practical calculations it is to implement this condition

 Since the conservation equation (\ref{main:eq.4}) is linear with respect to $C^i,$ any linear
 combination with constant coefficients of a finite number of
 conserved vectors is again a conserved vector. It is obvious that
 if the divergence of a vector (\ref{cons:vec.1}) vanishes
 identically, it is a conserved vector for any system of
 differential equations. This is a \textit{trivial} conserved
 vectors for all differential equations. Another type of \textit{trivial
 conserved vectors} for Eqs. (\ref{main:eq.1}) are provided
 by those vectors whose components $C^i$ vanish on the solutions of
 the system (\ref{main:eq.1}). One ignores both types of trivial
 conserved vectors. In other words, conserved vectors (\ref{cons:vec.1})
 are simplified by considering them up to addition of these trivial conserved vectors.

 The following less trivial operation with conserved
 vectors is particularly useful in practice. Let
   \begin{equation}
  \label{cons:vec.2}
C^1\big\vert_{(\ref{main:eq.1})} = \widetilde C^1 + D_2(H^2) +
\cdots + D_n(H^n)
 \end{equation}
 the conserved vector (\ref{cons:vec.1}) can be
 replaced with the equivalent conserved vector
  \begin{equation}
  \label{cons:vec.3}
 \widetilde C = (\widetilde C^1, \tilde C^2, \ldots, \widetilde C^n) =0
 \end{equation}
 with the components
  \begin{equation}
  \label{cons:vec.4}
 \widetilde C^1, \quad \widetilde C^2 = C^2 + D_1(H^2), \ \ldots, \quad \widetilde C^n = C^n +
 D_1(H^n).
 \end{equation}
The passage from (\ref{cons:vec.1}) to the vector (\ref{cons:vec.3})
is based on the commutativity of the total differentiations. Namely,
we have
 $$
 D_1 D_2 (H^2) = D_2 D_t (H^2), \quad D_1 D_n (H^n) = D_n D_t (H^n),
 $$
 and therefore the conservation equation (\ref{main:eq.4}) for the vector
 (\ref{cons:vec.1}) is equivalent to the conservation equation
 $$
 \left[D_i (\widetilde C^i)\right]_{(\ref{main:eq.1})} = 0
 $$
 for the vector (\ref{cons:vec.3}). If $n \geq 3,$ the
 simplification (\ref{cons:vec.4}) of the conserved vector can be
 iterated: if $\widetilde C^2$ contains the terms
 $$
D_3(\widetilde H^3) + \cdots + D_n(\widetilde H^n)
 $$
 one can subtract them from $\widetilde C^2$ and add to $\widetilde C^3, \ldots, \widetilde C^n$
 the corresponding terms
 $$
 D_2(\widetilde H^3), \ldots, D_2(\widetilde H^n).
 $$

Note that the conservation law  (\ref{main:eq.4}) for Eqs.
(\ref{main:eq.1}) can be written in the form
 \begin{equation}
 \label{coneq.1}
 D_i (C^i)  = \mu^{\bar \alpha} F_{\bar \alpha} \big(x, u, u_{(1)}, \ldots, u_{(s)}\big)
 \end{equation}
 with undetermined coefficients \ $\mu^{\bar \alpha} = \mu^{\bar \alpha} (x, u, u_{(1)}, \ldots)$
 depending on a finite number of variables $x, u, u_{(1)},
 \ldots\,.$  If $C^i$ depend on higher-order derivatives, Eq.
 (\ref{coneq.1}) is replaced with
 \begin{equation}
 \label{coneq.2}
 D_i (C^i)  = \mu^{\bar \alpha} F_{\bar \alpha}
 + \mu^{i \bar \alpha} D_i \big(F_{\bar \alpha}\big) + \mu^{i j \bar \alpha} D_i D_j \big(F_{\bar \alpha}\big) +
 \cdots\,.
 \end{equation}

It is manifest from Eq. (\ref{coneq.1}) or Eq. (\ref{coneq.2}) that
the total differentiations of a conserved vector (\ref{cons:vec.1})
provide again conserved vectors. Therefore, e.g. the vector
 \begin{equation}
 \label{coneq.3}
 D_1 (C)  = \big(D_1(C^1), \ldots, D_1(C^n)\big)
 \end{equation}
obtained from a known vector (\ref{cons:vec.1}) is not considered as
a new conserved vector.

If one of the independent variables is time, e.g. $x^1 = t,$ then
  the conservation equation
(\ref{main:eq.4}) is often written, using the divergence theorem, in
the integral form
 \begin{equation}
 \label{conser.integral}
  \frac{d}{d t} \int_{\R^{n-1}} C^1\, dx^2 \cdots d x^n  = 0.
 \end{equation}
 %An interesting discussion of differential and integral forms of
 %conservation laws can be found in \cite{rosh02} (see also
 %\cite{rosh06} and the references therein).
 %I would like to note here that
 But the differential form (\ref{main:eq.4})
of conservation laws carries, in general, more information than the
integral form (\ref{conser.integral}). Using the integral form
(\ref{conser.integral}) one may even lose some nontrivial
conservation laws. As an example, consider the two-dimensional
Boussinesq equations
 \begin{align}
 \label{dm:eq.10}
 \Delta \psi_t - g \rho_x - f v_z & =
 \psi_x \Delta \psi_z - \psi_z \Delta \psi_x\,, \notag\\[1ex]
 v_t + f \psi_z & = \psi_x v_z - \psi_z v_x\,,\\[1ex]
 \rho_t + \frac{N^2}{g}\, \psi_x  & = \psi_x \rho_z - \psi_z \rho_x \notag
 \end{align}
 used in geophysical fluid dynamics for investigating uniformly stratified incompressible fluid flows in the ocean.
 Here $\Delta$ is the two-dimensional Laplacian,
 $$
 \Delta = \frac{\partial^2}{\partial x^2} + \frac{\partial^2}{\partial
 z^2}\,,
 $$
 and $\psi$ is the stream function so that the $x, z$- components $u, w$ of the velocity
 $(u, v, w)$ of the fluid are given by
 \begin{equation}
 \label{dm:eq.11}
 u = \psi_z, \quad w = -\psi_x.
 \end{equation}
  Eqs. (\ref{dm:eq.10}) involve the physical constants:  $g$ is the gravitational
 acceleration, $f$ is the Coriolis parameter, and $N$ is responsible for the density stratification of the
 fluid. Each equation of the system (\ref{dm:eq.10}) has the
 conservation form (\ref{main:eq.4}), namely
 \begin{align}
 \label{dm:eq.10Cd}
  D_t(\Delta \psi) + D_x(- g \rho + \psi_z \Delta \psi)
   + D_z(- f v  - \psi_x \Delta \psi) & = 0, \notag\\[1ex]
  D_t(v) + D_x(v \psi_z) +D_z(f \psi - v \psi_x) = 0,\\[1ex]
  D_t(\rho) + D_x\left(\frac{N^2}{g}\, \psi + \rho\, \psi_z\right) +D_z(- \rho\, \psi_x) = 0. \notag
 \end{align}
In the  integral form (\ref{conser.integral}) these conservation
laws are written
 \begin{align}
 \label{dm:eq.10Ci}
   \frac{d}{d t} \int\!\int \Delta \psi\, dx d z  = 0, \notag\\[1ex]
  \frac{d}{d t} \int\!\int v\, dx d z  = 0,\\[1ex]
  \frac{d}{d t} \int\!\int \rho\, dx d z  = 0. \notag
 \end{align}
 We can rewrite the differential conservation equations (\ref{dm:eq.10Cd}) in an equivalent form by using the
 operations (\ref{cons:vec.2})-(\ref{cons:vec.4}) of the
 conserved vectors. Namely, let us apply these operations to the
 first equation (\ref{dm:eq.10Cd}), i.e. to the conserved vector
 \begin{equation}
 \label{dm:eq.10Cj}
 C^1 = \Delta \psi, \quad C^2 = - g \rho + \psi_z \Delta \psi, \quad
 C^3 = - f v  - \psi_x \Delta \psi.
 \end{equation}
Noting that
 $$
 C^1 = D_x(\psi_x) + D_z(\psi_z).
 $$
 and using the operations
 (\ref{cons:vec.2})-(\ref{cons:vec.4}) we transform the vector
 (\ref{dm:eq.10Cj}) to the  form
 \begin{equation}
 \label{dm:eq.10Ck}
  \widetilde C^1 = 0, \quad  \widetilde C^2 = - g \rho + \psi_{tx} + \psi_z \Delta \psi, \quad
  \widetilde C^3 = - f v + \psi_{tz}  - \psi_x \Delta \psi.
 \end{equation}
 The integral conservation equation (\ref{conser.integral}) for the vector for (\ref{dm:eq.10Ck}) is
 trivial, $ 0 = 0.$ Thus, after the transformation of the conserved vector
(\ref{dm:eq.10Cj}) to the equivalent form (\ref{dm:eq.10Ck}) we have
lost the first integral conservation law in (\ref{dm:eq.10Ci}). But
it does not mean that the conserved vector (\ref{dm:eq.10Ck}) has no
physical significance. Indeed, if write the differential
conservation equation with the vector (\ref{dm:eq.10Ck}), we again
obtain the first equation of the system (\ref{dm:eq.10}):
$$
 D_x(\widetilde C^2) + D_z(\widetilde C^3) =
 \Delta \psi_t - g \rho_x - f v_z -
 \psi_x \Delta \psi_z + \psi_z \Delta \psi_x\,.
$$

Let us assume that Eqs. (\ref{main:eq.1}) have a nontrivial local
conserved vector  satisfying  Eq. (\ref{coneq.1}). Then not all
$\mu^{\bar \beta}$ vanish simultaneously due to non-triviality of
the conserved vector. Furthermore, since $\mu^{\bar \beta} F_{\bar
\beta}$ depends on $x, u$ and a finite number of derivatives
$u_{(1)}, u_{(2)}, \ldots$ (i.e. it is a \textit{differential
function}) and has a divergence form, the following equations hold
(for a detailed discussion see \cite{ibr99}, Section 8.4.1):
 \begin{equation}
 \label{coneq.4}
 \frac{\delta}{\delta \alpha}\left[\mu^{\bar \beta} F_{\bar \beta} \big(x, u, u_{(1)}, \ldots,
 u_{(s)}\big)\right] = 0, \quad \alpha = 1, \ldots, m.
 \end{equation}
 Note that Eqs. (\ref{coneq.4}) are identical with Eqs. (\ref{gsa:adeqs}) where the differential substitution (\ref{gsa:eq4}) is made
 with $\varphi^{\bar \alpha} = \mu^{\bar \alpha}.$ Hence, the system (\ref{main:eq.1}) is nonlinearly self-adjoint.
 I formulate this simple observation as a theorem since it is useful in applications (see
 Section \ref{gas}).
  \begin{thm}
  \label{simple}
 Any system of differential equations (\ref{main:eq.1}) having a nontrivial
 local conserved vector satisfying Eq. (\ref{coneq.1})
 is nonlinearly self-adjoint.
  \end{thm}

 \subsection{Explicit formula for conserved vectors}

 Using Definition \ref{gsa:def1} of nonlinear self-adjointness and the theorem on
 conservation laws proved in \cite{ibr07a} by using the operator identity (\ref{opid.eq1}),
we obtain the explicit formula for constructing conservation laws
  associated  with symmetries of any nonlinearly
self-adjoint system of equations. The method is applicable
independently on the number of equations in the system and the
number of dependent variables. The result is as follows.
 \begin{thm}
 \label{main.thm}
 Let the system of differential equations (\ref{main:eq.1}) be nonlinearly
 self-adjoint. Specifically, let the adjoint system (\ref{gsa:adeqs}) to (\ref{main:eq.1}) be
  satisfied for all solutions
 of Eqs. (\ref{main:eq.1}) upon a substitution (\ref{gsa:eq1}),
  \begin{equation}
  \label{main:eq.2}
 v^{\bar \alpha} = \varphi^{\bar \alpha}(x, u), \quad \bar \alpha = 1, \ldots, \overline m.
 \end{equation}
  Then any Lie point, contact or Lie-B\"{a}cklund
 symmetry
 \begin{equation}
 \label{main:eq.3}
 X = \xi^i(x, u, u_{(1)}, \ldots) \frac{\partial}{\partial x^i} +
 \eta^\alpha (x,u, u_{(1)}, \ldots) \frac{\partial}{\partial u^\alpha}\,,
 \end{equation}
 as well as a nonlocal symmetry of Eqs. (\ref{main:eq.1}) leads to a conservation law
 (\ref{main:eq.4})
 constructed by the following formula:
 \begin{align}
 \label{main:eq.5a}
 & C^i  = \xi^i {\cal L}+W^\alpha\,
 \left[\frac{\partial {\cal L}}{\partial u_i^\alpha} -
 D_j \left(\frac{\partial {\cal L}}{\partial u_{ij}^\alpha}\right)
 + D_j D_k\left(\frac{\partial {\cal L}}{\partial u_{ijk}^\alpha}\right) -
 \ldots\right]\\[1.5ex]
 &+D_j\left(W^\alpha\right)\,
 \left[\frac{\partial {\cal L}}{\partial u_{ij}^\alpha} -
 D_k \left(\frac{\partial {\cal L}}{\partial
 u_{ijk}^\alpha}\right) + \ldots\right]
 + D_j D_k\left(W^\alpha\right)\left[\frac{\partial {\cal L}}{\partial u_{ijk}^\alpha} -
 \ldots\right],\notag
  \end{align}
where
 \begin{equation}
 \label{main:eq.6}
  W^\alpha = \eta^\alpha - \xi^j u_j^\alpha
 \end{equation}
  and ${\cal L}$ is the \textit{formal Lagrangian}  for the system
  (\ref{main:eq.1}),
 \begin{equation}
 \label{main:eq.7}
 {\cal L} = v^{\bar \beta} F_{\bar \beta}.
 \end{equation}
 In (\ref{main:eq.5a}) the formal Lagrangian ${\cal L}$ should be
 written in the symmetric form with respect to all mixed derivatives
 $u^\alpha_{ij}, \ u^\alpha_{ijk}, \ldots\,$ and  the ``non-physical variables" $v^{\bar \alpha}$ should be eliminated
via Eqs. (\ref{main:eq.2}).
 \end{thm}

% \begin{rem}
% \label{main.rem1}
 One can omit in (\ref{main:eq.5a}) the term $\xi^i {\cal L}$
 when it is convenient. This term provides a trivial conserved vector mentioned in Section \ref{cons.gen}
  because ${\cal L}$ vanishes  on
 the solutions of Eqs. (\ref{main:eq.1}). Thus,  the conserved vector (\ref{main:eq.5a})
 can be taken in the following form:
 %\end{rem}
 \begin{align}
 \label{main:eq.5}
 & C^i  = W^\alpha\,
 \left[\frac{\partial {\cal L}}{\partial u_i^\alpha} -
 D_j \left(\frac{\partial {\cal L}}{\partial u_{ij}^\alpha}\right)
 + D_j D_k\left(\frac{\partial {\cal L}}{\partial u_{ijk}^\alpha}\right) -
 \ldots\right]\\[1.5ex]
 &+D_j\left(W^\alpha\right)\,
 \left[\frac{\partial {\cal L}}{\partial u_{ij}^\alpha} -
 D_k \left(\frac{\partial {\cal L}}{\partial
 u_{ijk}^\alpha}\right) + \ldots\right]
 + D_j D_k\left(W^\alpha\right)\left[\frac{\partial {\cal L}}{\partial u_{ijk}^\alpha} -
 \ldots\right].\notag
  \end{align}
 \begin{rem}
 \label{main.rem1}
 One can use Eqs. (\ref{main:eq.5}) for constructing conserved
 vectors even
 if the system (\ref{main:eq.1}) is not self-adjoint, in particular, if one cannot find
 explicit formulae (\ref{main:eq.2}) or (\ref{gsa:eq4}) for
  point  or differential substitutions, respectively. The resulting conserved vectors will
 be \textit{nonlocal} in the sense that they involve the variables $v$ connected with the physical variables
 $u$ via differential equations, namely, adjoint equations to  (\ref{main:eq.1}).
 \end{rem}
 \begin{rem}
 \label{main.rem2}
 Theorem \ref{main.thm}, unlike Nother's theorem \ref{Noether},
 does not require additional restrictions such as the invariance condition (\ref{opid.eq9})
 or the divergence condition mentioned in Remark \ref{opid.rem2}.
 \end{rem}

 \section{A nonlinearly self-adjoint irrigation system}
 \label{irrig:cl}
 \setcounter{equation}{0}

 Let us apply Theorem \ref{main.thm} to Eq. (\ref{irrig.eq1}) satisfying the  condition
 (\ref{irrig.eq3}):
 \begin{align}
 C(\psi) \psi_t & = \left[K(\psi)\psi_{x}\right]_x
 + \left[K(\psi) \left( \psi_z - 1 \right)\right]_z - S(\psi),
 \label{irrig:cl.eq1}\\[1ex]
 S\,'(\psi) & = a C(\psi), \quad a = {\rm const.}  \label{irrig:cl.eq2}
\end{align}
 The formal Lagrangian (\ref{main:eq.7}) for Eq.
(\ref{irrig:cl.eq1})
 has the form
 \begin{equation}
 \label{irrig:cl.eq3}
 {\cal L} = \left[- C(\psi) \psi_t + K(\psi)(\psi_{xx} + \psi_{zz})
 + K'(\psi)(\psi_x^2 + \psi_z^2 - \psi_z) - S(\psi)\right] v.
 \end{equation}
 We will use the substitution (\ref{irrig.eq5}) of the particular form
  \begin{equation}
 \label{irrig:cl.eq4}
 v = {\rm e}^{at}.
 \end{equation}
Denoting $t = x^1, x = x^2, z = x^3$ we  write the conservation
equation (\ref{main:eq.4}) in the form
 \begin{equation}
 \label{irrig:cl.eq5}
 D_t (C^1) + D_x (C^2) + D_z (C^3) = 0.
 \end{equation}
This equation should be satisfied on the solutions of Eq.
(\ref{irrig:cl.eq1}).

  The formal Lagrangian (\ref{irrig:cl.eq3}) does not contain derivatives of order higher than two.
  Therefore in our case Eqs. (\ref{main:eq.5}) take the simple form
 \begin{equation}
 \label{irrig:cl.eq6}
  C^i  = W \left[\frac{\partial {\cal L}}{\partial \psi_i} -
  D_j \left(\frac{\partial {\cal L}}{\partial \psi_{ij}}\right)\right]
 + D_j(W) \frac{\partial {\cal L}}{\partial \psi_{ij}}
 \end{equation}
 and yield:
 \begin{align}
 & C^1 = W \frac{\partial {\cal L}}{\partial \psi_t}\,,\notag\\
 & C^2 = W \left[\frac{\partial {\cal L}}{\partial \psi_x} -
  D_x \left(\frac{\partial {\cal L}}{\partial \psi_{xx}}\right)\right]
 + D_x(W) \frac{\partial {\cal L}}{\partial \psi_{xx}}\,,\notag \\
  & C^3 = W \left[\frac{\partial {\cal L}}{\partial \psi_z} -
  D_z \left(\frac{\partial {\cal L}}{\partial \psi_{zz}}\right)\right]
 + D_z(W) \frac{\partial {\cal L}}{\partial \psi_{zz}}\,\cdot\notag
 \end{align}
 Substituting here the expression (\ref{irrig:cl.eq3}) for ${\cal L}$ we obtain
 \begin{align}
 & C^1 = - W C(\psi) v,\notag\\
 & C^2 = W [2 K'(\psi) v \psi_x - D_x(K(\psi) v)] + D_x(W) K(\psi) v,\notag \\
  & C^3 = W [K'(\psi) v (2\psi_z - 1) - D_z(K(\psi) v)] + D_z(W) K(\psi) v,\notag
 \end{align}
where $v$ should be eliminated by means of the substitution
 (\ref{irrig:cl.eq4}). So, we have:
 \begin{equation}
 \label{irrig:cl.eq7}
 \begin{split}
 & C^1 = - W C(\psi) {\rm e}^{a t},\\
 & C^2 = [W K'(\psi)\psi_x + D_x(W) K(\psi)] {\rm e}^{a t}, \\
  & C^3 = [W K'(\psi)(\psi_z - 1)+ D_z(W) K(\psi)] {\rm e}^{a t}.
 \end{split}
 \end{equation}

 Since  Eq. (\ref{irrig:cl.eq1}) does not explicitly involve the
 independent variables $t, x, z,$ it is invariant under the translations of
 these variables.
 Let us construct the conserved vector (\ref{irrig:cl.eq7})
 corresponding to the time translation group with the generator
 \begin{equation}
 \label{irrig:cl.eq8}
 X = \frac{\partial}{\partial t}\,\cdot
 \end{equation}
 For this operator Eq. (\ref{main:eq.6}) yields
 \begin{equation}
 \label{irrig:cl.eq9}
 W = - \psi_t.
 \end{equation}
 Substituting (\ref{irrig:cl.eq9}) in  Eqs. (\ref{irrig:cl.eq7}) we obtain
 \begin{equation}
 \label{irrig:cl.eq10}
 \begin{split}
 & C^1 =  C(\psi)\psi_t {\rm e}^{a t},\\
 & C^2 = - [K'(\psi)\psi_t \psi_x + K(\psi)\psi_{tx}] {\rm e}^{a t}, \\
  & C^3 = - [K'(\psi)\psi_t (\psi_z - 1)+ K(\psi) \psi_{tz}] {\rm e}^{a t}.
 \end{split}
 \end{equation}
 Now we replace in $C^1$ the term $C(\psi)\psi_t$ by the right-hand side of Eq.
 (\ref{irrig:cl.eq1}) to obtain:
  $$
 C^1 = - S(\psi) {\rm e}^{a t}
 + D_x \left(K(\psi)\psi_{x} {\rm e}^{a t}\right)
 + D_z\left(K(\psi)(\psi_z - 1){\rm e}^{a t}\right).
  $$
When we substitute this expression in the conservation equation
(\ref{irrig:cl.eq5}), we can write
 $$
D_t\left(D_x \left(K(\psi)\psi_{x} {\rm e}^{a t}\right)\right) =
D_x\left(D_t\left(K(\psi)\psi_{x} {\rm e}^{a t}\right)\right).
 $$
 Therefore we can transfer the terms $D_x(\ldots)$ and $D_z(\ldots)$ from $C^1$
 to $C^2$ and $C^3,$  respectively (see (\ref{cons:vec.4})). Thus, we rewrite the vector
 (\ref{irrig:cl.eq10}), changing its sign, as follows:
  \begin{align}
 & C^1 = S(\psi) {\rm e}^{a t},\notag\\
 & C^2 = [K'(\psi)\psi_t \psi_x + K(\psi)\psi_{tx}] {\rm e}^{a t} - D_t\left(K(\psi)\psi_x {\rm e}^{a t}\right),\notag \\
  & C^3 = [K'(\psi)\psi_t (\psi_z - 1)+ K(\psi) \psi_{tz}] {\rm e}^{a t} - D_t\left(K(\psi)(\psi_z - 1){\rm e}^{a t}\right).\notag
 \end{align}
Working out the differentiation $D_t$ in the last terms of $C^2$ and
$C^3$ we finally  arrive at the following vector:
  \begin{equation}
 \label{irrig:cl.eq11}
 \begin{split}
 & C^1 = S(\psi) {\rm e}^{a t},\\
 & C^2 = a K(\psi)\psi_x {\rm e}^{a t}, \\
  & C^3 = a K(\psi) (\psi_z - 1){\rm e}^{a t}.
 \end{split}
 \end{equation}
The reckoning shows that the vector (\ref{irrig:cl.eq11}) satisfies
the conservation equation (\ref{irrig:cl.eq5}) due to the condition
(\ref{irrig:cl.eq2}). Note that $C^1$ is the \textit{density} of the
conserved vector (\ref{irrig:cl.eq11}).

The use of the general substitution (\ref{irrig.eq5}) instead of its
particular case (\ref{irrig:cl.eq4}) leads to the conserved vector
with the density
 $$
C^1 = S(\psi) (b x + l){\rm e}^{a t}.
 $$

 This approach opens a new possibility to find a variety of conservation
 laws for the irrigation model (\ref{irrig.eq1}) by
 considering other self-adjoint cases of the model
 and using the extensions of symmetry Lie algebras (see \cite{ibr94-96}, vol. 2, Section 9.8).

 \section{Utilization of differential substitutions}
 \label{difsub}
 \setcounter{equation}{0}

  \subsection{Equation $u_{xy} = \sin u$}

 We return to Section \ref{gsa:rem2} and
 calculate the conservation laws for Eq. (\ref{gsa:eq5.r1}),
  \begin{equation}
  \label{difsub.eq1}
  u_{xy} = \sin u,
  \end{equation}
  using the differential substitution (\ref{gsa:eq5.r2}),
  \begin{equation}
  \label{difsub.eq2}
 v = A_1 [x u_x - y u_y] + A_2 u_x + A_3 u_y,
  \end{equation}
  and the  admitted three-dimensional Lie algebra with the basis
 \begin{equation}
 \label{difsub.eq3}
 X_1 = \frac{\partial}{\partial x}\,, \quad  X_2 = \frac{\partial}{\partial y}\,,
 \quad  X_3 = x \frac{\partial}{\partial x} - y \frac{\partial}{\partial
 y}\,\cdot
 \end{equation}
 The conservation equation for Eq. (\ref{difsub.eq1}) will
 be written in the form
  $$
 D_x(C^1) + D_y(C^2) = 0.
  $$
 We write the formal
 Lagrangian for Eq. (\ref{difsub.eq1}) in the symmetric form
 \begin{equation}
 \label{difsub.eq8}
  {\cal L} = \left(\frac{1}{2}\,u_{xy} + \frac{1}{2}\,u_{yx}
  - \sin u \right) v.
 \end{equation}

  Eqs. (\ref{main:eq.5})
  yield:
\begin{equation}
 \label{difsub.eq9}
  C^1 = \frac{1}{2}\,D_y (W) v - \frac{1}{2}\,W  v_y, \quad
  C^2 = \frac{1}{2}\,D_x (W) v - \frac{1}{2}\,W  v_x.
 \end{equation}
 where we have to eliminate the variable $v$ via the differential substitution (\ref{difsub.eq2}).

 Substituting in (\ref{difsub.eq9}) $W = - u_x$ corresponding to the operator
 $X_1$ from (\ref{difsub.eq3}), replacing $v$ with (\ref{difsub.eq2}) and
 $u_{xy}$ with $\sin u,$ then transferring the terms of the form $D_y (\ldots)$
 from $C^1$ to $C^2$ (see the simplification (\ref{cons:vec.4})) we obtain:
 $$
 C^1 = A_1 \cos u, \quad
  C^2 = \frac{1}{2}\,A_1 u_x^2.
 $$
 We let $A_1 = 1$ and conclude that the application of Theorem \ref{main.thm} to the symmetry $X_1$ yields the conserved
 vector
 \begin{equation}
 \label{difsub.eq5}
 C^1 = \cos u, \quad  C^2 = \frac{1}{2}\,u_x^2\,\cdot
 \end{equation}
 The similar calculations with the operator
 $X_2$ from (\ref{difsub.eq3}) lead to the conserved
 vector
 \begin{equation}
 \label{difsub.eq6}
 C^1 = \frac{1}{2}\,u_y^2\,, \quad  C^2 = \cos u.
 \end{equation}
The third symmetry,  $X_3$ from (\ref{difsub.eq3}), does not lead to
a new conserved vector. Indeed, in this case $W = y u_y - x u_x.$
 Substituting it in the first formula (\ref{difsub.eq9}) we obtain
 after simple calculations
 $$
 C^1 = \frac{1}{2} A_3 u_y ^2 - A_2 \cos u + D_y\left[(A_2 y + A_3 x)\left(\frac{1}{2} u_x u_y + \cos u\right)\right].
 $$
 Hence, upon transferring the term  $D_y (\ldots)$ from $C^1$ to
 $C^2$ the resulting $C^1$ will be a linear combination with
 constant coefficients of the components $C^1$ of the conserved
 vectors (\ref{difsub.eq5}) and (\ref{difsub.eq6}). The same will be
 true for $C^2.$ Therefore the conserved vector provided by the
 symmetry $X_3$ will be a linear combination with
 constant coefficients of the  conserved
 vectors (\ref{difsub.eq5}) and (\ref{difsub.eq6}).

 One can also use the Noether theorem because Eq. (\ref{difsub.eq1})
 has the classical Lagrangian, namely
 \begin{equation}
 \label{difsub.eq4}
  L = - \frac{1}{2}\,u_x u_y + \cos u.
 \end{equation}
 Then the symmetries $X_1$ and $X_2$
  provide again the conserved vectors (\ref{difsub.eq5}) and
 (\ref{difsub.eq6}), respectively. But now we obtain one more
 conserved vector using $X_3,$ namely
 \begin{equation}
 \label{difsub.eq7}
 C^1 = x \cos u - \frac{y}{2}\, u_y^2\,, \quad  C^2 = \frac{x}{2}\, u_x^2 - y \cos
 u.
 \end{equation}

 \subsection{Short pulse equation}
 \label{difsub.pulse}

 The differential equation (up to notation and appropriate scaling the physical variables)
 \begin{equation}
 \label{pulse}
  D_t D_x (u) = u + \frac{1}{6}\, D_x^2(u^3)
 \end{equation}
 was suggested in \cite{sch-way04} (see there Eq. (11), also \cite{sak-sak06}) as a mathematical model for the
 propagation of ultra-short light pulses in media with nonlinearities, e.g. in silica fibers.
 The mathematical model is derived in \cite{sch-way04} by considering the propagation of linearly
 polarized light in a one-dimensional medium and assuming that the light propagates in the infrared range.
 The final step in construction of the model is based on the method of multiple
 scales.

 Eq. (\ref{pulse}) is connected with Eq. (\ref{difsub.eq1}) by a non-point
 transformation which is constructed in \cite{sak-sak05} as a chain of differential
 substitutions (given also in \cite{sak-sak06} by Eqs. (2)). Using this connection,
 an exact solitary wave solution (a \textit{pulse solution})
 to Eq. (\ref{pulse}) is constructed in \cite{sak-sak06}.  One can also find in \cite{sak-sak05}
 a Lax pair and a recursion operator for Eq. (\ref{pulse}).

 Note that Eq. (\ref{pulse}) does not have a conservation form. I
 will find a conservation law of Eq. (\ref{pulse}) thus showing that it
 can be rewritten in a conservation form. A
 significance of this possibility is commonly known and is not discussed here.

 We write  the \textit{short pulse equation} (\ref{pulse}) in the expanded form
 \begin{equation}
  \label{pulse:2}
  u_{xt} = u + \frac{1}{2}\,u^2 u_{xx} + u u_x^2
  \end{equation}
 so that the formal Lagrangian is written
 \begin{equation}
 \label{difsub.eq11}
 {\cal L} = v \left[u_{xt} - u - \frac{1}{2}\,u^2 u_{xx}
 - u u_x^2 \right].
 \end{equation}
  Substituting (\ref{difsub.eq11}) in (\ref{gsa:adeqs})
 we obtain the following \textit{adjoint equation}  to Eq. (\ref{pulse:2}):
 \begin{equation}
 \label{difsub.eq12}
 v_{xt} = v + \frac{1}{2}\,u^2 v_{xx}.
 \end{equation}

  We first demonstrate the following statement.
 \begin{prop}
 \label{pulse.prop}
   Eq. (\ref{pulse})
  is not nonlinearly self-adjoint with a  substitution
 \begin{equation}
 \label{pulse.prop:eq1}
  v = \varphi (t, x, u)
 \end{equation}
  but it is nonlinearly
  self-adjoint with the differential substitution
  \begin{equation}
  \label{difsub.eq13}
  v = u_t - \frac{1}{2}\,u^2 u_x.
  \end{equation}
  \end{prop}
  \textbf{Proof.} We write the nonlinear self-adjointness condition
  (\ref{gsa:eq3}),
 $$
 \left[v_{xt} - v - \frac{1}{2}\,u^2 v_{xx}\right]_{(\ref{pulse.prop:eq1})}
 = \lambda [u_{xt} - u - \frac{1}{2}\,u^2 u_{xx}
 - u u_x^2],
 $$
  substitute here the expression (\ref{pulse.prop:eq1}) for
  $v$ and its derivatives
 \begin{equation}
 \label{pulse.prop:eq2}
 \begin{split}
 & v_{xx} = \varphi_u u_{xx} + \varphi_{uu} u_x^2 +  2 \varphi_{xu} u_x
  + \varphi_{xx},\\
  & v_{xt} = \varphi_u u_{xt} + \varphi_{uu} u_x u_t + \varphi_{xu}
  u_t + \varphi_{tu} u_x + \varphi_{xt}\,,
  \end{split}
 \end{equation}
 and first
obtain
 $\lambda = \varphi_u$ by
 comparing the terms with the second-order derivatives of $u.$
This reduces the  nonlinear self-adjointness condition to the
following equation:
 \begin{equation}
 \label{pulse.prop:eq3}
 \begin{split}
 & \varphi_{uu} u_x u_t + \varphi_{xu} u_t
 + \varphi_{tu} u_x + \varphi_{xt} - \varphi - \frac{1}{2}\,u^2(\varphi_{uu} u_x^2 +  2 \varphi_{xu} u_x
  + \varphi_{xx})\\[1ex]
  & = - \varphi_u [u  + u u_x^2].
  \end{split}
 \end{equation}
 The terms with $u_t$ in Eq. (\ref{pulse.prop:eq3}) yield $\varphi_{uu}  = \varphi_{xu} =
 0.$ Then we take the term with $u_x^2$ and obtain $\varphi_u = 0.$ Hence
 $$
 \varphi = a(t, x).
 $$
Now Eq. (\ref{pulse.prop:eq3}) gives  $a_{xx} = 0, \ a_{xt} - a =
0,$ whence $a= 0.$ Thus $$\varphi = 0,$$ i.e. the substitution
(\ref{pulse.prop:eq1}) is trivial.  This proves the first part of
Proposition \ref{pulse.prop}.
 Its second part  is proved by similar calculations
with the substitution $$v = \varphi (t, x, u, u_x, u_t).$$ I will
not reproduce these rather lengthy calculations,
 but instead
we will verify that the substitution (\ref{difsub.eq13})
  maps any solution of Eq. (\ref{difsub.eq1}) into a solution of the adjoint equation (\ref{difsub.eq12}).
   First we calculate
  $$
  v_x = u_{xt} - \frac{1}{2}\,u^2 u_{xx} - u u_x^2
  $$
 and see that on the solutions of Eq. (\ref{difsub.eq1}) we have $v_x = u.$ Now we
  calculate other derivatives and verify that on the solutions of Eq. (\ref{difsub.eq1})
   the following equations hold:
    \begin{equation}
  \label{difsub.eq14}
  v_x = u, \quad v_t = u_{tt} - \frac{1}{2}\,u^2 u_{xt} - u u_x u_t, \quad
   v_{xt} = u_t, \quad v_{xx} = u_x.
  \end{equation}
 It is easily seen that Eq. (\ref{difsub.eq12}) is satisfied. Namely,
 using (\ref{difsub.eq13}) and (\ref{difsub.eq14}) we have:
 $$
 v_{xt} - v - \frac{1}{2}\,u^2 v_{xx} = u_t - \left(u_t - \frac{1}{2}\,u^2
 u_x\right) - \frac{1}{2}\,u^2 u_x = 0.
 $$

 The maximal Lie algebra of point symmetries of Eq. (\ref{pulse}) is the three-dimensional algebra spanned by the operators
 \begin{equation}
  \label{pulse:3}
  X_1 = \frac{\partial}{\partial t}\,, \quad  X_2 = \frac{\partial}{\partial x}\,,
  \quad X_3 = u \frac{\partial}{\partial u} + x \frac{\partial}{\partial x} - t \frac{\partial}{\partial t}\,\cdot
  \end{equation}
 Let us construct the conservation laws
 \begin{equation}
  \label{pulse:1}
 D_t(C^1) + D_x(C^2) = 0
 \end{equation}
 for the basis operators (\ref{pulse:3}).

 Since the formal Lagrangian (\ref{difsub.eq11}) does not contain derivatives of order higher than two,
  Eqs. (\ref{main:eq.5}) are written
 (see (\ref{irrig:cl.eq6}))
  $$
  C^i  = W \left[\frac{\partial {\cal L}}{\partial u_i} -
  D_j \left(\frac{\partial {\cal L}}{\partial u_{ij}}\right)\right]
 + D_j(W) \frac{\partial {\cal L}}{\partial u_{ij}}\,\cdot
 $$
In our case we have:
 \begin{align}
 & C^1 = - W D_x \left(\frac{\partial {\cal L}}{\partial u_{tx}}\right) +
 D_x(W) \frac{\partial {\cal L}}{\partial u_{tx}}\,,
 \label{pulse:4}\\[1ex]
 & C^2 = W \left[\frac{\partial {\cal L}}{\partial u_x} -
  D_t \left(\frac{\partial {\cal L}}{\partial u_{xt}}\right)
  - D_x \left(\frac{\partial {\cal L}}{\partial u_{xx}}\right)\right]
  + D_t(W) \frac{\partial {\cal L}}{\partial u_{xt}}
 + D_x(W) \frac{\partial {\cal L}}{\partial u_{xx}}\,\cdot\notag
 \end{align}
 Substituting in (\ref{pulse:4}) the expression (\ref{difsub.eq11}) for ${\cal L}$ written in the symmetric form
 \begin{equation}
 \label{pulse:5}
 {\cal L} = v \left[\frac{1}{2}\,u_{tx} + \frac{1}{2}\,u_{xt} - u - \frac{1}{2}\,u^2 u_{xx}
 - u u_x^2 \right]
 \end{equation}
 we obtain
 \begin{align}
 \label{pulse.convec:1}
 & C^1 = - \frac{1}{2} W v_x + \frac{1}{2}\, v D_x (W),\\[1ex]
 & C^2 = - W \Big[u v u_x + \frac{1}{2}\, v_t - \frac{1}{2}\, u^2 v_x\Big] + \frac{1}{2}\, v D_t(W) - \frac{1}{2}\, u^2v D_x(W).\notag
 \end{align}
 Since $v$ should be eliminated via the differential
 substitution (\ref{difsub.eq13}),  we further simplify this
 vector by replacing $v_x$ with
 $u$ according to the first equation (\ref{difsub.eq14}) and obtain:
 \begin{align}
 \label{pulse.convec:2}
 & C^1 = - \frac{1}{2} W u + \frac{1}{2}\, v D_x (W),\\[1ex]
 & C^2 = - W \Big[u v u_x + \frac{1}{2}\, v_t - \frac{1}{2}\, u^3\Big] + \frac{1}{2}\, v D_t(W) - \frac{1}{2}\, u^2v D_x(W),\notag
 \end{align}
where $v$ and $v_t$ should be replaced with their values given in
Eqs. (\ref{difsub.eq13}), (\ref{difsub.eq14}).

 Let us construct the conserved vectors using the symmetries (\ref{pulse:3}). Their
 commutators are
 $$
 [X_1, X_3] = - X_1, \quad [X_2, X_3] = X_2.
 $$
 Hence, according to \cite{ibr83}, Section 22.4,
 the operator  $X_3$
  plays a distinguished role. Namely, the conserved vectors
  associated with $X_1$ and $X_2$ can be obtained from the conserved
  vector provided by $X_3$ using the adjoint actions of the
  operators $X_1$ and $X_2,$ respectively.
 Therefore we start with $X_3.$
  Substituting in (\ref{pulse.convec:2}) the expression
  $$
  W = u + t u_t - x u_x
  $$
  corresponding to the symmetry $X_3,$ eliminating the terms of
  the form $D_x(A)$ from $C^1$ and adding them to $C^2$ in the
  form $D_t(A)$ according to the
 simplification (\ref{cons:vec.4}), we obtain after routine calculations the following
  conserved vector:
 \begin{align}
 \label{pulse.convec:3}
 & C^1 = u^2,\\[.1ex]
 & C^2 = u^2 u_x u_t - u_t^2 - \frac{1}{4} u^4 - \frac{1}{4} u^4 u_x^2.\notag
 \end{align}
 The conservation equation (\ref{pulse:1}) for the vector  (\ref{pulse:7}) holds in the
 form
 \begin{equation}
 \label{pulse.convec:4}
 D_t(C^1) + D_x(C^2) = 2 \Big(u_t - \frac{1}{2}\,u^2 u_x\Big)\Big(u + \frac{1}{2}\,u^2 u_{xx} + u u_x^2 - u_{xt}\Big).
 \end{equation}

 Let us turn now to the operators $X_1$ and $X_2$ from
 (\ref{pulse:3}). To simplify the calculations it is useful to
 modify Eqs. (\ref{pulse.convec:2}) as follows.
  Noting that $$v D_x (W) = D_x( v W) - W v_x$$ we rewrite the vector (\ref{pulse.convec:1}) in the form
 \begin{align}
 & C^1 = - W v_x,\notag\\[.1ex]
 & C^2 = - W \Big[u v u_x - \frac{1}{2}\, u^2 v_x\Big] + v D_t(W) - \frac{1}{2}\, u^2v D_x(W).\notag
 \end{align}
 Then (\ref{pulse.convec:2}) is replaced with
 \begin{align}
 \label{pulse:6}
 & C^1 = - u W ,\\[.1ex]
 & C^2 = - W \Big[u v u_x - \frac{1}{2}\, u^3\Big] + v D_t(W) - \frac{1}{2}\, u^2v D_x(W).\notag
 \end{align}
 Substituting in the first formula (\ref{pulse:6}) to expression $W = -
 u_t$ corresponding the operator $X_1$ we obtain
 $C^1 = u u_t.$ This is the time derivative of $C^1$ from
 (\ref{pulse.convec:3}). Hence the symmetry $X_1$ leads to
 a trivial conserved vector  obtained from the  vector (\ref{pulse.convec:3})
 by the differentiation $D_t,$ in accordance with \cite{ibr83}. Likewise, it is manifest from
 (\ref{pulse:6}) that the operator $X_2$ leads to
 a trivial conserved vector obtained from the conserved vector (\ref{pulse.convec:3})
 by the differentiation $D_x.$
 Thus we have demonstrated the following statement.
 \begin{prop}
 \label{pulse.prop1}
 The Lie point symmetries (\ref{pulse:3}) of Eq. (\ref{pulse:2})
 yield one non-trivial conserved vector (\ref{pulse.convec:3}).
 Accordingly, the short pulse equation (\ref{pulse:2}) can be
 written in the following conservation form:
\begin{equation}
 \label{pulse:7}
D_t\Big(u^2\Big) + D_x \Big(u^2 u_x u_t - u_t^2 - \frac{1}{4} u^4 -
\frac{1}{4} u^4 u_x^2\Big) = 0.
 \end{equation}
 \end{prop}

\section{Gas dynamics}
 \label{gas}
 \setcounter{equation}{0}

 \subsection{Classical symmetries and conservation laws}
  \label{gas:1}

 Let us consider the polytropic gasdynamic equations
\begin{align}
& \bm v_t + (\bm v \cdot \nabla) \bm v+\frac{1}{\rho}\,\nabla
 p=0,\notag\\
 & \rho_t+\bm v\cdot \nabla \rho+\rho \nabla \cdot \bm v=0, \label{gas.eq1}\\
 & p_t + \bm v\cdot \nabla p + \gamma p \nabla \cdot \bm v=0,\notag
\end{align}
 where $ \gamma$ is a constant known as the polytropic (or
 adiabatic) exponent.
The independent variables are the time and the space coordinates:
 \begin{equation}
 \label{gas.eq2}
 t, \quad \bm x = (x^1,\ldots, x^n), \quad n \leq 3.
 \end{equation}
 The dependent variables are the velocity, the density and the
 pressure:
 \begin{equation}
 \label{gas.eq3}
 \bm v = (v^1,\ldots, v^n), \quad \rho, \quad p.
 \end{equation}

 Eqs. (\ref{gas.eq1}) with arbitrary $\gamma$ have
 the Lie algebra of point symmetries spanned by
\begin{align}
 \label{gas.eq4}
& X_0=\frac{\partial}{\partial t}\,, \quad
 X_i=\frac{\partial}{\partial x^i}\,, \quad
  Y_0=t\frac{\partial}{\partial
 t}+ x^i \frac{\partial}{\partial x^i}\,, \quad
 Y_i=t\frac{\partial}{\partial x^i}+\frac{\partial}{\partial v^i}\,,
\notag\\[1.5ex]
 & X_{ij}= x^j\frac{\partial}{\partial x^i}
  -x^i \frac{\partial}{\partial x^j} + v^j\frac{\partial}{\partial v^i} -
  v^i\frac{\partial}{\partial v^j}\,, \quad (i < j),\\[1.5ex]
 &  Z_0=\rho \frac{\partial}{\partial \rho}+
 p \frac{\partial}{\partial p}\,, \quad
 Z_1= t\frac{\partial}{\partial t}
 -v^i \frac{\partial}{\partial v^i} + 2\rho \frac{\partial}{\partial
 \rho}\,, \quad i, j =  1, \ldots, n,\notag
  \end{align}
and the following classical conservation laws:

\begin{tabular}{lll}
  ${\displaystyle \frac{d}{dt} \int\limits_{\Omega(t)}\rho d\omega =0}$ & -- & {\small Conservation of
  mass}  \\[4ex]
  ${\displaystyle \frac{d}{dt} \int\limits_{\Omega(t)} \left( \frac12 \,\rho
  |\bm v|^2 +\frac{p}{\gamma -1}\right)d\omega
  = - \int\limits_{S(t)} p\,\bm v \cdot\bm \nu dS}$ & -- & {\small   Energy}  \\[4ex]
  ${\displaystyle \frac{d}{dt}\int\limits_{\Omega(t)}\rho \bm v d\omega=
  - \int\limits_{S(t)}p \,\bm \nu dS}$ & -- & {\small  Momentum} \\[4ex]
  ${\displaystyle \frac{d}{dt}\int\limits_{\Omega(t)}\rho (\bm x \times \bm v)d\omega=
  - \int\limits_{S(t)} p (\bm x \times \bm \nu)dS }$ & --&{\small Angular momentum}    \\[4ex]
  ${\displaystyle \frac{d}{dt}\int\limits_{\Omega(t)}\rho (t\bm v -\bm x)d\omega=
  - \int\limits_{S(t)}tp\,\bm \nu dS }$ & -- & {\small Center-of-mass theorem.}  \\[4ex]
\end{tabular}

 \noindent
 The conservation laws are written in the integral form by
using the
 standard symbols:

 \begin{tabular}{lll}
   $\Omega(t)$ & - & arbitrary $n$-dimensional volume, moving with fluid,  \\
   $S(t)$ & - & boundary of the volume $\Omega (t),$ \\
   $\bm \nu$ & - & unit (outer) normal vector to the surface $S (t).$
   \\[1ex]
 \end{tabular}

 \noindent
 If we write the above conservation laws in the general form
    \begin{equation}
   \label{gas.eq5}
 \frac{d}{dt} \int\limits_{\Omega(t)} T d\omega =-\int\limits_{S(t)}
 (\bm{\chi}\cdot \bm{\nu}) dS,
     \end{equation}
 then the differential form of these  conservation laws will be
 \begin{equation}
 \label{gas.eq6}
  D_t (T) + \nabla  \cdot \left(\bm{\chi} + T \bm{v}\right) = 0.
 \end{equation}

 \subsection{Adjoint equations and self-adjointness when $n = 1$}
  \label{gas:2}

 Theorem \ref{simple} from Section \ref{cons.gen} shows that the
system of  gasdynamic equations (\ref{gas.eq1}) is nonlinearly
self-adjoint. Let us illustrate this statement in the
one-dimensional  case:
 \begin{align}
 & v_t +  v v_x +\frac{1}{\rho}\, p_x =0,\notag\\
 & \rho_t+ v \rho_x +\rho v_x =0, \label{gas.eq7}\\
 & p_t + v p_x + \gamma p v_x =0.\notag
\end{align}
We write the formal Lagrangian in the form
\begin{equation}
\label{gas.eq8}
 {\cal L} = U\Big(v_t +  v v_x +\frac{1}{\rho}\, p_x\Big)
 + R (\rho_t+ v \rho_x +\rho v_x) +
 P(p_t + v p_x + \gamma p v_x)
\end{equation}
 and  obtain the following adjoint system  for the new dependent variables $U, R, P:$
 \begin{align}
 & \frac{\delta {\cal L}}{\delta v} \equiv - U_t -  v U_x
 -  \rho R_x + (1-\gamma) P p_x - \gamma p P_x =0,\notag\\
 & \frac{\delta {\cal L}}{\delta \rho} \equiv - R_t - v R_x - \frac{1}{\rho^2}\,U p_x =0, \label{gas.eq9}\\
 & \frac{\delta {\cal L}}{\delta p} \equiv - P_t - \frac{1}{\rho}\,U_x +
  \frac{1}{\rho^2}\, U \rho_x + (\gamma - 1) P v_x - v P_x =0.\notag
\end{align}
 Let us take, e.g. the conservation of energy from Section
 \ref{cons.gen}. Then we have
 $$
 T = \frac{1}{2}\,\rho v^2 + \frac{p}{\gamma - 1}\,,\quad \chi= pv,
 $$
 and using the differential form (\ref{gas.eq6}) of the energy conservation
 we obtain the following equation (\ref{coneq.1}):
 \begin{align}
 \label{gas.eq10}
 & D_t \left(\frac{1}{2}\,\rho v^2 + \frac{p}{\gamma - 1}\right)
  + D_x \left(\frac{1}{2}\,\rho v^3 + \frac{\gamma}{\gamma - 1}\, pv\right)\\[1.5ex]
 & = \rho v \Big(v_t +  v v_x +\frac{1}{\rho}\, p_x\Big)
 + \frac{v^2}{2} (\rho_t+ v \rho_x +\rho v_x) +
 \frac{1}{\gamma - 1}(p_t + v p_x + \gamma p v_x).\notag
 \end{align}
Hence, the adjoint equations (\ref{gas.eq9}) are satisfied for all
solutions of the gasdynamic equations (\ref{gas.eq1}) upon the
substitution
\begin{equation}
\label{gas.eq11}
 U = \rho v, \quad R = \frac{v^2}{2}\,, \quad P = \frac{1}{\gamma -
 1}\,\cdot
\end{equation}
This conclusion can be easily verified by the direct substitution of
(\ref{gas.eq11}) in the adjoint system (\ref{gas.eq9}). Namely, we
have:
 \begin{align}
 \label{gas.eq12}
 & \frac{\delta {\cal L}}{\delta v}\bigg|_{(\ref{gas.eq11})}
 = - \rho \Big(v_t +  v v_x +\frac{1}{\rho}\, p_x\Big) - v (\rho_t+ v \rho_x +\rho v_x),\notag\\
 & \frac{\delta {\cal L}}{\delta \rho}\bigg|_{(\ref{gas.eq11})}
 = - v \Big(v_t +  v v_x +\frac{1}{\rho}\, p_x\Big),\notag\\
 & \frac{\delta {\cal L}}{\delta p}\bigg|_{(\ref{gas.eq11})} =0.
\end{align}

 \subsection{Adjoint system to equations (\ref{gas.eq1}) with $n\geq 2$}
  \label{gas:3}
 For gasdynamic equations (\ref{gas.eq1}) with two and three space variables $x^i$
  the formal Lagrangian (\ref{gas.eq8})  is  replaced by

\begin{align}
\label{gas.eq13}
 {\cal L} = \bm U \cdot \Big(\bm v_t + (\bm v \cdot \nabla) \bm v+\frac{1}{\rho}\,\nabla
  p\Big) & + R (\rho_t+\bm v\cdot \nabla \rho+\rho \nabla \cdot \bm v)\notag\\
  & +
 P(p_t + \bm v\cdot \nabla p + \gamma p \nabla \cdot \bm v),
\end{align}
where the vector $\bm U = (U^1, \ldots, U^n)$ and the scalars $R, P$
are new dependent variables. Using this formal Lagrangian, we obtain
 the following adjoint system  instead of (\ref{gas.eq9}):
 \begin{align}
  \frac{\delta {\cal L}}{\delta \bm v} \equiv & - \bm U_t -(\bm v \cdot \nabla)\bm U
 + (\bm U \cdot \nabla)\bm v - (\nabla \cdot \bm v)\bm U\notag\\ & - \rho \nabla R + (1-\gamma) P \nabla p - \gamma p \nabla P =0,\notag\\
  \frac{\delta {\cal L}}{\delta \rho} \equiv & - R_t - \bm v \cdot \nabla R - \frac{1}{\rho^2}\,\bm U \cdot \nabla p =0, \label{gas.eq14}\\
  \frac{\delta {\cal L}}{\delta p} \equiv & - P_t - \frac{1}{\rho}\,(\nabla \cdot \bm U) +
  \frac{1}{\rho^2}\, \bm U \cdot \nabla \rho + (\gamma - 1) P  (\nabla \cdot \bm v) - \bm v \cdot \nabla P =0.\notag
\end{align}
 The nonlinear self-adjointness of the system (\ref{gas.eq1}) can be
 demonstrated as in the one-dimensional case discussed  in Section \ref{gas:2}.

 \subsection{Application to nonlocal symmetries of the Chaplygin gas}
  \label{gas:4}

 The Chaplygin gas is described by the one-dimensional gasdynamic
 equations (\ref{gas.eq7}) with $\gamma = - 1:$
 \begin{align}
 & v_t +  v v_x +\frac{1}{\rho}\, p_x =0,\notag\\
 & \rho_t+ v \rho_x +\rho v_x =0, \label{gas.eq15}\\
 & p_t + v p_x - p v_x =0.\notag
 \end{align}
 Eqs. (\ref{gas.eq15}) have the same maximal Lie algebra of Lie point symmetries
 as Eqs. (\ref{gas.eq7}) with arbitrary $\gamma.$
 This algebra is
 spanned by the symmetries (\ref{gas.eq4}) in the one-dimensional case, namely
 \begin{align}
 \label{gas.eq16}
& X_1=\frac{\partial}{\partial t}\,, \quad
 X_2=\frac{\partial}{\partial x}\,, \quad
 X_3=t\frac{\partial}{\partial x}+\frac{\partial}{\partial v}\,,
  \quad  X_4=t\frac{\partial}{\partial t}
 + x \frac{\partial}{\partial x}\,, \notag\\[1.5ex]
 & X_5 =\rho \frac{\partial}{\partial \rho}+
  p \frac{\partial}{\partial p}\,, \quad
  X_6= t\frac{\partial}{\partial t}
 -v \frac{\partial}{\partial v} + 2\rho \frac{\partial}{\partial
  \rho}\,\cdot
  \end{align}
  But the Chaplygin gas has more symmetries than an arbitrary
  one-dimensional polytropic gas upon rewriting it in Lagrange's variables
  obtained by replacing $x$ and $\rho$ with $y$ and $q,$ respectively, obtained by the following
  \textit{nonlocal transformation}:
\begin{equation}
\label{gas.eq17}
 \tau = \int \rho dx, \quad q = \frac{1}{\rho}\,\cdot
\end{equation}
 Then the system (\ref{gas.eq15}) becomes
\begin{align}
 & q_t - v_\tau =0,\notag\\
 & v_t + p_\tau =0, \label{gas.eq18}\\
 & p_t - \frac{p}{q}\, v_\tau =0\notag
\end{align}
and admits the 8-dimensional Lie algebra with the basis
 \begin{align}
 \label{gas.eq19}
& Y_1=\frac{\partial}{\partial t}\,, \quad
 Y_2=\frac{\partial}{\partial \tau}\,, \quad
 Y_3= \frac{\partial}{\partial v}\,,
  \quad  Y_4= t\frac{\partial}{\partial t}
 + \tau \frac{\partial}{\partial \tau}\,, \notag\\[1.5ex]
 & Y_5 = \tau \frac{\partial}{\partial \tau}+
  p \frac{\partial}{\partial p} - q \frac{\partial}{\partial q}\,, \quad
  Y_6= v \frac{\partial}{\partial v}
 +p \frac{\partial}{\partial p} + q \frac{\partial}{\partial q}\,,\\[1.5ex]
 & Y_7 = \frac{\partial}{\partial p} + \frac{q}{p}\, \frac{\partial}{\partial q}\,, \quad
  Y_8= t \frac{\partial}{\partial v}
  - y \frac{\partial}{\partial p} - \frac{y q}{p}\, \frac{\partial}{\partial
  q}\,\cdot\notag
  \end{align}
 It is shown in \cite{akh-gaz89} that the operators
 $Y_7, Y_8$ from (\ref{gas.eq19}) lead to the following \textit{nonlocal symmetries} for
 Eqs. (\ref{gas.eq15}):
 \begin{equation}
 \label{gas.eq20}
\begin{split}
&  X_7 = \sigma \frac{\partial}{\partial x} -
  \frac{\partial}{\partial p} + \frac{\rho}{p}\,
  \frac{\partial}{\partial \rho}\,, \\[1.5ex]
 & X_8= \left(\frac{t^2}{2} + s\right) \frac{\partial}{\partial x}
 + t \frac{\partial}{\partial v} - \tau \frac{\partial}{\partial p}
 + \frac{\rho\, \tau}{p}\, \frac{\partial}{\partial  \rho}\,,
 \end{split}
\end{equation}
 where $\tau, s, \sigma$ are the following \textit{nonlocal
 variables}:
 \begin{equation}
 \label{nlv}
 \tau = \int \rho dx, \quad  s = - \int \frac{\tau}{p}\, dx, \quad
 \sigma = - \int \frac{dx}{p}\,\cdot
 \end{equation}
They can be equivalently defined by the compatible over-determined
systems
 \begin{align}
 & \tau_x = \rho, \qquad \tau_t + v \tau_x = 0,\notag\\
 & s_x = - \frac{\tau}{p}\,, \quad s_t + v s_x = 0, \label{gas.eq21}\\
 & \sigma_x = - \frac{1}{p}\,, \quad \sigma_t + v \sigma_x = 0,\notag
\end{align}
 or
\begin{align}
 & \tau_x = \rho, \qquad \tau_t = - v \rho,\notag\\
 & s_x = - \frac{\tau}{p}\,, \quad s_t  = \frac{v \tau}{p}\,, \label{gas.eq22}\\
 & \sigma_x = - \frac{1}{p}\,, \quad \sigma_t  = \frac{v}{p}\,\cdot\notag
\end{align}

Let us verify that the operator $X_7$ is admitted by Eqs.
(\ref{gas.eq15}). Its first prolongation is obtained by applying the
usual prolongation procedure and eliminating the partial derivatives
$\sigma_x$ and $\sigma_t$ via Eqs. (\ref{gas.eq22}). It has the form
\begin{align}
  X_7 & = \sigma \frac{\partial}{\partial x} -
  \frac{\partial}{\partial p} + \frac{\rho}{p}\, \frac{\partial}{\partial \rho} -
  \frac{v v_x}{p}\, \frac{\partial}{\partial v_t} + \frac{v_x}{p}\, \frac{\partial}{\partial
  v_x} - \frac{v p_x}{p}\, \frac{\partial}{\partial p_t} + \frac{p_x}{p}\, \frac{\partial}{\partial
  p_x}\notag\\[1ex]
 & + \left(\frac{\rho_t}{p} - \frac{\rho p_t}{p^2} - \frac{v \rho_x}{p}\right)\,\frac{\partial}{\partial \rho_t}
 + \left(2 \frac{\rho_x}{p} - \frac{\rho p_x}{p^2}\right)\,\frac{\partial}{\partial
 \rho_x}\,\cdot \label{gas.eq20P}
\end{align}
The calculation shows that the invariance condition is satisfied in
the following form:
 \begin{align}
 & X_7 \left(v_t +  v v_x +\frac{1}{\rho}\, p_x\right) =0,\notag\\
 & X_7(\rho_t+ v \rho_x +\rho v_x) = \frac {1}{p}\,(\rho_t+ v \rho_x +\rho v_x)
 - \frac {\rho}{p^2}\,(p_t + v p_x - p v_x) ,\notag\\
 & X_7(p_t + v p_x - p v_x) =0.\notag
\end{align}
One can  verify likewise that the invariance test for the operator
$X_8$ is satisfied in the following form:
 \begin{align}
 & X_8 \left(v_t +  v v_x +\frac{1}{\rho}\, p_x\right) =0,\notag\\
 & X_8(\rho_t+ v \rho_x +\rho v_x) = \frac {\tau}{p}\,(\rho_t+ v \rho_x +\rho v_x)
 - \frac {\rho \tau}{p^2}\,(p_t + v p_x - p v_x) ,\notag\\
 & X_8(p_t + v p_x - p v_x) =0.\notag
 \end{align}

The operators $Y_1, \ldots, Y_6$ from (\ref{gas.eq19}) do not add to
the operators (\ref{gas.eq16}) new
 symmetries of the system (\ref{gas.eq15}).

Thus, the Chaplygin gas described by Eqs. (\ref{gas.eq15}) admits
the eight-dimensional vector space spanned by the operators
(\ref{gas.eq16}) and (\ref{gas.eq20}). However this vector space is
not a Lie algebra. Namely,  the commutators of the dilation
generators $X_4, X_5, X_6$ from (\ref{gas.eq16}) with the operators
(\ref{gas.eq20}) are not  linear combinations of the operators
(\ref{gas.eq16}), (\ref{gas.eq20}) with constants coefficients. The
reason is that the operators $X_4, X_5, X_6$ are not admitted by
 the differential equations (\ref{gas.eq21}) for the nonlocal
 variables $\tau, s, \sigma.$ Therefore I will extend the action of the dilation
 generators to $\tau, s, \sigma$ so that the extended operators will be admitted
 by Eqs. (\ref{gas.eq21}).

 Let us take the operator $X_4.$ We write it in the extended form
 $$
  X'_4=t\frac{\partial}{\partial t} + x \frac{\partial}{\partial x}
  + \alpha \frac{\partial}{\partial \tau} + \beta \frac{\partial}{\partial s}
   + \mu \frac{\partial}{\partial \sigma}\,,
 $$
 where $\alpha, \beta, \mu$ are unknown functions of $t, x, v, \rho,
 p, \tau, s, \sigma.$ Then we make the prolongation of $X'_4$ to the
first-order partial derivatives of the nonlocal variables with
respect to $t$ and $x$ by treating $\tau, s, \sigma$ as
 new dependent variables and obtain
 \begin{align}
  X'_4 & =t\frac{\partial}{\partial t} + x \frac{\partial}{\partial x}
  + \alpha \frac{\partial}{\partial \tau} + \beta \frac{\partial}{\partial s}
  + \mu \frac{\partial}{\partial \sigma}\notag\\[1ex]
  & + [D_t(\alpha) - \tau_t]\frac{\partial}{\partial \tau_t}
   + [D_x(\alpha) - \tau_x]\frac{\partial}{\partial \tau_x}\notag\\[1ex]
  & + [D_t(\beta) - s_t]\frac{\partial}{\partial s_t}
   + [D_x(\beta) - s_x]\frac{\partial}{\partial s_x}\notag\\[1ex]
  & + [D_t(\mu) - \sigma_t]\frac{\partial}{\partial \sigma_t}
   + [D_x(\mu) - \sigma_x]\frac{\partial}{\partial
   \sigma_x}\notag\,\cdot
 \end{align}
 Now we require the invariance of Eqs.
 (\ref{gas.eq21}):
 \begin{align}
 & X'_4(\tau_x - \rho) = 0, \qquad X'_4(\tau_t + v \tau_x) =
 0,\notag\\[1ex]
 & X'_4\left(s_x + \frac{\tau}{p}\right) = 0, \quad X'_4(s_t + v s_x) = 0, \label{gas.eq23}\\[1ex]
 & X'_4\left(\sigma_x + \frac{1}{p}\right) = 0, \quad X'_4(\sigma_t + v \sigma_x) = 0.\notag
\end{align}
As usual, Eqs. (\ref{gas.eq23}) should be satisfied on the solutions
of Eqs. (\ref{gas.eq21}). Let us solve the equations
 $X'_4(\tau_x - \rho) = 0, \quad X'_4(\tau_t + v \tau_x) = 0.$ They
 are written
 \begin{equation}
 \label{gas.eq24}
 \left[D_x(\alpha) - \tau_x\right]_{(\ref{gas.eq21})} = 0, \quad
 \left[D_t(\alpha) - \tau_t + v \left(D_x(\alpha) - \tau_x\right)\right]_{(\ref{gas.eq21})} = 0.
 \end{equation}
Since $\tau_x = D_x(\alpha),$ the first equation in (\ref{gas.eq24})
is satisfied if we take
 $$
 \alpha = \tau
 $$
With this $\alpha$ the second equation in (\ref{gas.eq24}) is also
satisfied because $\tau_t + v \tau_x = 0.$ Now the first equation in
the second line of Eqs. (\ref{gas.eq23}) becomes
$$
 \left[D_x(\beta) - s_x + \frac{\tau}{p}\right]_{(\ref{gas.eq21})} = D_x(\beta) - 2 s_x = 0
$$
 and yields
$$
\beta = 2 s.
$$
 The second equation in the second line
of Eqs. (\ref{gas.eq23}) is also satisfied with this $\beta.$
Applying the same approach to the third line of Eqs.
(\ref{gas.eq23}) we obtain
 $$
 \mu = \sigma.
 $$

 After similar calculations with $X_5$
 and $X_6$ we obtain the following extensions of the dilation
generators:
 \begin{align}
 \label{gas.eq25}
 & X'_4=t\frac{\partial}{\partial t}
 + x \frac{\partial}{\partial x}
  + \tau \frac{\partial}{\partial \tau} + 2 s \frac{\partial}{\partial s}
   + \sigma \frac{\partial}{\partial \sigma}\,, \notag\\[1.5ex]
 & X'_5 =\rho \frac{\partial}{\partial \rho}+
  p \frac{\partial}{\partial p}
  + \tau \frac{\partial}{\partial \tau} -\sigma \frac{\partial}{\partial \sigma}\,, \\[1.5ex]
 & X'_6= t\frac{\partial}{\partial t} -v \frac{\partial}{\partial v}
 + 2\rho \frac{\partial}{\partial \rho}
 + 2 \tau \frac{\partial}{\partial \tau} + 2 s \frac{\partial}{\partial s}\,\cdot\notag
  \end{align}

 The operators (\ref{gas.eq20}), (\ref{gas.eq25}) together with the operators $X_1, X_2, X_3$
 from (\ref{gas.eq16}) span the eight-dimensional Lie algebra $L_8$ admitted by
 Eqs. (\ref{gas.eq15}) and Eqs. (\ref{gas.eq21}). The algebra
$L_8$ has the following commutator table:
$$
 %\begin{equation}
% \label{gas.eq26}
 \begin{tabular}{||c||c|c|c|c|c|c|c|c||}
 \hline\hline &&&&&&&&\\
 & $X_1$ & $X_2$ & $X_3$ & $X'_4$ & $X'_5$ & $X'_6$ & $X_7$& $X_8$\\[1ex]
\hline\hline
$X_1$ & $0$ & $0$ & $X_2$ & $X_1$ & $0$ & $X_1$ & $0$ & $X_3$\\[1ex]
\hline
$X_2$ & $0$ & $0$ & $0$ & $X_2$ & $0$ & $0$& $0$ & $0$\\[1ex]
\hline
$X_3$ & $- X_2$ & $0$ & $0$ & $0$ & $0$ & $-X_3$& $0$ & $0$\\[1ex]
\hline
$X'_4$ & $-X_1$ & $- X_2$ & $0$ & $0$ & $0$ & $0$& $0$ & $X_8$\\[1ex]
\hline
$X'_5$ & $0$ & $0$ & $0$ & $0$ & $0$ & $0$& $- X_7$ & $0$\\[1ex]
\hline
$X'_6$ & $- X_1$ & $0$ & $X_3$ & $0$ & $0$ & $0$& $0$ & $2X_8$\\[1ex]
\hline
$X_7$ & $0$ & $0$ & $0$ &$0$ & $X_7$ & $0$& $0$ & $0$\\[1ex]
\hline
$X_8$ & $- X_3$ & $0$ & $0$ & $- X_8$ & $0$ & $- 2 X_8$& $0$ & $0$\\[1ex]
\hline\hline
\end{tabular}
$$

Let us apply Theorem \ref{main.thm} to the nonlocal symmetries
(\ref{gas.eq20}) of the Chaplygin gas. The formal Lagrangian
(\ref{gas.eq8}) for Eqs. (\ref{gas.eq15}) has the form
\begin{equation}
\label{gas.eq26}
 {\cal L} = U\Big(v_t +  v v_x +\frac{1}{\rho}\, p_x\Big)
 + R (\rho_t+ v \rho_x +\rho v_x) +
 P(p_t + v p_x - p v_x).
\end{equation}
 Accordingly, the adjoint system (\ref{gas.eq9}) for the Chaplygin
 gas is  written
 \begin{align}
 & \frac{\delta {\cal L}}{\delta v} \equiv - U_t -  v U_x
 -  \rho R_x + 2 P p_x + p P_x =0,\notag\\
 & \frac{\delta {\cal L}}{\delta \rho} \equiv - R_t - v R_x - \frac{1}{\rho^2}\,U p_x =0, \label{gas.eq27}\\
 & \frac{\delta {\cal L}}{\delta p} \equiv - P_t - \frac{1}{\rho}\,U_x +
  \frac{1}{\rho^2}\, U \rho_x -2 P v_x - v P_x =0.\notag
\end{align}
Let us proceed as in Section \ref{gas:2}. Namely, let us first
construct solutions to the adjoint system (\ref{gas.eq27}) by using
the known conservation laws given in Section \ref{cons.gen}. Since
 the one-dimensional does not have the conservation of
angular momentum, we use the conservation of mass, energy, momentum
and center-of-mass and obtain the respective differential
conservation equations (see the derivation of  Eq.
(\ref{gas.eq10})):
 \begin{align}
 & D_t (\rho) + D_x(\rho v)
   =\rho_t+ v \rho_x +\rho v_x,\label{gas.eq28}\\
 & D_t (\rho v^2 - p) + D_x(p v + \rho v^3)
   =2 \rho v \Big(v_t +  v v_x +\frac{1}{\rho}\, p_x\Big)\notag\\
   & \qquad \qquad \quad + v^2 (\rho_t+ v \rho_x +\rho v_x)
   - (p_t + v p_x - p v_x), \label{gas.eq29}\\
 & D_t (\rho v) + D_x(p + \rho v^2)
   =\rho \Big(v_t +  v v_x +\frac{1}{\rho}\, p_x\Big)\notag\\
   & \qquad \qquad \qquad \qquad \qquad + v (\rho_t+ v \rho_x +\rho v_x), \label{gas.eq30}\\
  & D_t (t \rho v - x \rho) + D_x(t p + t \rho v^2 - x \rho v)\notag\\
   & \quad =t \rho \Big(v_t +  v v_x +\frac{1}{\rho}\, p_x\Big)
    + (tv - x) (\rho_t+ v \rho_x +\rho v_x). \label{gas.eq31}
\end{align}
Eqs. (\ref{gas.eq28})- (\ref{gas.eq31}) give the following solutions
to the adjoint equations (\ref{gas.eq27}):
 \begin{align}
 & U = 0, \qquad R = 1, \qquad \ \ P = 0,\label{gas.eq32}\\
 & U = 2 \rho v, \quad R = v^2, \qquad P = - 1, \label{gas.eq33}\\
 & U = \rho, \qquad R = v, \qquad \ \ P = 0, \label{gas.eq34}\\
  & U = t \rho, \quad \ R = tv - x, \quad P = 0. \label{gas.eq35}
\end{align}

The formal Lagrangian (\ref{gas.eq26}) contains the derivatives only
of the first order. Therefore Eqs. (\ref{main:eq.5}) for calculating
the conserved vectors take the simple form
 \begin{equation}
 \label{gas.eq36}
 C^i  = W^\alpha\,\frac{\partial {\cal L}}{\partial
 u_i^\alpha}\,, \quad i = 1, 2.
 \end{equation}
We denote
 $$
 t = x^1, \quad x = x^2, \quad v = u^1, \quad \rho = u^2,
 \quad p = u^3.
 $$
 In this notation conservation equation (\ref{main:eq.4}) will be written in the form
 \begin{equation}
 \label{gas.eq37}
 \left[D_t (C^1) + D_x (C^2) \right]_{(\ref{gas.eq15})} = 0.
 \end{equation}
 Writing (\ref{gas.eq36}) in the form
 \begin{align}
 & C^1 = W^1\frac{\partial {\cal L}}{\partial v_t} + W^2\frac{\partial {\cal L}}{\partial \rho_t}
  + W^3\frac{\partial {\cal L}}{\partial p_t}\,,\notag\\[1.5ex]
 & C^2 = W^1\frac{\partial {\cal L}}{\partial v_x} + W^2\frac{\partial {\cal L}}{\partial \rho_x}
  + W^3\frac{\partial {\cal L}}{\partial p_x} \notag
\end{align}
 and substituting the expression (\ref{gas.eq26}) for ${\cal L}$ we obtain the following
 final expressions for computing the components of conserved vectors:
 \begin{align}
 & C^1 = U W^1 + R W^2 + P W^3,\label{gas.eq38}\\[1.5ex]
 & C^2 = (v U + \rho R - p P) W^1 + v R W^2 + \left(\frac{1}{\rho}\, U + v P\right) W^3, \label{gas.eq39}
\end{align}
where
 \begin{equation}
 \label{gas.eq40}
 W^\alpha = \eta^\alpha - \xi^i u_i^\alpha, \quad \alpha = 1, 2, 3.
 \end{equation}

 We will apply Eqs. (\ref{gas.eq38})-(\ref{gas.eq39}) to the
 nonlocal symmetries (\ref{gas.eq20}). First we write the expressions
 (\ref{gas.eq40}) for the operator $X_7$ from (\ref{gas.eq20}):
 \begin{equation}
 \label{gas.eq41}
 W^1 = - \sigma v_x, \quad W^2 = \frac{\rho}{p} - \sigma \rho_x,
 \quad W^3 = - (1 + \sigma p_x).
 \end{equation}
 Then we substitute (\ref{gas.eq41}) in
 (\ref{gas.eq38})-(\ref{gas.eq39}) and obtain four conserved
 vectors  by replacing $U, R, P$ with each of four different solutions
 (\ref{gas.eq32})-(\ref{gas.eq35}) of the adjoint system
 (\ref{gas.eq27}). Some of these conserved vectors may be trivial.
 We select only the nontrivial ones.

 Let us calculate the conserved vector obtained by eliminating
 $U, R, P$ by using the solution (\ref{gas.eq32}),
 $U = 0, \ R = 1, \ P = 0.$ In this case
 (\ref{gas.eq38})-(\ref{gas.eq39}) and (\ref{gas.eq41}) yield
 \begin{align}
 & C^1 =  W^2 = \frac{\rho}{p} - \sigma \rho_x,\label{gas.eq42}\\
 & C^2 =  \rho W^1 + v W^2 =- \sigma \rho v_x + \frac{\rho}{p}\,v
 - \sigma v \rho_x. \notag
\end{align}
We write $$- \sigma \rho_x = - D_x(\sigma \rho) + \rho \sigma_x,$$
replace $\sigma_x$ with $- 1/p$ according to Eqs. (\ref{gas.eq21})
and obtain
$$C^1 = - D_x(\sigma \rho).$$
Therefore application of the operations
(\ref{cons:vec.2})-(\ref{cons:vec.4}) yields $\widetilde C^1 = 0$
and
 \begin{align}
 & \widetilde C^2 = - \sigma \rho v_x + \frac{\rho}{p}\,v - \sigma v
\rho_x - D_t (\sigma \rho)\notag\\
 &  =  - \sigma \rho v_x + \frac{\rho}{p}\,v - \sigma v
\rho_x -\sigma \rho_t - \sigma_t \rho\notag\\
 &  =  - \sigma (\rho_t + v  \rho_x + \rho v_x).\notag
\end{align}
We have replaced $\sigma_t$  with $v/p$ according to Eqs.
(\ref{gas.eq22}). The above expression for $\widetilde C^2$ vanishes
on Eqs. (\ref{gas.eq15}). Hence, the conserved vector
(\ref{gas.eq42}) is trivial.

Utilization of the solutions (\ref{gas.eq33}) and (\ref{gas.eq34})
also leads to trivial conserved vectors only. Finally, using the
solution (\ref{gas.eq35}),
 $$
 U = t \rho, \quad \ R = tv - x, \quad P = 0,
 $$
we obtain, upon simplifying by using the operations
(\ref{cons:vec.2})-(\ref{cons:vec.4}), the following nontrivial
conserved vector:
 \begin{equation}
 \label{gas.eq43}
 C^1 = \sigma \rho, \quad C^2 = \sigma \rho v + t.
 \end{equation}
The conservation equation (\ref{gas.eq37}) is satisfied in the
following form:
 \begin{equation}
 \label{gas.eq44}
 D_t (C^1) + D_x (C^2) = \sigma (\rho_t + v  \rho_x + \rho v_x).
 \end{equation}

Note that we can write $C^2$ in (\ref{gas.eq43}) without $t$ since
it adds only the trivial conserved vector with the components $C^1 =
0, \ C^2 = t.$ Thus, removing $t$ in (\ref{gas.eq43}) and using the
definition of $\sigma$ given in (\ref{nlv}) we formulate the result.
 \begin{prop}
 \label{chap:prop1}
 The nonlocal symmetry $X_7$ of the Chaplygin gas gives the following nonlocal conserved
vector:
 \begin{equation}
 \label{gas.eq46}
 C^1 = - \rho \int \frac{dx}{p}\,, \quad C^2 = - \rho v \int
 \frac{dx}{p}\,\cdot
 \end{equation}
 \end{prop}

 Mow we use the operator $X_8$ from (\ref{gas.eq20}). In this case
 \begin{equation}
 \label{gas.eq47}
 \begin{split}
 & W^1 = t - \left(\frac{t^2}{2} + s\right) v_x, \\
 & W^2 = \frac{\rho \tau}{p} - \left(\frac{t^2}{2} + s\right)
 \rho_x,\\
 & W^3 = - \tau - \left(\frac{t^2}{2} + s\right) p_x.
 \end{split}
 \end{equation}

 Substituting in (\ref{gas.eq38})-(\ref{gas.eq39}) the expressions (\ref{gas.eq47}) and the
 solution (\ref{gas.eq32}) of the adjoint system, i.e.
 letting $U = 0, \ R = 1, \ P = 0,$  we obtain
 \begin{align}
 & C^1 =  W^2 = \frac{\rho \tau}{p} - \left(\frac{t^2}{2} + s\right)
 \rho_x,\notag\\
 & C^2 =  \rho W^1 + v W^2 = t \rho + \frac{\rho v \tau}{p} -
  \left(\frac{t^2}{2} + s\right)(\rho v_x +  v \rho_x). \notag
\end{align}
Noting that
 $$
- \left(\frac{t^2}{2} + s\right) \rho_x = - \frac{\rho \tau}{p} -
D_x\left(\frac{t^2}{2}\,\rho + \rho s\right)
 $$
 we reduce the above vector to the trivial conserved vector $\widetilde C^1 = 0, \
 \widetilde C^2 = 0.$

 Taking the  solution (\ref{gas.eq33}) of the adjoint system, i.e.
 letting $$U = 2 \rho v, \ R = v^2, \ P = - 1,$$  we obtain
  \begin{align}
  C^1 &  =  2  \rho v W^1 + v^2 W^2  - W^3 \notag \\
   & =
 2 t \rho v + \frac{\rho \tau v^2}{p} + \tau -
  \left(\frac{t^2}{2} + s\right) D_x \left(\rho v^2 -
 p\right),\notag\\[1.5ex]
 C^2 & =  (3 \rho v^2 + p) W^1 + v^3 W^2 + v W^3\notag \\
   & = t (3\rho v^2 + p) + \frac{\rho \tau v^3}{p} - v \tau\notag\\
   & -
  \left(\frac{t^2}{2} + s\right)(3 \rho v^2 v_x +  v^3 \rho_x  + p v_x
  + v p_x). \notag
\end{align}
Then, upon rewriting  $C^1$ in the form
 $$
 C^1 = 2 t \rho v + 2 \tau -
 D_x \left[\left(\frac{t^2}{2} + s\right) (\rho v^2 - p)\right]
 $$
and applying the operations (\ref{cons:vec.2})-(\ref{cons:vec.4}) we
arrive at the following conserved vector:
 \begin{equation}
 \label{gas.eq48}
 C^1 =  t \rho v + \tau, \quad C^2 = t (\rho v^2 +  p).
 \end{equation}
The conservation equation (\ref{gas.eq37}) is satisfied for
(\ref{gas.eq48}) in the following form:
 \begin{equation}
 \label{gas.eq49}
 D_t (C^1) + D_x (C^2) = t \rho \left( v_t +  v v_x +\frac{1}{\rho}\, p_x\right) + t v (\rho_t + v  \rho_x + \rho v_x).
 \end{equation}

 Taking the  solution (\ref{gas.eq34}) of the adjoint system, i.e.
 letting $$U = \rho, \ R = v, \ P = 0,$$  we obtain
 $$
  C^1 =  \rho W^1 + v W^2, \quad C^2 =  2 \rho v W^1 + v^2 W^2 +
  W^3.
  $$
 Substituting the expressions (\ref{gas.eq47}) for $W^1, W^2, W^3$ and simplifying
 as in the previous case we obtain the conserved vector
 \begin{equation}
 \label{gas.eq50}
 C^1 = t \rho, \quad C^2 = t \rho v - \tau.
 \end{equation}
The conservation equation (\ref{gas.eq37}) is satisfied for
(\ref{gas.eq48}) in the following form:
 \begin{equation}
 \label{gas.eq51}
 D_t (C^1) + D_x (C^2) =  t (\rho_t + v  \rho_x + \rho v_x).
 \end{equation}

Finally, we take the solution (\ref{gas.eq35}), \
 $
 U = t \rho, \ R = tv - x, \ P = 0,
 $
 and obtain
$$
 C^1 = t \rho W^1 + (tv - x) W^2, \quad C^2 (2 t \rho v - x \rho)
 W^1 + (tv^2 - x v) W^2 + t W^3.
$$
Simplifying as above, we arrive at
 the conserved vector
 \begin{equation}
 \label{gas.eq52}
 C^1 = \left(\frac{t^2}{2} - s\right)\rho, \quad C^2 =\left(\frac{t^2}{2} - s\right)\rho v -  t \tau.
 \end{equation}
The conservation equation (\ref{gas.eq37}) is satisfied for
(\ref{gas.eq48}) in the following form:
 \begin{equation}
 \label{gas.eq53}
  D_t (C^1) + D_x (C^2) = \left(\frac{t^2}{2} - s\right) (\rho_t + v  \rho_x + \rho v_x).
 \end{equation}

 Substituting in the conserved vectors (\ref{gas.eq48}), (\ref{gas.eq50}) and (\ref{gas.eq52}) the definition
(\ref{nlv}) of the nonlocal variables we formulate the result.
 \begin{prop}
 \label{chap:prop2}
 The nonlocal symmetry $X_8$ of the Chaplygin gas gives the following nonlocal conserved
vectors:
 \begin{align}
 & C^1 =  t \rho v + \int \rho dx, \quad C^2 = t (\rho v^2 +
 p);\label{gas.eq54}\\
 & C^1 = t \rho, \qquad \quad C^2 = t \rho v - \int \rho dx;\label{gas.eq55}
 \end{align}
 \begin{equation}
 \label{gas.eq56}
 \begin{split}
 & C^1 = \left[\frac{t^2}{2} +
 \int \frac{1}{p}\left(\int\rho dx\right) dx \right]\rho,\\[1.5ex]
 & C^2 =\left[\frac{t^2}{2} + \int \frac{1}{p}\left(\int\rho dx\right) dx \right]\rho v -  t \int \rho dx.
 \end{split}
 \end{equation}
 \end{prop}
 \begin{thm}
 \label{gas:thm}
 Application of Theorem \ref{main.thm} to \textit{two} nonlocal symmetries (\ref{gas.eq20})
 gives \textit{four} nonlocal conservation
laws (\ref{gas.eq46}), (\ref{gas.eq54})-(\ref{gas.eq56})  for the
Chaplygin gas (\ref{gas.eq15}).
 \end{thm}

 \subsection{The operator identity for nonlocal symmetries}
  \label{gas:5}

 \begin{exa}
 \label{gas:5.exa}
 Let us verify that the operator identity (\ref{opid.eq1}) is satisfied for
 the nonlocal symmetry $X_7$ of the Chaplygin gas. Specifically, let us check
 that the coefficients of
 \begin{equation}
 \label{gas.eq58}
 \frac{\partial}{\partial v}\,, \quad
 \frac{\partial}{\partial \rho}\,, \quad \frac{\partial}{\partial
p}\,, \quad \frac{\partial}{\partial v_t}\,, \quad
\frac{\partial}{\partial v_x}\,, \quad
 \frac{\partial}{\partial \rho_t}\,, \quad \frac{\partial}{\partial \rho_x}\,,
 \quad \frac{\partial}{\partial p_t}\,, \quad
 \frac{\partial}{\partial p_x}
 \end{equation}
 in both sides of  (\ref{opid.eq1}) are equal. Using the first prolongation (\ref{gas.eq20P})
 of $X_7$ and the definition of the nonlocal variable $\sigma$ given in Eqs. (\ref{gas.eq22})
 we see that the left-hand side of the identity (\ref{opid.eq1}) is written
 \begin{align}
  X_7 & + D_i (\xi^i) = \sigma \frac{\partial}{\partial x} -
  \frac{\partial}{\partial p} + \frac{\rho}{p}\, \frac{\partial}{\partial \rho} -
  \frac{v v_x}{p}\, \frac{\partial}{\partial v_t} + \frac{v_x}{p}\, \frac{\partial}{\partial
  v_x} - \frac{v p_x}{p}\, \frac{\partial}{\partial p_t}\notag\\[1ex]
 &  + \frac{p_x}{p}\, \frac{\partial}{\partial p_x} + \left(\frac{\rho_t}{p}
 - \frac{\rho p_t}{p^2} - \frac{v \rho_x}{p}\right)\,\frac{\partial}{\partial \rho_t}
 + \left(2 \frac{\rho_x}{p} - \frac{\rho p_x}{p^2}\right)\,\frac{\partial}{\partial
 \rho_x} - \frac{1}{p}\,\cdot \label{gas.eq59}
\end{align}
 Then we use the expressions (\ref{gas.eq41})  of  $W^\alpha$ for the
 operator $X_7,$ substitute them in the definition (\ref{opid.eq5})
 of ${\sf N}^i$
 and obtain  in our approximation:
 \begin{align}
 & {\sf N}^1 =  - \sigma v_x \frac{\partial}{\partial v_t}
 + \left(\frac{\rho}{p} - \sigma \rho_x\right)\frac{\partial}{\partial \rho_t}
 - (1  + \sigma p_x) \frac{\partial}{\partial p_t}\,, \notag\\[1ex]
 & {\sf N}^2 = \sigma - \sigma v_x \frac{\partial}{\partial v_x}
 + \left(\frac{\rho}{p} - \sigma \rho_x\right)\frac{\partial}{\partial \rho_x}
 - (1  + \sigma p_x) \frac{\partial}{\partial p_t}\,\cdot \notag
 \end{align}
 Now the right-hand side of (\ref{opid.eq1}) is written:
 \begin{align}
 & W^1 \frac{\delta}{\delta v}  +  W^2 \frac{\delta}{\delta \rho} +  W^3 \frac{\delta}{\delta p}
 + D_t {\sf N}^1 + D_x {\sf N}^2\notag\\[1ex]
 &  =  - \sigma v_x \left[\frac{\partial}{\partial v} - D_t \frac{\partial}{\partial v_t}
 - D_t \frac{\partial}{\partial v_x}\right]\notag\\[1ex]
 &  + \left(\frac{\rho}{p} - \sigma \rho_x\right)\left[\frac{\partial}{\partial \rho}
 - D_t\frac{\partial}{\partial \rho_t} - D_x\frac{\partial}{\partial \rho_x}\right] \label{gas.eq60}\\[1ex]
 &  - (1  + \sigma p_x)\left[\frac{\partial}{\partial p}
 - D_t\frac{\partial}{\partial p_t} - D_x\frac{\partial}{\partial p_x}\right]\notag\\[1ex]
 & + D_t \left[- \sigma v_x \frac{\partial}{\partial v_t}
 + \left(\frac{\rho}{p} - \sigma \rho_x\right)\frac{\partial}{\partial \rho_t}
 - (1  + \sigma p_x) \frac{\partial}{\partial p_t}\right]\notag\\[1ex]
 & + D_x \left[\sigma - \sigma v_x \frac{\partial}{\partial v_x}
 + \left(\frac{\rho}{p} - \sigma \rho_x\right)\frac{\partial}{\partial \rho_x}
 - (1  + \sigma p_x) \frac{\partial}{\partial p_t}\right]\,\cdot\notag
\end{align}
 Making the changes in two last lines of Eq. (\ref{gas.eq60}) such as
 $$
 D_t \left[- \sigma v_x \frac{\partial}{\partial v_t}\right] =
 - \sigma v_x D_t \frac{\partial}{\partial v_t} -  D_t(\sigma v_x) \frac{\partial}{\partial v_t}
 = - \sigma v_x D_t \frac{\partial}{\partial v_t} - \left(\frac{v}{p}\, v_x + \sigma v_{tx}\right) \frac{\partial}{\partial v_t}
 $$
 one can see that the coefficients of the differentiations
 (\ref{gas.eq58}) in (\ref{gas.eq59}) and (\ref{gas.eq60}) coincide.
 Inspection of the coefficients of the differentiations in higher
derivatives $v_{tt}, v_{tx}, v_{xx}, \ldots$ requires  the
 higher-order prolongations of the operator $X_7.$

 \end{exa}
 \begin{exer}
\label{gas:5.exer}
 % \label{chap:exer}
 Verify that the operator identity (\ref{opid.eq1}) is satisfied in the same
 approximation as in Example \ref{gas:5.exa} for
 the nonlocal symmetry operator  $X_8$ from (\ref{gas.eq20}).
 \end{exer}

 \section{Comparison  with the ``direct method"}
 \label{dm}
 \setcounter{equation}{0}

 \subsection{General discussion}

  Theorem \ref{main.thm} allows to construct conservation laws for equations with known symmetries
 simply by substituting in Eqs. (\ref{main:eq.5}) the expressions
 $W^\alpha$ and ${\cal L}$ given by Eqs. (\ref{main:eq.6})
 and (\ref{main:eq.6}), respectively.

 The ``direct method" means the determination of the conserved
 vectors (\ref{cons:vec.1}) by solving Eq. (\ref{main:eq.4})
 for $C^i.$
 Upon restricting the highest order of derivatives of $u$ involved in  $C^i,$
 Eq. (\ref{main:eq.4}) splits into several equations. If one can solve
 the resulting system, one obtains the desired conserved vectors.
 Existence of symmetries is not required.

 To the best of my knowledge, the direct method was used for the first
 time
 in 1798 by Laplace \cite{lap98}. He
applied the method to Kepler's problem in celestial mechanics and
found a new vector-valued conserved quantity (see \cite{lap98}, Book
II, Chap. III, Eqs. (P)) known as Laplace's vector.

The application of the direct method to the gasdynamic equations
(\ref{gas.eq1})
 allowed to demonstrate in \cite{ter-shm75} that  all conservation
 laws involving only the independent and
dependent variables (\ref{gas.eq2}), (\ref{gas.eq3})
 were
 provided by the classical conservation laws (mass, energy, momentum, angular momentum and
 center-of-mass) given in Section \ref{gas:1}
 and the following two special conservation laws
\begin{align}
 & \frac{d}{dt} \int\limits_{\Omega(t)} \left\{ t (\rho |\bm v|^2 +np
)-\rho \, \bm x \cdot \bm v
 \right\}d\omega=- \int\limits_{S(t)} p \,(2 t\bm v-\bm x)\cdot \bm
 \nu dS,\notag\\[1ex]
 & \frac{d}{dt} \int\limits_{\Omega(t)} \left\{ t^2 (\rho |\bm v|^2 +np )
 -\rho \bm x \cdot ( 2t \bm v
-\bm x \right\}d\omega=- \int\limits_{S(t)} 2tp\, ( t\bm v-\bm
x)\cdot \bm \nu dS\notag
\end{align}
 that were found in \cite{ibr73} in the case $\gamma =(n+2)/n$
  by using the symmetry ideas.

All local conservation laws for the
 heat equation $u_t - u_{xx} = 0$ have been found  by
the direct method in \cite{dor-svir83} (see in \cite{ibr94-96}, vol.
1, Section 10.1; see also \cite{ste-wol81}). Namely it has been
shown by considering the conservation equations of the form
 $$
 D_t[\tau (t, x, u, u_x, u_{xx}, \ldots)] + D_x[\psi (t, x, u, u_x, u_{xx},
\ldots)] = 0
 $$
 that all such conservation laws are given by
 $$
 D_t[\varphi(t, x) u] + D_x[u \varphi_x (t, x) - u_x \varphi (t, x)] = 0,
 $$
 where $v = \varphi(t, x)$ is an arbitrary solution of the
 adjoint equation $v_t + v_{xx} = 0$ to the heat equation.
 Similar result can be obtained by applying Theorem \ref{main.thm} for any linear equation, e.g. for the heat equation
$
 u_t - \Delta u = 0
 $
with any number of spatial variables $x = (x^1, \ldots, x^n).$
 Namely, applying formula (\ref{main:eq.5}) to the scaling symmetry $X = u \partial/\partial u$
 we obtain the conservation law
 $$
 D_t[\varphi (t, x) u] + \nabla \cdot [u \nabla \varphi (t, x) - \varphi (t, x) \nabla u] =
0,
$$
 where $v = \varphi (t, x)$ is an arbitrary solution of the adjoint equation $v_t + \Delta v = 0$ to
 the heat equation. This conservation law embraces
the conservation laws associated with all other symmetries
 of the heat equation.

 Various mathematical models for describing the geological process of
  segregation and migration of large volumes of molten rock were
 proposed in the geophysical literature (see the papers \cite{sco-ste84},
 \cite{bar-ric86}, \cite{sco-ste91}, \cite{har96}, \cite{mal-mas07} and the references therein).
 One of them is known as the
 \textit{generalized magma equation} and has the form
 \begin{equation}
 \label{magma}
 u_t+D_z\left[u^n  - u^n D_z \left(u^{-m}
 u_t\right)\right]=0, \quad n, m = {\rm const.}
 \end{equation}
 It is accepted as a reasonable mathematical model
 for describing melt migration through the Earth's mantle. Several
 conservation laws for this model have been calculated by the direct method in
 \cite{bar-ric86}, \cite{har96} and interpreted from symmetry point of view in \cite{mal-mas07}.
 It is shown in \cite{kha09} that Eq. (\ref{magma}) is quasi self-adjoint
 with the substitution (\ref{qsa0.eq1}) given by $v = u^{1 - n -
 m}$ if $m+n \not= 1$ and  $v = \ln |u|$ if $m+n = 1.$ These
substitutions show
 that Eq. (\ref{magma}) is strictly self-adjoint (Definition
 \ref{saeq:def}) if $m+n = 0.$ Using the quasi self-adjointness, the
 conservation laws are easily computed in \cite{kha09}.

Some simplification of the direct method was suggested in
\cite{anc-blu97}. Namely, one writes the conservation equation in
the form (\ref{coneq.1}),
 \begin{equation}
 \tag{\ref{coneq.1}}
 D_i (C^i)  = \mu^{\bar \alpha} F_{\bar \alpha} \big(x, u, u_{(1)}, \ldots,
 u_{(s)}\big),
 \end{equation}
 and
first finds the undetermined coefficients $\mu^{\bar \alpha}$ by
satisfying the integrability condition of Eqs. (\ref{coneq.1}), i.e.
by solving the equations (see Proposition \ref{test:div} in Section
\ref{test})
 \begin{equation}
 \label{dm:eq.2}
  \frac{\delta}{\delta u^\alpha}\left[\mu^{\bar \beta} (x, u, u_{(1)},
 \ldots)\, F_{\bar \beta} \big(x, u, u_{(1)}, \ldots, u_{(s)}\big)\right] =0, \quad \alpha = 1, \ldots, m.
 \end{equation}
Then, for each solution $\mu^{\bar \alpha}$ of Eqs. (\ref{dm:eq.2}),
the components $C^i$ of the corresponding conserved vector are
computed from Eq. (\ref{coneq.1}). In simple situations $C^i$ can be
detected merely by looking at the  right-had side of Eq.
(\ref{coneq.1}), see further Example \ref{dm:KdV}.
 \begin{rem}
 \label{dm:rem1}
 Note that Eq. (\ref{dm:eq.2}) should be satisfied on the solutions of
 Eqs. (\ref{main:eq.1}). Then the left-hand side of
 (\ref{dm:eq.2}) can be written as
 $$
 F^*_\alpha \big(x, u, v, \ldots, u_{(s)}, v_{(s)}\big)\Big|_{v
 = \mu (x, u, u_{(1)}, \ldots)}
 $$
 with $F^*_\alpha$  defined by Eq. (\ref{gsa:adeqs}).
 \end{rem}

 The reader can find a detailed discussion of the direct method in the recent book \cite{blu-che-anc10}.
 I will compare two methods by considering few examples and exercises.

 \subsection{Examples and exercises}

 \begin{exa}
 \label{dm:KdV}
 (See \cite{blu-che-anc10}, Sec. 1.3). Let us consider the KdV equation (\ref{gsa:KdV.eq1}),
 \begin{equation}
 \tag{\ref{gsa:KdV.eq1}}
 u_t = u_{xxx} + u u_x,
 \end{equation}
and write the condition (\ref{dm:eq.2}) for $\mu = \mu(t, x, u).$ We
have:
 \begin{align}
 & \frac{\delta}{\delta u}\left[\mu(t, x, u) (u_t - u_{xxx} - u
 u_x)\right]\notag\\[1ex]
  & = - D_t(\mu) + D_x^3(\mu) + D_x (u \mu) - \mu u_x
  + (u_t - u_{xxx} - u u_x) \frac{\partial \mu}{\partial u} \notag\\[1ex]
  & = - D_t(\mu) + D_x^3(\mu) + u D_x (\mu)
  + (u_t - u_{xxx} - u u_x) \frac{\partial \mu}{\partial u}\,\cdot \notag
 \end{align}
 In accordance with Remark \ref{dm:rem1}, we consider this expression
 on the solutions of the KdV equation and see that
 Eq. (\ref{dm:eq.2}) coincides with the adjoint equation (\ref{gsa:KdV.eq2}) to  (\ref{gsa:KdV.eq1}):
 \begin{equation}
 \label{dm:eq.3}
  D_t(\mu) = D_x^3(\mu) + u D_x (\mu).
  \end{equation}
 Its solution is given in Example \ref{gsa:KdV} and has the form
 (\ref{gsa:KdV.eq3}),
 $$
 \mu = A_1 + A_2 u + A_3 (x + t u), \quad A_1, A_2, A_3 = {\rm
 const.}
 $$
 Thus, we have the following  three linearly independent solutions
 of Eq. (\ref{dm:eq.3}):
 $$
 \mu_1 = 1, \quad \mu_2 = u, \quad \mu_3 = (x + t u).
 $$
 and the corresponding three equations  (\ref{coneq.1}):
 \begin{align}
 & D_t(C^1) + D_x (C^2) = u_t - u_{xxx} - u u_x, \label{dm:eq.4}\\
 & D_t(C^1) + D_x (C^2) = u (u_t - u_{xxx} - u u_x), \label{dm:eq.5}\\
 & D_t(C^1) + D_x (C^2) = (x + t u) (u_t - u_{xxx} - u u_x). \label{dm:eq.6}
 \end{align}

In this simple example the components $C^1, C^2$ of the conserved
vector can be easily seen from the right-hand sides of Eqs.
(\ref{dm:eq.4})-(\ref{dm:eq.6}). In the case of (\ref{dm:eq.4}),
 (\ref{dm:eq.5}) it is obvious. Therefore let us consider the right-hand side of Eq. (\ref{dm:eq.6}).
 We see that
 \begin{align}
 (x + t u) u_t & =   D_t\left(x u + \frac{1}{2}\,
 tu^2\right) - \frac{1}{2}\,u^2,\notag\\[1ex]
 - (x + t u) u u_x & = -  D_x\left(\frac{1}{2}\,x u^2 + \frac{1}{3}\,t u^3\right) + \frac{1}{2}\,u^2,\notag\\[1ex]
 - (x + t u) u_{xxx} & = -  D_x\left(x u_{xx} + t u u_{xx}\right) + u_{xx} + t u_x
 u_{xx},\notag\\[1ex]
 & =  D_x\left(u_x + \frac{1}{2}\,t u_x ^2 - x u_{xx} - t u u_{xx}\right).\notag
 \end{align}
Hence, the right-hand side of Eq. (\ref{dm:eq.6}) can be written in
the divergence form:
 \begin{align}
 & (x + t u) (u_t - u_{xxx} - u u_x)\notag\\[1ex]
 & = D_t \left( t\frac{u^2}{2} +xu \right) +D_x \left[u_x +
 t\left(\frac{u^2_x}{2} -uu_{xx} -\frac{u^3}{3} \right)
   -x\left(\frac{u^2}{2} + u_{xx}\right)\right].\notag
 \end{align}
 The expressions under $D_t(\cdots)$ and $D_x(\cdots)$ give $C^1$ and $C^2,$ respectively, in (\ref{dm:eq.6}).
 Note that the corresponding  conservation law
 \begin{equation}
 \label{dm:eq.7}
 D_t \left( t\frac{u^2}{2} +xu \right) +D_x \left[u_x +
 t\left(\frac{u^2_x}{2} -uu_{xx} -\frac{u^3}{3} \right)
   -x\left(\frac{u^2}{2} + u_{xx}\right)\right]=0.
 \end{equation}
 was derived from the Galilean invariance of
 the KdV equation (see \cite{ibr83}, Section 22.5) and by the direct method (see \cite{blu-che-anc10}, Section 1.3.5).

 The similar treatment of the right-hand sides of the equations (\ref{dm:eq.4}) and (\ref{dm:eq.5}) leads
 to Eq. (\ref{gsa:KdV.eq1}) and  to the conservation law
 \begin{equation}
 \label{dm:eq.8}
 D_t(u^2) +D_x \left(u^2_x - 2uu_{xx} -\frac{2}{3}\,u^3\right) = 0,
 \end{equation}
 respectively.  Theorem \ref{main.thm} associates the conservation
 law (\ref{dm:eq.8}) with the scaling symmetry of the KdV equation.
 \end{exa}

  \begin{exer}
 \label{dm:pulse}
 Apply the direct method to the \textit{short pulse equation} (\ref{pulse:2})
  using the differential substitution
 (\ref{difsub.eq13}). In this case Eq.
 (\ref{coneq.1}) is written
 \begin{equation}
 \label{dm:eq.9}
  \begin{split}
 & D_t(C^1) + D_x(C^2) =  u_t u_{xt} - \frac{1}{2}\,u^2 u_x
 u_{xt}\\[1ex]
 & - \left(u + \frac{1}{2}\,u^2 u_{xx} + u u_x^2 \right) u_t
  + \frac{1}{2}\,u^3 u_x + \frac{1}{4}\,u^4 u_x u_{xx}
 + \frac{1}{2}\,u^3 u_x^3.
 \end{split}
 \end{equation}
 \end{exer}
  \begin{exer}
 \label{dm:stratified}
 Consider  the Boussinesq equations (\ref{dm:eq.10}).
Taking its formal Lagrangian
  \begin{align}
 {\cal L} & = \omega \left[\Delta \psi_t - g \rho_x - f v_z -
 \psi_x \Delta \psi_z + \psi_z \Delta \psi_x\right]\notag\\[1.5ex]
 & + \mu \left[v_t + f \psi_z - \psi_x v_z + \psi_z v_x\right]
 + r \left[\rho_t + (N^2/g)\, \psi_x - \psi_x \rho_z + \psi_z
 \rho_x\right],\notag
 \end{align}
 where $\omega, \mu, r$ are new dependent variables, we obtain the adjoint system to Eqs.
 (\ref{dm:eq.10}):
 \begin{equation}
 \label{dm:eq.12}
 \frac{\delta {\cal L}}{\delta \psi}= 0, \quad \frac{\delta {\cal L}}{\delta v}= 0,
 \quad \frac{\delta {\cal L}}{\delta \rho} = 0.
 \end{equation}
 It is shown in \cite{ibr-ibr09b} that the system (\ref{dm:eq.10})
 is self-adjoint. Namely, the substitution
 \begin{equation}
 \label{dm:eq.13}
 \omega = \psi, \quad \mu = - v, \quad r = - (g^2/N^2)\,\rho
 \end{equation}
 maps the adjoint system (\ref{dm:eq.12}) into the system
 (\ref{dm:eq.10}). Using the self-adjointness,  nontrivial conservation laws
 were constructed via Theorem \ref{main.thm}.
 Apply the direct method to the system (\ref{dm:eq.10}). Note that knowledge of the substitution (\ref{dm:eq.13})
 gives the following equation Eq. (\ref{coneq.1}):
 \begin{equation}
 \label{dm:eq.14}
  \begin{split}
 & D_t(C^1) + D_x(C^2) + D_z(C^3)\\[1ex]
 & = \psi \big[\psi_{txx} + \psi_{tzz} -  g \rho_x -  f v_z
  - \psi_x \big(\psi_{zxx} + \psi_{zzz}\big)
 + \psi_z \big(\psi_{xxx} + \psi_{xzz}\big)\big]\\[.5ex]
 & - v \left[v_t + f \psi_z - \psi_x v_z + \psi_z v_x\right]
 - \frac{g^2}{N^2}\,\rho \Big[\rho_t + \frac{N^2}{g}\, \psi_x - \psi_x \rho_z + \psi_z
 \rho_x\Big].
 \end{split}
 \end{equation}
 \end{exer}
 \begin{exa}
 \label{dm:exa2}
 Let us consider the conservation equation (\ref{gas.eq44}),
 $$
 D_t (C^1) + D_x (C^2) = \sigma (\rho_t + v  \rho_x + \rho v_x),
 $$
 where $\sigma$ is connected with the velocity $v$ and the pressure
 $p$ of the Chaplygin gas by Eqs. (\ref{gas.eq21}),
 $$
 \sigma_x = - \frac{1}{p}\,, \quad \sigma_t + v \sigma_x = 0.
 $$
 In this example
Eqs. (\ref{dm:eq.2}) are not satisfied. Indeed, we have
 \begin{align}
 & \frac{\delta}{\delta v} \left[\sigma (\rho_t + v  \rho_x + \rho
 v_x)\right] = \sigma \rho_x - D_x (\sigma \rho) = - \rho \sigma
= \rho \int \frac{dx}{p} \not= 0,\notag\\[1ex]
 & \frac{\delta}{\delta \rho} \left[\sigma (\rho_t + v  \rho_x + \rho
 v_x)\right] = \sigma_t - D_x (\sigma v) + \sigma v_x = - (\sigma_t + v \sigma_x) = 0,\notag\\[1ex]
 & \frac{\delta}{\delta p} \left[\sigma (\rho_t + v  \rho_x + \rho
 v_x)\right] = 0.\notag
 \end{align}
 \end{exa}
 \begin{exa}
 \label{dm:exa3}
 Let us consider the conservation equation (\ref{gas.eq49}),
 $$
 D_t (C^1) + D_x (C^2) = t \rho \left( v_t +  v v_x +\frac{1}{\rho}\, p_x\right) + t v (\rho_t + v  \rho_x + \rho v_x).
 $$
 Here Eqs. (\ref{dm:eq.2}) are not satisfied. Namely, writing
 $$
 t \rho \left( v_t +  v v_x +\frac{1}{\rho}\, p_x\right) + t v (\rho_t + v  \rho_x + \rho v_x)
 = t \rho v_t + 2 t \rho v v_x + t p_x + t v \rho_t + t v^2  \rho_x
 $$
 we obtain:
 \begin{align}
 & \frac{\delta}{\delta v} \left[t \rho v_t + 2 t \rho v v_x + t p_x + t v \rho_t + t v^2  \rho_x\right] =  - \rho,\notag\\[1ex]
  & \frac{\delta}{\delta \rho}\left[t \rho v_t + 2 t \rho v v_x + t p_x + t v \rho_t + t v^2  \rho_x\right] =  - v,\notag\\[1ex]
 & \frac{\delta}{\delta p}\left[t \rho v_t + 2 t \rho v v_x + t p_x + t v \rho_t + t v^2  \rho_x\right] = 0.\notag
 \end{align}
 \end{exa}
  \begin{exer}
 \label{dm:exer2}
 Check if  Eqs. (\ref{dm:eq.2}) are satisfied for the conservation equations (\ref{gas.eq51}) and
 (\ref{gas.eq53}).\\
 \end{exer}

 \newpage
 \begin{center}
 \section*{{\sc Part 3}\protect\\ Utilization of conservation laws\\ \qquad for constructing solutions of PDEs}
 \end{center}
 \addcontentsline{toc}{chapter}{Part 3. Utilization of conservation laws for constructing solutions of PDEs}
 \label{uclsol}

 %\section{Preliminaries}
 \section{General discussion of the method}
 %\section{Utilization of an operator identity}
 \label{uclsol.1}
 \setcounter{equation}{0}

 As mentioned in Section \ref{int.f}, one can integrate or reduce the order of
 linear ordinary differential equations by rewriting them in a
 conservation form (\ref{opid.eq38}). Likewise one can
 integrate or reduce the order of a  nonlinear ordinary differential
equation as well as a system of ordinary differential
 equations using their  conservation laws. Namely,
a conservation law
 \begin{equation}
 \label{uclsol.1:eq1}
 D_x \left(\psi (x, y, y', \ldots, y^{(s-1)})\right) = 0
 \end{equation}
 for a nonlinear ordinary differential equation
 \begin{equation}
 \label{uclsol.1:eq2}
 F(x, y, y',\ldots, y^{(s)}) = 0
 \end{equation}
 yields the first integral
 \begin{equation}
 \label{uclsol.1:eq3}
 \psi (x, y, y', \ldots, y^{(s-1)}) = C_1.
 \end{equation}

 We will discuss now an extension of this idea to partial differential equations. Namely,
 we will apply conservation laws for constructing particular
 exact solutions of systems of partial differential equations.
 Detailed calculations are given in examples considered in the next
sections.

 Let us assume that the system (\ref{main:eq.1}),
 \begin{equation}
 \label{uclsol.1:eq4}
 F_{\bar \alpha} \left(x, u, u_{(1)}, \ldots, u_{(s)}\right) = 0, \quad
 \bar \alpha = 1, \ldots, \overline m,
 \end{equation}
 has a conservation law (\ref{main:eq.4}),
 \begin{equation}
 \label{uclsol.1:eq5}
 \left[D_i (C^i)\right]_{(\ref{uclsol.1:eq4})} = 0,
 \end{equation}
 with a known conserved vector
 \begin{equation}
 \label{uclsol.1:eq6}
  C = \left(C^1, \ \ldots\,, \ C^n\right),
 \end{equation}
 where
 $$
 C^i = C^i\left(x, u, u_{(1)}, \ldots \right), \quad i = 1, \ldots, n.
 $$
 We write the conservation equation (\ref{uclsol.1:eq5}) in the form
 (\ref{coneq.1}),
 \begin{equation}
 \label{uclsol.1:eq7}
 D_i (C^i)  = \mu^{\bar \alpha} F_{\bar \alpha} \big(x, u, u_{(1)}, \ldots,
 u_{(s)}\big).
 \end{equation}
 For a given conserved vector (\ref{uclsol.1:eq6}) the coefficients \ $\mu^{\bar \alpha}$ in
 Eq. (\ref{uclsol.1:eq7}) are known functions \ $\mu^{\bar \alpha} = \mu^{\bar \alpha}
(x, u, u_{(1)}, \ldots).$

 We will construct particular solutions of
 the system (\ref{uclsol.1:eq4}) by requiring that \textit{on these
 solutions the vector} (\ref{uclsol.1:eq6}) \textit{reduces to the following
 trivial conserved vector}\,:
 \begin{equation}
 \label{uclsol.1:eq8}
  C = \left(C^1 (x^2, \ldots, x^n), \ \ldots\,, \ C^n (x^1, \ldots, x^{n-1})\right).
 \end{equation}
 In other words, we look for particular solutions of the system (\ref{uclsol.1:eq4}) by adding to
 Eqs. (\ref{uclsol.1:eq4}) the \textit{differential constraints}
 \begin{equation}
 \label{uclsol.1:eq9}
 \begin{split}
 & C^1\left(x, u, u_{(1)}, \ldots \right) = h^1 (x^2, x^3, \ldots, x^n), \\
 & C^2\left(x, u, u_{(1)}, \ldots \right) = h^2 (x^1, x^3, \ldots, x^n), \\
 & \cdots \cdots \cdots \cdots \cdots \cdots \cdots \cdots \cdots \cdots \cdots \cdots\\
 & C^n\left(x, u, u_{(1)}, \ldots \right) = h^n (x^1, \ldots, x^{n-1}),
 \end{split}
 \end{equation}
 where $C^i\left(x, u, u_{(1)}, \ldots \right)$ are the components
 of the known conserved vector (\ref{uclsol.1:eq6}). Due to the
constraints (\ref{uclsol.1:eq9}), the left-hand side of Eq.
(\ref{uclsol.1:eq7}) vanishes identically. Hence the number of
 equations in the system (\ref{uclsol.1:eq4}) will be reduced by one.

 The differential constraints (\ref{uclsol.1:eq9}) can be equivalently
written as follows:
 \begin{equation}
 \label{uclsol.1:eq10}
 \begin{split}
 & D_1\left[C^1\left(x, u, u_{(1)}, \ldots \right)\right] = 0, \\
 & D_2\left[C^2\left(x, u, u_{(1)}, \ldots \right)\right] = 0, \\
 & \cdots \cdots \cdots \cdots \cdots \cdots \cdots \cdots \cdot\\
 & D_n\left[C^n\left(x, u, u_{(1)}, \ldots \right)\right] = 0.
 \end{split}
 \end{equation}
 \begin{rem}
 \label{uclsol.rem1}
 The overdetermined system of $\overline m + n$ equations (\ref{uclsol.1:eq4}),
 (\ref{uclsol.1:eq10}) reduces to $\overline m + n - 1$ equations
 due to the conservation law (\ref{uclsol.1:eq5}).
 \end{rem}

 \section{Application to the Chaplygin gas}
 %\section{Discussion of the operator identity}
 %\section{Utilization of an operator identity}
 \label{uclsol.2}
 \setcounter{equation}{0}

 \subsection{Detailed discussion of one case}

 Let us  apply the method to the Chaplygin gas equations (\ref{gas.eq15}),
 \begin{align}
 & v_t +  v v_x +\frac{1}{\rho}\, p_x =0,\notag\\
 & \rho_t+ v \rho_x +\rho v_x =0, \label{uclsol.2:eq1}\\
 & p_t + v p_x - p v_x =0.\notag
 \end{align}
We will construct a particular solution of the system
(\ref{uclsol.2:eq1}) using the simplest conservation law
(\ref{gas.eq28}),
 \begin{equation}
 \label{uclsol.2:eq2}
 D_t (\rho) + D_x(\rho v) =\rho_t+ v \rho_x +\rho v_x.
 \end{equation}
 The conservation equation (\ref{uclsol.2:eq2}) is written in the form
 (\ref{uclsol.1:eq7}) with the following conserved vector
 (\ref{uclsol.1:eq6}):
 \begin{equation}
 \label{uclsol.2:eq3}
 C^1 = \rho, \quad  C^2 = \rho v.
 \end{equation}
 The differential constraints
 (\ref{uclsol.1:eq9}) are written as follows:
 \begin{equation}
 \label{uclsol.2:eq4}
 \rho = g (x), \quad \rho v = h (t).
 \end{equation}
Thus we look for solutions of the form
 \begin{equation}
 \label{uclsol.2:eq5}
 \rho = g (x), \quad v = \frac{h (t)}{g (x)}\,\cdot
 \end{equation}
  The functions (\ref{uclsol.2:eq5}) solve the second equation in (\ref{uclsol.2:eq1})
 because the conservation law (\ref{uclsol.2:eq2}) coincides with the second equation
(\ref{uclsol.2:eq1}) (see Remark \ref{uclsol.rem1}). Therefore it
remains to substitute (\ref{uclsol.2:eq5}) in the first and third
equations of the system (\ref{uclsol.2:eq1}). The result of this
substitution can be  solved for the derivatives of $p:$
 \begin{equation}
 \label{uclsol.2:eq6}
 \begin{split}
 & p_x = - h' + \frac{h^2 g'}{g^2}\,,\\[1ex]
 & p_t = - \frac{h g'}{g^2}\,p + \frac{h h'}{g} - \frac{h^3 g'}{g^3}\,\cdot
 \end{split}
 \end{equation}
 The compatibility condition $p_{xt} = p_{tx}$ of the system
 (\ref{uclsol.2:eq6}) gives the equation
 \begin{equation}
 \label{uclsol.2:eq7}
 \left(g'' - 2\, \frac{g'^2}{g}\right) p = g^2\,\frac{h''}{h} - 2 g'h'
  - h^2\,\frac{ g''}{g} + 2 h^2\, \frac{g'^2}{g^2}\,\cdot
 \end{equation}
 For illustration purposes I will simplify further calculations by
 considering the particular case when the coefficient for $p$ in Eq.
 (\ref{uclsol.2:eq6}) vanishes:
 \begin{equation}
 \label{uclsol.2:eq8}
 g'' - 2\, \frac{g'^2}{g} = 0.
 \end{equation}
 The solution of Eq. (\ref{uclsol.2:eq8})  is
 \begin{equation}
 \label{uclsol.2:eq9}
  g (x) = \frac{1}{ax + b}\,, \quad a, b = {\rm const.}
 \end{equation}
 Substituting (\ref{uclsol.2:eq9}) in Eq. (\ref{uclsol.2:eq7}) we
 obtain
 \begin{equation}
 \label{uclsol.2:eq10}
 h'' + 2 a h h' = 0,
 \end{equation}
 whence
 \begin{equation}
 \label{uclsol.2:eq11}
 h (t) = k \tan (c - a k t)
 \end{equation}
if $a \not= 0,$ and
 \begin{equation}
 \label{uclsol.2:eq12}
 h (t) = A\, t  + B
 \end{equation}
 if $a = 0.$

If the constant $a$ in (\ref{uclsol.2:eq9}) does not vanish, we
substitute (\ref{uclsol.2:eq9}) and (\ref{uclsol.2:eq11}) in Eqs.
(\ref{uclsol.2:eq6}), integrate them and obtain
 \begin{equation}
 \label{uclsol.2:eq13}
  p = k^2 (a x + b) + Q \cos (c - a k t), \quad Q = {\rm
const.}
 \end{equation}
In the case  $a = 0$ the similar calculations yield
 \begin{equation}
 \label{uclsol.2:eq14}
  p = - A x + \frac{b}{2}\,A^2 t^2 + ABb t + Q, \quad Q = {\rm
const.}
 \end{equation}

 Thus, using the conservation law (\ref{uclsol.2:eq2}) we have
arrived at the solutions
 \begin{equation}
 \label{uclsol.2:eq15}
 \begin{split}
 & \rho = \frac{1}{ax + b}\,, \\
 & v = k (ax + b) \tan (c - a k t), \\
 & p = k^2 (a x + b) + Q \cos (c - a k t)
 \end{split}
 \end{equation}
and
 \begin{equation}
 \label{uclsol.2:eq16}
 \begin{split}
 & \rho = \frac{1}{b}\,, \\
 & v =  b  (A\, t  + B), \\
 & p = - A x + \frac{b}{2}\,A^2 t^2 + ABb t + Q.
 \end{split}
 \end{equation}

 \subsection{Differential constraints provided by other conserved vectors}

 The conservation laws (\ref{gas.eq29})-(\ref{gas.eq31}) give the
following differential constraints (\ref{uclsol.1:eq9}):
 \begin{align}
 & \rho v^2 - p = g(x), \quad \ \  p v + \rho v^3 = h(t),
\label{uclsol.2:eq17}\\[1ex]
 & \rho v = g(x), \qquad \quad \ \ \ p + \rho v^2 = h(t), \label{uclsol.2:eq18}\\[1ex]
 & t \rho v - x \rho= g(x), \quad t p + t \rho v^2 - x \rho v = h(t). \label{uclsol.2:eq19}
\end{align}

 The nonlocal conserved vectors (\ref{gas.eq48}), (\ref{gas.eq50})
 and (\ref{gas.eq52}) lead to the
following differential constraints (\ref{uclsol.1:eq9}):
 \begin{align}
 &  t \rho v + \tau = g(x), \qquad \quad p + \rho v^2 = h(t),
 \label{uclsol.2:eq20}\\[1ex]
 & t \rho = g(x), \qquad \qquad \quad t \rho v - \tau = h(t), \label{uclsol.2:eq21}\\[1ex]
 &\left(\frac{t^2}{2} - s\right)\rho = g(x), \quad \left(\frac{t^2}{2} -
s\right)\rho v -  t \tau = h(t). \label{uclsol.2:eq22}
\end{align}
 The constraints (\ref{uclsol.2:eq20}) are not essentially different from
the constraints (\ref{uclsol.2:eq18}). It is manifest if we write
them in the form (\ref{uclsol.1:eq10}).

 \section{Application to nonlinear equation describing an irrigation system}
 %\section{Application to the model of an irrigation system}
 %\section{Discussion of the operator identity}
 %\section{Utilization of an operator identity}
 \label{uclsol.3}
 \setcounter{equation}{0}

 The method of Section \ref{uclsol.1} can be used for constructing
particular  solutions not only of a system, but of a single partial
differential equations as well.

 Let us consider the nonlinear equation (\ref{irrig.eq1}),
 \begin{equation}
 \label{uclsol.3:eq1}
 C(\psi) \psi_t  = \left[K(\psi)\psi_{x}\right]_x
 + \left[K(\psi) \left( \psi_z - 1 \right)\right]_z - S(\psi),
\end{equation}
 satisfying the nonlinear self-adjointness condition
 (\ref{irrig.eq3}),
 \begin{equation}
 \label{uclsol.3:eq2}
 S\,'(\psi)  = a C(\psi), \quad a = {\rm const.}
\end{equation}
 and apply the method of Section \ref{uclsol.1} to the conserved
 vector (\ref{irrig:cl.eq11}),
  \begin{equation}
 \label{uclsol.3:eq3}
 C^1 = S(\psi) {\rm e}^{a t},\quad C^2 = a K(\psi)\psi_x {\rm e}^{a t},
 \quad C^3 = a K(\psi) (\psi_z - 1){\rm e}^{a t}.
 \end{equation}
 The conditions (\ref{uclsol.1:eq9}) are written:
 $$
 S(\psi) {\rm e}^{a t} = f(x, z), \quad a K(\psi)\psi_x {\rm e}^{a t} = g(t, z),
 \quad a K(\psi) (\psi_z - 1){\rm e}^{a t} = h(t, x).
 $$
 These conditions mean that the left-hand sides of the first, second and third  equation do not depend on
 $t,x$ and $z,$ respectively. Therefore they can be equivalently written as the following
 differential constraints (see Eqs. (\ref{uclsol.1:eq10})):
 \begin{equation}
 \label{uclsol.3:eq4}
 \begin{split}
 & a S(\psi) + S'(\psi) \psi_t = 0,\\[1ex]
 & \left[K(\psi)\psi_{x}\right]_x = 0, \\[1ex]
 & \left[K(\psi) \left( \psi_z - 1 \right)\right]_z = 0.
 \end{split}
 \end{equation}
 The constraints (\ref{uclsol.3:eq4}) reduce Eq. (\ref{uclsol.3:eq1}) to
 Eq. (\ref{uclsol.3:eq2}). Hence, the particular solutions of Eq. (\ref{uclsol.3:eq1})
 provided by the conserved vector (\ref{uclsol.3:eq3}) are described
by the system
 \begin{equation}
 \label{uclsol.3:eq5}
 \begin{split}
 & a C(\psi) - S\,'(\psi) = 0,\\[1ex]
 & a S(\psi) + S'(\psi) \psi_t = 0,\\[1ex]
 & \left[K(\psi)\psi_{x}\right]_x = 0, \\[1ex]
 & \left[K(\psi) \left( \psi_z - 1 \right)\right]_z = 0.
 \end{split}
 \end{equation}

 \newpage
 \begin{center}
 \section*{{\sc Part 4}\protect\\ Approximate self-adjointness and approximate conservation laws}
 \end{center}
 \addcontentsline{toc}{chapter}{Part 4. Approximate self-adjointness and approximate conservation laws}
 \label{apprself}

 The methods developed in this paper can be extended to differential equations with a small parameter
 in order to construct approximate conservation laws using approximate
 symmetries.  I will illustrate this possibility by examples. The reader interested in approximate symmetries can
 find enough material in \cite{ibr94-96}, vol. 3, Chapters 2 and 9. A brief
 introduction to the subject can be found also in \cite{ibr-kov09}.\\

 \section{The van der Pol equation}
 \label{apprself.1}
 \setcounter{equation}{0}

 The van der Pol equation has the form
 \begin{equation}
 \label{apprself.1:eq1}
 F \equiv y'' + y + \varepsilon (y'^3- y') = 0, \quad \varepsilon = {\rm
 const. \not= 0.}
 \end{equation}

 \subsection{Approximately adjoint equation}
 \label{apprself.1.1}

 We have:
 $$
 \frac{\delta}{\delta y}\left\{z \left[y'' + y + \varepsilon \left(y'^3 - y'\right)\right]\right\} =
 z'' + z + \varepsilon D_x\left(z - 3 z y'^2\right).
 $$
 Thus, the adjoint equation to the van der Pol equation is
 $$
 F^* \equiv z'' + z + \varepsilon \left(z' - 3 z' y'^2 - 6 z y' y''\right) = 0.
 $$
 We eliminate here $y''$ by using Eq. (\ref{apprself.1:eq1}),
  consider $\varepsilon$ as a small parameter and write $F^*$ in the first order
  of precision with respect to $\varepsilon.$ In other words, we write
 \begin{equation}
 \label{apprself.1:eq2}
   y'' \approx - y.
 \end{equation}
   Then we obtain the following \textit{approximately adjoint equation}
  to Eq. (\ref{apprself.1:eq1}):
 \begin{equation}
 \label{apprself.1:eq3}
 F^* \equiv z'' + z + \varepsilon \left(z' - 3 z' y'^2 + 6 z y y'\right) = 0.
 \end{equation}

 \subsection{Approximate self-adjointness}
 \label{apprself.1.2}

 Let us investigate Eq. (\ref{apprself.1:eq1}) for approximate
 self-adjointness. Specifically, I will call Eq. (\ref{apprself.1:eq1})
 \textit{approximately self-adjoint} if there exists a non-trivial (not vanishing identically) approximate substitution
 \begin{equation}
 \label{apprself.1.2:eq1}
  z \approx f(x, y, y') + \varepsilon g(x, y, y')
 \end{equation}
 such that $F$ given by Eq. (\ref{apprself.1:eq1}) and $F^*$ defined by Eq. (\ref{apprself.1:eq3})
 approximately satisfy the condition (\ref{gsa:eq3}) of
 nonlinear self-adjointness. In other words, the following equation
 is satisfied  in the first-order of precision in $\varepsilon:$
 \begin{equation}
 \label{apprself.1.2:eq2}
 F^*\big|_{z = f + \varepsilon g} = \lambda F.
 \end{equation}

 Note, that the unperturbed equation $y'' + y = 0$ is nonlinearly
 self-adjoint. Namely it coincides with the adjoint equation $z'' + z = 0$
 upon the substitution
 \begin{equation}
 \label{apprself.1.2:eq3}
  z = \alpha y + \beta \cos x + \gamma \sin x, \quad \alpha, \beta, \gamma = {\rm
  const.}
 \end{equation}
 Therefore we will consider the substitution
 (\ref{apprself.1.2:eq1}) of the following restricted form:
 \begin{equation}
 \label{apprself.1.2:eq4}
 z \approx f(x, y) + \varepsilon g(x, y, y').
 \end{equation}
 In differentiating $g(x, y, y')$ we will use Eq.
 (\ref{apprself.1:eq2}) because we make out calculations in the first order
 of precision with respect to $\varepsilon.$
 Then  we obtain:
 \begin{equation}
 \label{apprself.1.2:eq5}
 \begin{split}
  z' & = D_x(f) + \varepsilon D_x(g)\big|_{y'' = - y} \equiv  f_x + y' f_y + \varepsilon (g_x + y' g_y - y g_{y'}), \\[1ex]
  z'' & = D_x^2(f) + \varepsilon D_x^2(g)\big|_{y'' = - y} \equiv f_{xx} + 2 y' f_{xy} + y'^2 f_{yy} + y'' f_y \\[1ex]
   & +
  \varepsilon (g_{xx} + 2 y' g_{xy} - 2 y g _{x y'} + y'^2 g_{yy}
   - 2 y y' g _{y y'} + y^2 g _{y' y'} - y g_y - y' g_{y'}).
 \end{split}
 \end{equation}
 Substituting (\ref{apprself.1.2:eq5}) in (\ref{apprself.1:eq3}) and
 solving Eq. (\ref{apprself.1.2:eq2}) with $\varepsilon = 0$ we
 see that $f$ is given by Eq. (\ref{apprself.1.2:eq3}). Then $\lambda = C$ and
 the terms with $\varepsilon$ in Eq. (\ref{apprself.1.2:eq2}) give
 the following second-order linear partial differential equation for
 $g(x, y, y'):$
 \begin{equation}
 \label{apprself.1.2:eq7}
 \begin{split}
  & g + D_x^2(g)\big|_{y'' = - y}
  = \alpha \left(4 y'^3 - 6 y^2 y' - 2 y'\right) \\[1ex]
  & + \beta \left(\sin x - 3 y'^2 \sin x - 6 y y' \cos x \right)
  + \gamma \left(3 y'^2 \cos x - \cos x - 6 y y' \sin x \right).
 \end{split}
 \end{equation}

 The standard existence theorem guarantees that  Eq. (\ref{apprself.1.2:eq7})
 has a solution. It is manifest that the solution does not vanish because $g = 0$ does not satisfy Eq.
 (\ref{apprself.1.2:eq7}). We conclude that the van der  Pol
 equation (\ref{apprself.1:eq1}) with a small parameter
 $\varepsilon$ is approximately self-adjoint. The substitution (\ref{apprself.1.2:eq4})
 satisfying the approximate self-adjointness
 condition (\ref{apprself.1.2:eq2}) has the form
 \begin{equation}
 \label{apprself.1.2:eq8}
 z \approx \alpha y + \beta \cos x + \gamma \sin x + \varepsilon g(x, y, y'),
 \end{equation}
 where $\alpha, \ \beta, \ \gamma$ are arbitrary constants and $g (x, y, y')$ solves Eq.
 (\ref{apprself.1.2:eq7}).

% \newpage

 \subsection{Exact and approximate symmetries}
 \label{apprself.1.3}

 If $\varepsilon$ is treated as an arbitrary constant, Eq. (\ref{apprself.1:eq1}) has only one
 point symmetry, namely the one-parameter group of translations of the
 independent variable $x.$ We will write the generator $X_1 = \partial/\partial x$ of this
 group in the form (\ref{opid.eq14}):
 \begin{equation}
 \label{apprself.1.3:eq1}
 X_1 = y' \frac{\partial}{\partial y}\,\cdot
 \end{equation}
 If $\varepsilon$ is a small parameter, then Eq. (\ref{apprself.1:eq1})
 has, along with the exact symmetry (\ref{apprself.1.3:eq1}),
  the following 7 approximate symmetries (\cite{ibr94-96}, vol. 3, Section
 9.1.3.3):
 \begin{equation}
 \label{apprself.1.3:eq2}
 \begin{split}
  X_2 & = \left\{4 y - \varepsilon \left[y^2 y' + 3 x y \left(y^2
  + y'^2\right)\right]\right\} \frac{\partial}{\partial
  y}\,,\\[.1ex]
  X_3 & = \left\{8 \cos x + \varepsilon \left[\left(4 - 3 y'^2 - 9 y^2\right) x \cos x
  + 3 (x y^2)' \sin x \right]\right\} \frac{\partial}{\partial
  y}\,,\\[1ex]
   X_4 & = \left\{8 \sin x + \varepsilon \left[\left(4 - 3 y'^2 - 9 y^2\right) x \sin x
  - 3 (x y^2)' \cos x \right]\right\} \frac{\partial}{\partial y}\,,\\[1ex]
  X_5 & = \big\{24 y^2\cos x - 24 y y' \sin x + \varepsilon \big[\big(12 yy' + 9 y y'^3
  + 9 y^3 y'\big) x \sin x\\[.5ex]
 &  + \big(12 y^2 - 9 y^2 y'^2 - 6 y^4\big) \sin x  - \big(12 y^2 - 9 y^2 y'^2 - 9 y^4\big) x \cos
 x\\[.5ex]
  & - 3 y^3 y' \cos x \big]\big\} \frac{\partial}{\partial y}\,,\\[1ex]
  X_6 & = \big\{24 y^2 \sin x + 24 y y' \cos x - \varepsilon \big[\big(12 yy' + 9 y y'^3
  + 9 y^3 y'\big) x \cos x\\[.5ex]
 &  + \big(12 y^2 - 9 y^2 y'^2 - 6 y^4\big) \cos x  + \big(12 y^2 + 9 y^2 y'^2
 + 9 y^4\big) x \sin x\\[.5ex]
  & + 3 y^3 y' \sin x \big]\big\} \frac{\partial}{\partial y}\,,\\[1ex]
  X_7 & = \big\{4 y \cos 2 x - 4 y' \sin 2 x + \varepsilon \big[3 \big(yy'^2 - y^3\big) x \cos 2 x\\[.5ex]
 &  - 3 y^2 y' \cos 2 x  + 6 y^2 y' x \sin 2 x + 2 (y - y^3) \sin 2 x \big]\big\} \frac{\partial}{\partial y}\,,\\[1ex]
  X_8 & = \big\{4 y \sin 2 x + 4 y' \cos 2 x - \varepsilon \big[3 \big(y^3 -yy'^2\big) x \sin 2 x\\[.5ex]
 &  + 3 y^2 y' \sin 2 x  + 6 y^2 y' x \cos 2 x + 2 (y - y^3) \cos 2 x \big]\big\} \frac{\partial}{\partial
 y}\,\cdot
 \end{split}
 \end{equation}

 \subsection{Approximate conservation laws}
 \label{apprself.1.4}

 We can construct now approximate
 conserved quantities for the van der Pol equation using the formula
 (\ref{main:eq.5}) and the approximate substitution (\ref{apprself.1.2:eq8}).
 Inserting in (\ref{main:eq.5}) the formal Lagrangian
 $${\cal L} = z \left[y'' + y + \varepsilon \left(y'^3 -
 y'\right)\right]$$ we obtain
 \begin{equation}
 \label{apprself.1.4:eq1}
 C = W \left[- z' + \varepsilon \left(3 y'^2 z - z\right)\right] + W' z.
 \end{equation}

 Let us calculate the conserved quantity (\ref{apprself.1.4:eq1}) for
 the operator $X_1$ given by Eq. (\ref{apprself.1.3:eq1}). In this
 case $W = y', \ W' = y'',$ and therefore (\ref{apprself.1.4:eq1})
 has the form
 $$
 C = - y' z' + \varepsilon \left(3 y'^3 - y'\right) z + y'' z.
 $$
 We eliminate here $y''$ via Eq. (\ref{apprself.1:eq1}),
  use the approximate substitution (\ref{apprself.1.2:eq8}) and
  and obtain (in the first order
  of precision with respect to $\varepsilon$) the following approximate  conserved quantity:
 \begin{equation}
 \label{apprself.1.4:eq2}
  \begin{split}
  C = & - \alpha \left(y^2 + y'^2\right) + \beta \left(y' \sin x - y \cos x\right)
  - \gamma \left(y' \cos x + y \sin x\right)\\[1.5ex]
  & + \varepsilon \left(2 \alpha y y'^3  + 2 \beta y'^3 \cos x
  + 2 \gamma y'^3 \sin x - y g - y' D_x(g)\big|_{y'' = - y} \right).
 \end{split}
 \end{equation}
 Differentiating it and using the equations (\ref{apprself.1:eq1})
 and (\ref{apprself.1:eq2}) we obtain
 \begin{equation}
 \label{apprself.1.4:eq3}
  \begin{split}
   D_x(C)& = \varepsilon y' \Big[\alpha \left(4 y'^3 - 6 y^2 y' - 2 y'\right)
  + \beta \left(\sin x - 3 y'^2 \sin x - 6 y y' \cos x \right)\\[1.5ex]
  & + \gamma \left(3 y'^2 \cos x - \cos x - 6 y y' \sin x \right) - g - D_x^2(g)\big|_{y'' = - y} \Big] +
  o(\varepsilon),
 \end{split}
 \end{equation}
 where $o(\varepsilon)$ denotes the higher-order terms in $\varepsilon.$ The equations (\ref{apprself.1.2:eq7}) and (\ref{apprself.1.4:eq3})
 show that the quantity (\ref{apprself.1.4:eq2}) satisfies the
 approximate conservation law
 \begin{equation}
 \label{apprself.1.4:eq4}
  D_x(C)\big|_{(\ref{apprself.1:eq1})} \approx 0.
 \end{equation}

 Let us consider the operator $X_2$ from  (\ref{apprself.1.3:eq2}). In this case we have
 \begin{equation}
 \label{apprself.1.4:eq5}
  \begin{split}
 &  W = 4 y - \varepsilon \left[y^2 y' + 3 x y \left(y^2 +
 y'^2\right)\right],\\[1ex]
 &  W' \approx 4 y' - \varepsilon \left[2 y^3 + 5 y y'^2 + 3 x \left(y^2 y' +
  y'^3\right)\right].
 \end{split}
 \end{equation}
Proceeding as above we obtain the following approximate  conserved
quantity:
 \begin{equation}
 \label{apprself.1.4:eq6}
  \begin{split}
  C & =  4 y' (\beta \cos x + \gamma \sin x) -
  4 y (\gamma \cos x - \beta \sin x)\\[1.5ex]
  & + \varepsilon \Big\{2 \alpha y^2 \left(4 y'^2 - y^2 - 2\right) + 4 y' g - 4 y D_x(g)\big|_{y'' = - y}\\[1.5ex]
  & + \left[7 yy'^2 - 3 xy'(y^2 + y'^2) - 2 y^3 - 4 y \right](\beta \cos x + \gamma \sin x)\\[1.5ex]
  & + \left[y^2 y' + 3 xy(y^2 + y'^2)\right](\gamma \cos x - \beta \sin x) \Big\}.
 \end{split}
 \end{equation}
 The calculation shows that the quantity (\ref{apprself.1.4:eq6}) satisfies the
 approximate conservation law (\ref{apprself.1.4:eq4})
  in the following form:
  \begin{align}
 \label{apprself.1.4:eq7}
   D_x(C)& = 4 (\beta \cos x + \gamma \sin x)\left[y'' + y + \varepsilon \left(y'^3 -
  y'\right)\right]\notag\\[1.5ex]
   & + 4 \varepsilon y \Big[\alpha \left(4 y'^3 - 6 y^2 y' - 2 y'\right)
  + \beta \left(\sin x - 3 y'^2 \sin x - 6 y y' \cos x \right)\notag\\[1.5ex]
  & + \gamma \left(3 y'^2 \cos x - \cos x - 6 y y' \sin x \right) - g - D_x^2(g)\big|_{y'' = - y} \Big] + o(\varepsilon).
 \end{align}

 Continuing this procedure, one can construct approximate
 conservation laws for the remaining approximate symmetries
 (\ref{apprself.1.3:eq2}).

 \section{Perturbed KdV equation}
 \label{apprself.2}
 \setcounter{equation}{0}

 Let us consider again the KdV equation (\ref{gsa:KdV.eq1}),
 \begin{equation}
 \tag{\ref{gsa:KdV.eq1}}
 \label{apprself.2:eq1}
 u_t = u_{xxx} + u u_x,
 \end{equation}
 and the following perturbed equation :
 \begin{equation}
 \label{apprself.2:eq2}
F \equiv u_t - u_{xxx} - u u_x - \varepsilon u = 0.
 \end{equation}
 We will follow the procedure described in Section \ref{apprself.1}.

 \subsection{Approximately adjoint equation}
 \label{apprself.2.1}

 Let us write the formal Lagrangian for Eq. (\ref{apprself.2:eq2})
 in the form
 \begin{equation}
 \label{apprself.2.1:eq1}
 {\cal L} = v\left[- u_t + u_{xxx} + u u_x + \varepsilon u\right].
 \end{equation}
 Then
 $$
\frac{\delta {\cal L}}{\delta u} = v_t - v_{xxx} - D_x(uv) + v u_x +
\varepsilon v = v_t - v_{xxx} - u v_x + \varepsilon v.
 $$
 Hence, the approximately adjoint equation to Eq. (\ref{apprself.2:eq2})
 has the form
 \begin{equation}
 \label{apprself.2.1:eq2}
 F^* \equiv v_t - v_{xxx} - u v_x + \varepsilon v = 0.
 \end{equation}

 \subsection{Approximate self-adjointness}
 \label{apprself.2.2}

 As mentioned in Section \ref{gsa:1}, Example \ref{gsa:KdV}, the KdV equation
 (\ref{apprself.2:eq1}) is nonlinearly self-adjoint with the
 substitution (\ref{gsa:KdV.eq3}),
 \begin{equation}
 \tag{\ref{gsa:KdV.eq3}}
v =  A_1 + A_2 u + A_3 (x + t u).
 \end{equation}
 Therefore in the case of the perturbed equation  (\ref{apprself.2:eq2})
 we look for the substitution
 $$
 v  = \phi(t, x, u)+ \varepsilon \psi (t, x, u),
 $$
 satisfying the nonlinear self-adjointness condition
 \begin{equation}
 \label{apprself.2.2:eq1}
 F^*\big|_{v = \phi + \varepsilon \psi} = \lambda F
 \end{equation}
in the first-order of precision in $\varepsilon,$ in the following
form:
 \begin{equation}
 \label{apprself.2.2:eq2}
 v  = A_1 + A_2 u + A_3 (x + t u) + \varepsilon \psi (t, x, u).
 \end{equation}
When we substitute the expression (\ref{apprself.2.2:eq2}) in the
definition (\ref{apprself.2.1:eq2}) of $F^*,$  the terms without
$\varepsilon$ in Eq. (\ref{apprself.2.2:eq1}) disappear by
construction of the
 substitution (\ref{gsa:KdV.eq3}) and give $\lambda = A_2 + A_3
t.$ Then we write Eq. (\ref{apprself.2.2:eq1}), rearranging the
terms, in the form
 \begin{align}
 \label{apprself.2.2:eq1A}
 &\varepsilon \psi_u [u_t - u_{xxx} - u u_x] - 3 \varepsilon u_{xx} [u_x \psi_{uu} +
 \psi_{xu}] - \varepsilon u_x [u_x^2 \psi_{uuu} + 3 u_x \psi_{xuu} + 3
 \psi_{xxu}]\notag\\[1ex]
 & + \varepsilon [\psi_t - \psi_{xxx} - u \psi_x + A_1 + A_2 u + A_3 (x + t u)] = - \varepsilon (A_2 + A_3 t)
 u.
 \end{align}
 In view Eq. (\ref{apprself.2:eq2}), the first term in
 the first line of Eq. (\ref{apprself.2.2:eq1A}) is written $\varepsilon^2 u \psi_u.$
 Hence, this term vanishes in our approximation. The terms with
 $u_{xx}$ in the first line of Eq. (\ref{apprself.2.2:eq1A}) yield
 $$
 \psi_{uu} = 0, \quad \psi_{xu} = 0,
 $$
 whence $$\psi = f(t) u + g (t, x).$$ The third bracket in the first line of Eq.
 (\ref{apprself.2.2:eq1A}) vanishes, and Eq. (\ref{apprself.2.2:eq1A}) becomes
 $$
 [f'(t) - g_x (t, x)] u + g_t (t, x) - g_{xxx} (t, x) + 2 [A_2 + A_3 t] u + A_1 + A_3 x=
 0.
 $$
 After rather simple calculations we solve this equation and obtain
 $$
 g(t, x) = A_4 - A_1 t + (A_5 + 2 A_2 - A_3t)x, \quad f(t) = A_6 + A_5 t - \frac{3}{2}\,A_3 t^2.
 $$

 We conclude that the perturbed KdV equation  (\ref{apprself.2:eq2}) is approximately self-adjoint.
 The approximate substitution (\ref{apprself.2.2:eq2}) has the following form:
 \begin{align}
 \label{apprself.2.2:eq3}
  v & \approx A_1 + A_2 u + A_3 (x + t u) \\[1ex]
  & + \varepsilon \left[\left(A_6 + A_5 t - \frac{3}{2}\,A_3 t^2\right) u + A_4 - A_1 t + (A_5 + 2 A_2 - A_3t)x \right].\notag
 \end{align}

 \subsection{Approximate symmetries}
 \label{apprself.2.3}

 Recall that the Lie algebra of point symmetries of the KdV equation
 (\ref{apprself.2:eq1}) is spanned by the following operators:
 \begin{align}
 \label{apprself.2.3:eq1}
   & X_1 = \frac{\partial}{\partial t}\,, \quad X_2 = \frac{\partial}{\partial x}\,,
   \quad X_3 = t \frac{\partial}{\partial x} - \frac{\partial}{\partial u}\,,\\[1ex]
  & X_4 = 3 t \frac{\partial}{\partial t} + x \frac{\partial}{\partial x}
  - 2 u \frac{\partial}{\partial u}\,\cdot\notag
 \end{align}

 Following the method for calculating approximate symmetries and using the terminology presented in \cite{ibr94-96}, vol. 3, Chapter
 2, we can prove that all symmetries (\ref{apprself.2.3:eq1}) are
 stable. Namely the perturbed equation (\ref{apprself.2:eq2})
 inherits the symmetries (\ref{apprself.2.3:eq1}) of the KdV
 equation in the form of the following approximate symmetries:
  \begin{align}
 \label{apprself.2.3:eq2}
   & X_1 = \frac{\partial}{\partial t}\,, \quad X_2 = \frac{\partial}{\partial x}\,,
   \quad X_3 = t \frac{\partial}{\partial x} - \frac{\partial}{\partial u}
   + \varepsilon \left(\frac{1}{2}\,t^2 \frac{\partial}{\partial x} - t \frac{\partial}{\partial u}\right)\,,\\[1.5ex]
  & X_4 = 3 t \frac{\partial}{\partial t} + x \frac{\partial}{\partial x}
  - 2 u \frac{\partial}{\partial u}
  - \varepsilon \left[\frac{9}{2}\,t^2 \frac{\partial}{\partial t} +
  3 t x \frac{\partial}{\partial x} - (6 t u + 3 x)\frac{\partial}{\partial u}\right]\,\cdot\notag
 \end{align}

 \subsection{Approximate conservation laws}
 \label{apprself.2.4}

 We can construct now the approximate
 conservation laws
 \begin{equation}
 \label{apprself.2.4:eq1}
 \left[D_t (C^1) + D_x (C^2)\right]_{(\ref{apprself.2:eq2})} \approx 0
 \end{equation}
 for the perturbed KdV equation (\ref{apprself.2:eq2}) using its
 approximate symmetries (\ref{apprself.2.3:eq2}),
 the general formula
 (\ref{main:eq.5}) and the approximate substitution (\ref{apprself.2.2:eq3}).
 Inserting in (\ref{main:eq.5}) the formal Lagrangian (\ref{apprself.2.1:eq1})
 we obtain
 \begin{equation}
 \label{apprself.2.4:eq2}
 \begin{split}
 & C^1 = - W v,\\
 & C^2 = W \left[u v + v_{xx}\right] - v_x D_x(W) + v D_x^2(W).
 \end{split}
 \end{equation}

  I will calculate here the conserved vector (\ref{apprself.2.4:eq2}) for
 the operator $X_4$ from (\ref{apprself.2.3:eq2}). In this
 case we have
 \begin{equation}
 \label{apprself.2.4:eq3}
 W = - 2 u - 3 t u_t - x u_x + \varepsilon \left(6 t u + 3 x + \frac{9}{2}\,t^2 u_t + 3 t x u_x\right).
 \end{equation}
We further simplify the calculations by taking the particular
substitution (\ref{apprself.2.2:eq3}) with $A_2 = 1, \ A_1 = A_3 =
\cdots = A_6 = 0.$
 Then
 \begin{equation}
 \label{apprself.2.4:eq4}
 v = u + 2 \varepsilon x.
 \end{equation}
 Substituting (\ref{apprself.2.4:eq3}), (\ref{apprself.2.4:eq4}) in the first component of the vector (\ref{apprself.1.4:eq2})
 and then eliminating $u_t$ via Eq. (\ref{apprself.2:eq2}) we obtain:
 \begin{align}
  C^1 & \approx (2 u + 3 t u_t + x u_x)(u + 2 \varepsilon x)
  - \varepsilon \Big(6 t u + 3 x + \frac{9}{2}\,t^2 u_t + 3 t x u_x\Big) u\notag\\[1ex]
 & = 2 u^2 + 3 t u u_{xxx} + 3 t u^2 u_x + x u u_x + \varepsilon \Big(x u + 6 t x u_{xxx}
  + 3 t x u u_x \notag\\[1ex]
 & + 2 x^2 u_x - 3 t u^2 - \frac{9}{2}\,t^2 u u_{xxx}
 - \frac{9}{2}\,t^2 u^2 u_x \Big).\notag
 \end{align}
 Upon singling out the total derivatives in $x,$ it is written:
 \begin{align}
 \label{apprself.2.4:eq5}
  C^1 & \approx \frac{3}{2}\, u^2 - 3 \varepsilon \Big(x u + \frac{3}{2}\,t u^2\Big)  + D_x\Big[\frac{1}{2}\,x u^2
  + t u^3 - \frac{3}{2}\,t u_x^2 + 3 t u u_{xx}\quad \null\\[1ex]
  & + \varepsilon \Big(2 x^2 u  + \frac{3}{2}\, t x u^2 - \frac{3}{2}\,t^2 u^3
  - 6 t u_x  + 6 t x u_{xx} + \frac{9}{4}\,t^2 u_x^2
  - \frac{9}{2}\,t^2 u u_{xx}\Big)\Big].\qquad \null \notag
 \end{align}

Then we substitute (\ref{apprself.2.4:eq3}),
(\ref{apprself.2.4:eq4}) in the second component of the vector
(\ref{apprself.1.4:eq2}), transfer the term $D_x(\ldots)$ from $C^1$
to $C^2,$ multiply the resulting vector $(C^1, C^2)$ by 2/3 and
arrive at the following vector:
 \begin{align}
 \label{apprself.2.4:eq6}
 & C^1  =  u^2 - 2 \varepsilon \Big[x u + \frac{3}{2}\,t u^2\Big],\\[1ex]
 & C^2  =  u_x^2 - \frac{2}{3}\, u^3 - 2 u u_{xx}+ \varepsilon \left[x u^2 - 2 u_x
  + 2 x u_{xx} + 2 t u^3
  - 3 t u_x^2 + 6 t u u_{xx}\right].\qquad \null \notag
 \end{align}
 The approximate conservation law (\ref{apprself.2.4:eq1}) for the vector (\ref{apprself.2.4:eq6}) is
 satisfied in the following form:
  \begin{align}
 \left[D_t (C^1) + D_x (C^2)\right] & = 2 u (u_t - u_{xxx} - u u_x - \varepsilon u)\notag\\
  & - 2 \varepsilon (x + 3 t u)(u_t - u_{xxx} - u u_x) + o(\varepsilon).\notag
 \end{align}

 \textbf{ Acknowledgements}\\[3ex]
 I thank Sergey Svirshchevskii for fruitful discussion which
provided me with an incentive to write Section \ref{opid}.
 %regarding Section \ref{opid}.
 I am cordially grateful to my wife
 Raisa for reading the  manuscript carefully. Without her help and questions the
work would be less palatable and contain essentially more misprints.

% \begin{thebibliography}{99}

 \end{document}